\documentclass[12pt]{aastex631}
\usepackage{natbib}
\usepackage{amsmath}
\usepackage{verbatim}
\usepackage{amssymb}
\usepackage{hyperref}
\usepackage{float}
\usepackage{graphicx}
\usepackage{mathtools}
\usepackage{empheq}
\usepackage{appendix}
\usepackage{microtype}
\usepackage{commath}
\usepackage{bm}
\usepackage{bold-extra}

\DeclareGraphicsExtensions{.pdf,.png,.jpg}

\newcommand{\sy}{\scriptstyle}

\def\arcsec{\ensuremath{^{\prime\prime}}}

\newcommand{\bq}{\begin{equation}} 
\newcommand{\eq}{\end{equation}}   
 
\newcommand{\kms}{km s$^{-1}$ Mpc$^{-1}$}
\newcommand{\kmss }{km s$^{-1}$ Mpc$^{-1}$\ }
\newcommand{\hbase}{$ 73.04 \pm 1.01 $ km s$^{-1}$ Mpc$^{-1}$}
\newcommand{\hsbase}{$ 73.04 \pm  1.04  $ km s$^{-1}$ Mpc$^{-1}$}

\newcommand{\hsbaseq}{\bm{73.04 \pm 1.04} \textbf{\ km\ s}\bm{^{-1}}\ \textbf{Mpc}\bm{^{-1}}}
\newcommand{\chisqbase}{$1.03$}
\newcommand{\bslopebase}{$ -3.299 \pm 0.015$ mag/dex}
\newcommand{\gammabase}{$ -0.217 \pm 0.046$ mag/dex}
\newcommand{\mwbase}{$ -5.894 \pm 0.017$ mag}
\newcommand{\mbbase}{$ -19.253 \pm 0.027$ mag}
\newcommand{\hmaser}{$ 72.51 \pm 1.54 $ km s$^{-1}$ Mpc$^{-1}$}

\newcommand{\hwtrgb}{$ 72.76 \pm 0.95 $ km s$^{-1}$ Mpc$^{-1}$}

\newcommand{\hwtrgbcchp}{$ 72.29 \pm 0.94 $ km s$^{-1}$ Mpc$^{-1}$}

\newcommand{\hlow}{$ 71.93 \pm  1.19  $ km s$^{-1}$ Mpc$^{-1}$} 
\newcommand{\hhigh}{$ 74.85 \pm  2.33  $ km s$^{-1}$ Mpc$^{-1}$} 
\newcommand{\hopt}{$ 72.70 \pm  1.07  $ km s$^{-1}$ Mpc$^{-1}$} 
\newcommand{\hsopt}{$ 72.70 \pm  1.30  $ km s$^{-1}$ Mpc$^{-1}$} 
\newcommand{\hwtrgbmean}{$ 72.53 \pm 0.95 $ km s$^{-1}$ Mpc$^{-1}$}
\newcommand{\hswtrgbmean}{$ 72.53 \pm 0.99 $ km s$^{-1}$ Mpc$^{-1}$}
\newcommand{\hhiz}{$ 73.30 \pm 1.04 $ km s$^{-1}$ Mpc$^{-1}$}
\newcommand{\qhiz}{$ -0.51 \pm 0.024 $}
\newcommand{\faceplanck}{$ 5 $}

\newcommand{\beq}{\begin{equation}}
\newcommand{\eeq}{\end{equation}}
\newcommand{\beqa}{\begin{eqnarray}}
\newcommand{\eeqa}{\end{eqnarray}}
\newcommand{\om}{\Omega_m}

\newcommand{\PL}{$P$--$L$}
\newcommand{\PLs}{$P$--$L$\ }

\newcommand{\nd}{\multicolumn{1}{c}{$\dots$}}
\newcommand{\td}{..\!\!}
\newcommand{\oh}{{\rm [O/H]}}
\newcommand{\lp}{\log P}

\shorttitle{A Comprehensive Measurement of H$_0$ from SH0ES}
\shortauthors{Riess et al.}

\begin{document} 

\title{A Comprehensive Measurement of the Local Value of the Hubble Constant with 1 \kms $\,$  Uncertainty from the {\it Hubble Space Telescope} and the SH0ES Team} 

\author[0000-0002-6124-1196]{Adam G.~Riess}
\affiliation{Space Telescope Science Institute, 3700 San Martin Drive, Baltimore, MD 21218, USA}
\affiliation{Department of Physics and Astronomy, Johns Hopkins University, Baltimore, MD 21218, USA}

\author[0000-0001-9420-6525]{Wenlong Yuan}
\affiliation{Department of Physics and Astronomy, Johns Hopkins University, Baltimore, MD 21218, USA}

\author[0000-0002-1775-4859]{Lucas M.~Macri}
\affiliation{George P.\ and Cynthia W.\ Mitchell Institute for Fundamental Physics and Astronomy,\\ Department of Physics \& Astronomy, Texas A\&M University, College Station, TX 77843, USA}

\author[0000-0002-4934-5849]{Dan Scolnic}
\affiliation{Department of Physics, Duke University, Durham, NC 27708, USA}

\author[0000-0001-5201-8374]{Dillon Brout}
\affiliation{Center for Astrophysics, Harvard \& Smithsonian, 60 Garden St., Cambridge, MA 02138, USA}

\author{Stefano Casertano}
\affiliation{Space Telescope Science Institute, 3700 San Martin Drive, Baltimore, MD 21218, USA}

\author[0000-0002-6230-0151]{David O.~Jones}
\affiliation{Einstein Fellow, Department of Astronomy \& Astrophysics, University of California, Santa Cruz, CA 95064, USA}

\author[0000-0002-8342-3804]{Yukei Murakami}
\affiliation{Department of Physics and Astronomy, Johns Hopkins University, Baltimore, MD 21218, USA}

\author[0000-0002-5259-2314]{ Gagandeep S.~Anand}
\affiliation{Space Telescope Science Institute, 3700 San Martin Drive, Baltimore, MD 21218, USA}

\author[0000-0003-3889-7709]{Louise Breuval}
\affiliation{Department of Physics and Astronomy, Johns Hopkins University, Baltimore, MD 21218, USA}
\affiliation{LESIA, Observatoire de Paris, Universit\'e PSL, CNRS, Sorbonne Universit\'e, Universit\'e de Paris, 5 place Jules Janssen, F-92195 Meudon, France}

\author[0000-0001-5955-2502]{Thomas G.~Brink}
\affiliation{Department of Astronomy, University of California, Berkeley, CA 94720-3411, USA}

\author[0000-0003-3460-0103]{Alexei V.~Filippenko}
\affiliation{Department of Astronomy, University of California, Berkeley, CA 94720-3411, USA}
\affiliation{Miller Institute for Basic Research in Science, University of California, Berkeley, CA 94720, USA}

\author[0000-0002-4312-7015]{Samantha Hoffmann}
\affiliation{Space Telescope Science Institute, 3700 San Martin Drive, Baltimore, MD 21218, USA}

\author[0000-0001-8738-6011]{Saurabh W.~Jha}
\affiliation{Department of Physics and Astronomy, Rutgers, the State University of New Jersey, Piscataway, NJ 08854, USA}

\author[0000-0002-5153-5983]{W.~D'arcy Kenworthy}
\affiliation{Department of Physics and Astronomy, Johns Hopkins University, Baltimore, MD 21218, USA}

\author[0000-0002-4312-7015]{John Mackenty}
\affiliation{Space Telescope Science Institute, 3700 San Martin Drive, Baltimore, MD 21218, USA}

\author[0000-0002-3169-3167]{Benjamin E.~Stahl}
\affiliation{Department of Astronomy, University of California, Berkeley, CA 94720-3411, USA}

\author[0000-0002-2636-6508]{WeiKang Zheng}
\affiliation{Department of Astronomy, University of California, Berkeley, CA 94720-3411, USA}

\begin{abstract}
We report observations from the {\it Hubble Space Telescope} ({\it HST}) of Cepheid variables in the host galaxies of 42 Type Ia supernovae (SNe~Ia) used to calibrate the Hubble constant (H$_0$). These include the complete sample of all suitable SNe~Ia discovered in the last four decades at redshift $z \leq 0.01$, collected and calibrated from $\geq 1000$  {\it HST}  orbits, more than doubling the sample whose size limits the precision of the direct determination of H$_0$. The Cepheids are calibrated geometrically from {\it Gaia} EDR3 parallaxes, masers in NGC$\,$4258 (here tripling that sample of Cepheids), and detached eclipsing binaries in the Large Magellanic Cloud. All Cepheids in these anchors and SN~Ia hosts were measured with the same instrument (WFC3) and filters ({\it F555W}, {\it F814W}, {\it F160W}) to negate zeropoint errors.

We present multiple verifications of Cepheid photometry and six tests of background determinations that show Cepheid measurements are accurate in the presence of crowded backgrounds. The SNe~Ia in these hosts calibrate the  magnitude--redshift relation from the revised Pantheon+ compilation, accounting here for covariance between all SN data and with host properties and SN surveys matched throughout to negate systematics. We decrease the uncertainty in the local determination of H$_0$ to 1 \kmss including systematics. We present results for a comprehensive set of nearly 70 analysis variants to explore the sensitivity of H$_0$ to selections of anchors, SN surveys, redshift ranges, the treatment of Cepheid dust, metallicity, form of the period--luminosity relation, SN color, peculiar-velocity corrections, sample bifurcations, and simultaneous measurement of the expansion history.

Our baseline result from the Cepheid--SN~Ia sample is H$_0$ = \hsbase, which includes systematic uncertainties and lies near the median of all analysis variants.  We demonstrate consistency with measures from {\it HST} of the TRGB between SN~Ia hosts and NGC$\,$4258, and include them {\it simultaneously}  to yield \hswtrgbmean.  The inclusion of high-redshift SNe~Ia yields H$_0$ = \hhiz\ and $q_0$ = \qhiz. We find a \faceplanck $\sigma$ difference with the prediction of H$_0$ from {\it Planck} CMB observations under $\Lambda$CDM, with no indication that the discrepancy arises from measurement uncertainties or analysis variations considered to date.  The source of this now long-standing discrepancy between direct and cosmological routes to determining H$_0$ remains unknown.

\ \par

\ \par

\ \par

\end{abstract}

\clearpage

\tableofcontents

\clearpage 

\section{Introduction}

The present expansion rate of the Universe, the Hubble constant (H$_0$), sets its size and age scale, relating redshift (the direct consequence of expansion) to distance and time.  The value of H$_0$ may be determined locally with measurements of distances and redshifts, and it can also be predicted from a cosmological model calibrated in the early Universe (i.e., pre-recombination at redshift $z \geq 1100$) with measurements of the cosmic microwave background (CMB).  The comparison of measured and predicted values of H$_0$ thus provides a crucial ``end-to-end'' test of the widest available range of the validity of cosmological models, from early times when the Universe is dense and dominated by dark matter and radiation, to the present when it is dilute and dominated by dark energy.  

\subsection{The SH0ES Program\label{sc:1.1}}

The Hubble constant is the most accessible parameter in the cosmological model.  It can be estimated with a wide range of approaches and accuracies from limited knowledge of many types of astronomical sources, nearly all of which have been utilized in this endeavor over the past century.  There have been $>1000$ estimates published since 1980, with $1/3$ of those in the last five years and 20\% in the last two years --- a recent quadrupling of the effort indicating the accelerating interest in H$_0$ \citep{Steer:2020}.  However, past discrepancies internal to the body of local measurements reveal that systematic errors can dominate determinations of H$_0$, and that there is no reason to believe all efforts will regress to the mean or that a more accurate result can be derived from their median \citep{Chen:2011}.  Rather, to keep systematic uncertainties in check it is necessary to pursue the most powerful, simplest, and most reliable tools, with strict attention paid to understanding, mitigating, and accounting for sources of measurement error.  

Since the launch of the {\it Hubble Space Telescope (HST)} with a design goal of achieving a 10\% determination of H$_0$, the leading approach to measuring it in the local Universe (as indicated by the observing time competitively awarded by the community) has relied on imaging of Cepheid variable stars in the host galaxies of recent, nearby Type Ia supernovae (SNe~Ia).  Cepheids have been favored as primary distance indicators because they are very luminous ($M_V \approx -6$~mag at $P \approx 30$~d), are easy to identify thanks to their periodicity \citep{Leavitt:1912}, obey a tight period--luminosity relation (\PL, the ``Leavitt Law'') that yields extremely precise distances \citep[3\% per source;][hereafter R19]{Riess:2019b}, and have well-understood physics \citep{Eddington:1917,Bono:1999}.  Other primary distance indicators that have been measured in the hosts of SNe~Ia to determine H$_0$ include the tip of the red giant branch \citep[TRGB;][]{Freedman:2019} and Mira variable stars \citep{Huang:2020}.

Cepheids also offer the most opportunities for obtaining strictly {\it differential} flux measurements --- i.e., the use of the same facility to measure calibrator and source, a key requirement for eliminating zeropoint errors.  This is feasible through the use of {\it HST} to directly observe Cepheids in a large set of SN hosts and in geometric calibrators of Cepheid luminosities: the megamaser host NGC$\,$4258, the Milky Way (hereafter MW) with plentiful parallaxes, and the Large Magellanic Cloud (hereafter LMC) via detached eclipsing binaries (DEBs). SNe~Ia are favored to measure the Hubble expansion owing to their high precision (5\% in distance per source), ubiquity, and deep reach, which reduces the impact of local flows.

The SH0ES program (Supernovae and H$_0$ for the Equation of State of dark energy) began in 2005 with a proposal in {\it HST} Cycle 15 to break the degeneracy among cosmological parameters used to model CMB data and the equation-of-state parameter $w = P/(\rho c^2)$, where $P$ is the pressure and $\rho$ is the mass density of dark energy. Its stated ambitious goal, based on the recommendation by \citet{hu05}, was to eventually reach a percent-level measurement of H$_0$, a goal approached if not fully reached in this work.  This project was a ``second-generation'' effort to measure H$_0$ with  {\it HST}  from a distance ladder of Cepheids and SNe~Ia using the then recently-installed ACS (and later WFC3) instruments, following successful efforts during the 1990s by the ``first-generation'' {\it HST} Key Project on the Extragalactic Distance Scale \citep{freedman01} and the SNe~Ia Luminosity Calibration Program \citep{sandage06}, both of  which primarily used WFPC2.  The former searched for Cepheids in the hosts of numerous secondary distance indicators (excluding SNe~Ia) while the latter focused on SN~Ia hosts. The types of targets suitable for Cepheid searches and the observing sequences for use with  {\it HST}  were developed and first implemented by these ground-breaking programs.  

The precision of the first-generation programs was ultimately limited by the lack of a precise geometric calibration of Cepheid variables, by the limited practical range of WFPC2 to measure Cepheids in SN~Ia hosts (distance $D \lesssim 20$--25~Mpc), by the impact of reddening in the optical, and by the limited characterization of the Cepheid metallicity dependence at those wavelengths.  The ability of SNe~Ia to measure individual distances with $\sim 5$--10\% precision and to sharply delineate the Hubble flow began with the use of light-curve vs.~luminosity relations \citep{Phillips:1993}, SN~Ia colors \citep{Riess:1996,Tripp:1998,Phillips:1999}, and modern, digital samples \citep{Hamuy:1996,Riess:1999}. 
    
Unfortunately, SNe~Ia at $D \leq 20$~Mpc are rare, occurring about once per decade, with most of the few objects in this range observed up to a century ago using photographic technology. Such observations lacked the photometric precision, well-characterized bandpasses, and accurate determinations of host-galaxy backgrounds, SN light-curve shapes, and SN colors to take advantage of the new standardization methods.  The tendency for intrinsically brighter SNe~Ia with broader light curves to occur in (late-type) Cepheid hosts would also bias H$_0$ lower without light-curve standardization.  A number of systematic differences in the first-generation calibration of SNe~Ia by \citet{sandage06} were quantified by \citet[][Table 16]{riess05}.  These differences, totaling about 20\%, arose from several effects which were amplified by small sample statistics: problematic SN~Ia data such as photographic photometry, highly extinguished objects, and poorly sampled light curves; from photometric anomalies in WFPC2, such as the ``long vs.~short effect’’ \citep{Holtzman:1995} and charge-transfer efficiency \citep[CTE; e.g.,][]{whitmore99}; and from limited knowledge of the slope of the Cepheid \PLs relation. The present geometric calibration of the distance to the LMC by \citet{Pietrzynski:2019} using DEBs is also 7\% smaller than the value assumed by \citet{sandage06} to calibrate Cepheids.
    
The SH0ES program has been designed to improve upon past determinations of H$_0$ by (1) extending the range of Cepheid observations with ACS and WFC3 to reach the hosts of a large sample of ``ideal'' SNe~Ia, free from the preceding problems; (2) using near-infrared (NIR) observations of all Cepheids in SN~Ia hosts with NICMOS and WFC3 to reduce the systematic uncertainty associated with the reddening laws for Cepheids and their hosts and the Cepheid metallicity dependence; and (3) calibrating Cepheids with new, geometric distances tied directly with {\it HST} to the Cepheids in SN~Ia hosts to nullify zeropoint uncertainties.  ``Ideal'' or suitable SNe~Ia for calibrating H$_0$ (given limited {\it HST} time)  were defined by \cite{riess05} to be (1) observed before maximum light, (2) through low interstellar extinction ($A_V < 0.5$~mag), (3) with the same instruments and filters as the SNe~Ia in the Hubble flow (at that time obtained by the Cal\'an/Tololo and CfA surveys), and (4) to have typical light-curve shapes\footnote{These color and shape requirements translate in the Pantheon SN standardization \citep{Scolnic:2018} as $\abs{c}<0.2$ and $\abs{x1} < 2$}. These characteristics are necessary to provide low dispersion in the Hubble flow, but they applied to only three Cepheid-calibrated SNe~Ia from the first-generation projects (SNe 1981B, 1990N, and 1998aq).

\begin{figure}[b]
\begin{center}
\includegraphics[width=0.85\textwidth]{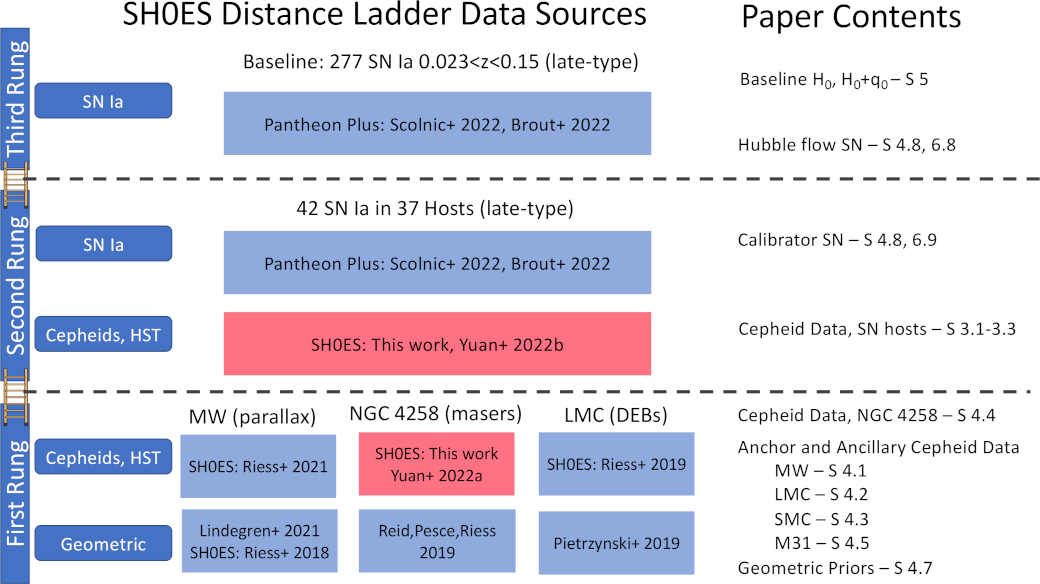}
\end{center}
\caption{\label{fg:datasrc} Sources of data for distance ladder. Red block shows data from this work.}
\end{figure}

\begin{figure}[b]
\includegraphics[width=\textwidth]{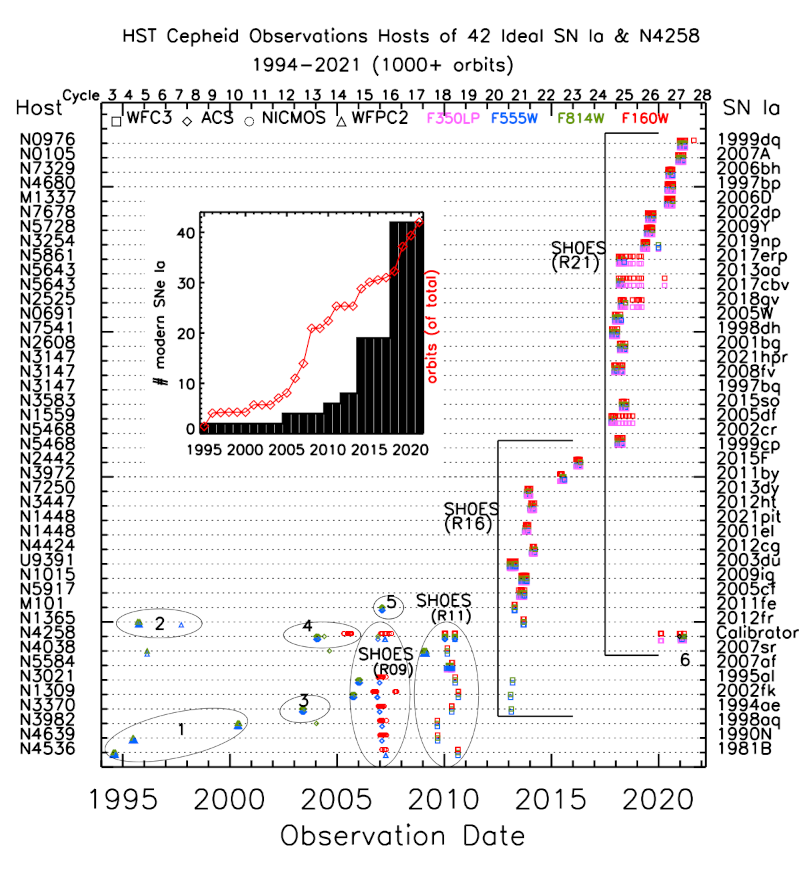}
\caption{\label{fg:hstobs} {\it HST} observations of Cepheids in 37 hosts of 42 ``ideal'' SNe~Ia and NGC$\,$4258, collected over 20~yr with 4 cameras and $>1000$ orbits of {\it HST} time.  In most cases, 60 to 90-day campaigns in {\it F555W} and {\it F814W} or in {\it F350LP} were used to identify Cepheids from their light curves with occasional observations years later to identify longer-period Cepheids.  NIR follow-up observations in {\it F160W} are used to reduce the effects of host-galaxy extinction, sensitivity to metallicity, and breaks in the \PLs relation. Data sources: (1) {\it HST} Key Project, \citet{freedman01}; (2) {\it HST} SN~Ia Luminosity Calibration Program, \citet{sandage06}; (3) \citet{riess05}; (4) \citet{macri06}; (5) \citet{Mager:2013}; (6) \citet{Yuan:2021_N4258}.}
\end{figure}

\clearpage

Unlike the first-generation programs, which were granted long-term status and a large initial allocation of observing time with {\it HST}, the SH0ES program was proposed to the STScI time-allocation committee year-by-year. The cumulative result, after fifteen years (cycles), has been to collect Cepheid observations in 37 hosts of 42 SNe~Ia and calibrate them geometrically to Cepheids in the MW, the LMC, and NGC$\,$4258, with a total of 18 individual {\it HST} proposals utilizing $> 1000$ orbits.  The source of the distance-ladder data is shown in Fig.~\ref{fg:datasrc} and the source and sequence of all observations in SN~Ia hosts are shown in Fig.~\ref{fg:hstobs} and listed in Table~\ref{tb:wfc3}.  Observing Cepheids in hosts at $D \approx 40$--50~Mpc, double the range of first-generation observations and a factor of 8 increase in targets, was feasible owing to two features of WFC3: significantly better sampling of the point-spread function (PSF), with pixel sizes a factor of 2.5 smaller than the wide channel of WFPC2 which greatly reduced the impact of crowded backgrounds, and a white-light filter ({\it F350LP}) that, combined with the better sensitivity of WFC3, reduced the observing time required to identify Cepheids and measure their periods by a factor of $\sim 4$. Calibrating those Cepheids differentially to bright Cepheids in the MW and LMC became feasible through the development of spatial scanning of {\it HST} and a rapid slew and guiding mode under gyroscopic control.

\begin{deluxetable}{llrrrrr}[t]
\tablecaption{Cepheid Observations with  {\it HST}  \label{tb:wfc3}}
\tabletypesize{\normalsize}
\tablewidth{0pc}
\tablehead{\colhead{} & \colhead{} & \multicolumn{3}{c}{Exposure time [s]} &\colhead{} & \colhead{}\\[-0.25cm]
\colhead{Galaxy} & \colhead{SN(e)~Ia} & \multicolumn{2}{c}{WFC3} & \multicolumn{1}{c}{All} & \multicolumn{1}{c}{NIR}& \multicolumn{1}{c}{UT}\\[-0.25cm]
\colhead{} & \colhead{} & \multicolumn{1}{c}{NIR$^a$} & \multicolumn{1}{c}{UVIS$^b$} & \multicolumn{1}{c}{opt.$^c$}  &\multicolumn{1}{c}{Prop ID(s)} &\multicolumn{1}{c}{Date$^d$}}
\startdata
M101&2011fe&4846&3776&53072&   12880&2013-03-03\\
Mrk1337&2006D&13823&25994&25994&   15640&2020-04-15\\
N0105&2007A&13270&34302&34302&   16269&2020-10-23\\
N0691&2005W&9647&31413&31413&   15145&2017-11-10\\
N0976&1999dq&15482&34312&34312&   16269&2020-11-21\\
N1015&2009ig&14364&39336&39336&   12880&2013-06-30\\
N1309&2002fk&6991&30020&89282&   11570,12880&2010-07-24\\
N1365&2012fr&3617&31800&91200&   12880&2013-08-06\\
N1448&2001el,2021pit&6035&17562&17562&   12880&2013-09-15\\
N1559&2005df&10058&22245&22245&   15145&2017-09-09\\
N2442&2015F&6035&20976&20976&   13646&2016-01-21\\
N2525&2018gv&9821&21177&21177&   15145&2018-02-14\\
N2608&2001bg&9647&26942&26942&   15145&2018-02-04\\
N3021&1995al&4426&29620&88722&   11570,12880&2010-06-03\\
N3147&1997bq,2008fv,2021hpr&14470&37426&37426&   15145&2017-10-28\\
N3254&2019np&8441&22106&22106&   15640&2019-03-11\\
N3370&1994ae&4376&29820&88222&   11570,12880&2010-04-04\\
N3447&2012ht&4529&19114&19114&   12880&2013-12-15\\
N3583&2015so&9647&27001&27001&   15145&2018-03-06\\
N3972&2011by&6635&19932&19932&   13647&2015-04-19\\
N3982&1998aq&4017&14000&90840&   11570&2009-08-04\\
N4038&2007sr&6794&20640&85684&   11577&2010-01-22\\
N4258&Anchor&40234&10120&103690&   11570&2020-01-02\\
N4424&2012cg&3623&17782&17782&   12880&2014-01-08\\
N4536&1981B&2564&26000&95000&   11570&2010-07-19\\
N4639&1990N&5379&16000&77480&   11570&2009-08-07\\
N4680&1997bp&9647&25217&25217&   15640&2020-04-24\\
N5468&1999cp,2002cr&14470&36566&36566&   15145&2017-12-22\\
N5584&2007af&4929&74940&74940&   11570&2010-04-04\\
N5643&2013aa,2017cbv&9052&24741&24741&   15145&2018-01-16\\
N5728&2009Y&13823&26111&26111&   15640&2019-05-05\\
N5861&2017erp&10058&21798&21798&   15145&2018-01-13\\
N5917&2005cf&7235&23469&23469&   12880&2013-05-20\\
N7250&2013dy&5435&18158&18158&   12880&2013-11-08\\
N7329&2006bh&9044&24665&24665&   15640&2020-05-06\\
N7541&1998dh&9647&26766&26766&   15145&2017-09-21\\
N7678&2002dp&12058&32060&32060&   15640&2019-05-25\\
U9391&2003du&13711&39336&39336&   12880&2012-12-14
\enddata
\tablecomments{(a) Obtained with WFC3/NIR and {\it F160W}; (b) obtained with WFC3/UVIS and {\it F555W}, {\it F814W}, or {\it F350LP} used to find and measure the flux of Cepheids; (c) includes time-series data from an earlier program and a different camera --- see Fig.~\ref{fg:hstobs}; (d) date of first WFC3/NIR observation; }
\end{deluxetable}

The first SH0ES results \citep[][hereafter R09]{riess09a} were based on Cepheids observed in the hosts of 6 ideal SN~Ia calibrators using ACS (optical) and NICMOS (NIR), and one geometric anchor (NGC$\,$4258) with a maser-based distance of 3\% precision \citep{Humphreys13} and a large sample of Cepheids \citep{macri06}.  The result was a 5\% measurement of H$_0$, $74.2\pm3.6$~\kms, which combined with the 5-year WMAP results \citep{Komatsu:2009} yielded $w=-1.12\pm0.12$ and was consistent with a value of H$_0=71.9\pm2.6$ determined from WMAP and $\Lambda$CDM alone\footnote{To improve readability, the conventional units of H$_0$ will frequently be omitted in the rest of this paper. $\Lambda$CDM refers to the standard\\ cosmological model with the cosmological constant ($\Lambda$) and cold dark matter (CDM).}.  The second iteration \citep[][hereafter R11]{riess11} increased the calibrator sample to 8 SNe~Ia, observed and measured all Cepheids with WFC3 (both optical and NIR), and expanded the geometric calibration of Cepheids beyond NGC$\,$4258 by including two additional independent anchors: the LMC, through various DEB-based distances \citep[e.g.,][]{Pietrzynski:2009}, and the MW via parallaxes measured with the Fine Guidance Sensor on {\it HST} \citep{benedict07}. This resulted in H$_0=73.8\pm2.4$, which coupled with the 7-year WMAP results \citep{komatsu11} yielded $w=-1.08\pm0.10$, or an estimate of the effective number of relativistic species of $N_{\rm eff}=4.2\pm0.7$.  This result was closely matched by a recalibration of the final {\it HST} Key Project results using the same MW parallaxes, different Cepheid measurements, and an updated Hubble diagram of SNe~Ia which yielded H$_0=74.4\pm2.2$ \citep[][hereafter F12]{freedman12}.  Indications of any tension between the early and late Universe at that time were $<2\sigma$ in significance.

\subsection{The Hubble Tension}   
 
The first release of CMB data from the ESA {\it Planck} mission \citep{Planck:2013} yielded H$_0=67.2\pm1.2$ in the context of $\Lambda$CDM, a then $2\sigma$ reduction relative to WMAP and a difference of $3\sigma$ from the results of R11 and F12.  Reanalyses of the R11 data \citep{Fiorentino:2013,Efstathiou:2014,Zhang:2017} produced essentially the same results as R11, with H$_0$ ranging from 72.5 to 76.0. The third iteration of SH0ES \citep[][hereafter R16]{Riess:2016} more than doubled the calibrator sample to 19 SNe~Ia, used refined distance estimates to NGC$\,$4258 and the LMC \citep{Pietrzynski:2013}, and new Cepheid parallaxes measured by the SH0ES team using spatial scanning with WFC3 \citep{riess14, Casertano:2016} to reach a 2.4\% determination of H$_0=73.2\pm1.7$, $3.4\sigma$ greater than the refined value from {\it Planck}$+\Lambda$CDM of $66.9\pm0.6$ \citep{Planck:2016}, a difference of $\sim 9$\% or 0.2~mag in units of 5\,log\,H$_0$.  An extensive number of reanalyses of the R16 data with many variations were undertaken \citep{Cardona:2017,Feeney:2017,Follin:2017,Bovy:2018,Burns:2018,Dhawan:2018,Avelino:2019}, resulting in H$_0$ values ranging from 73 to 74 and uncertainties from 2\% to 2.5\%.  These analyses explored varying reddening laws, use of NIR SN data, use of alternative SN light-curve fitting, and hierarchical Bayesian statistics for data fitting, resulting in little change in H$_0$ or its uncertainty. \citet{Javanmardi:2021} selected the Cepheid host from R16 with the largest sample of Cepheids (NGC$\,$5584, also near the median sample host distance) and remeasured these starting from the archived {\it HST} pixels and using different methods to do photometry, finding agreement with the R16 measurement to 1\% precision and ruling out a significant methodological error in these measurements.  

Since then, additional sources of MW parallax calibration of Cepheid luminosities have come from further use of {\it HST}  WFC3 spatial scanning \citep{Riess:2018a}, from {\it Gaia} DR2 Cepheid parallaxes with {\it HST}  photometry \citep{Riess:2018b}, from {\it Gaia} DR2 Cepheid binary companions and cluster hosts \citep{Breuval:2020}, and from {\it Gaia} EDR3 parallaxes coupled with additional {\it HST} photometry \citep{Riess:2021}.  Likewise, {\it HST}  observations of 70 long-period Cepheids in the LMC \citep{Riess:2019} and improved distance estimates to the LMC \citep{Pietrzynski:2019} and NGC$\,$4258 \citep{Reid:2019} resulted in H$_0=73.2\pm1.3$, raising the difference with {\it Planck}$+\Lambda$CDM to $4.2\sigma$.  Other precise measures of H$_0$ in the local Universe from the distance--redshift relation generally range from 70 to 75, and those grounded in the pre-recombination version of $\Lambda$CDM range from 67 to 68, and have been extensively reviewed \citep{Verde:2019,Divalentino:2021,Shah:2021}.  Because the tension is seen between different routes which are comparable only via an accurate cosmological model, numerous possible theoretical explanations for the emergent ``Hubble tension'' have been proposed but no consensus has yet emerged \citep{Divalentino:2021}.   Indeed, theoretical priors weigh heavily on these proposals or whether it may be considered ``extraordinary'' for the $\Lambda$CDM model to fail or pass this cosmic test.  The test itself, however, is empirical and few would conclude it has yet been satisfactorily passed.  	

\subsection{This Work\label{sc:1.3}}

In this publication, we more than double the sample of SN~Ia calibrators from 19 in R16 to reach 42 objects in 37 hosts.  This increase is a milestone for a sample whose size has limited the precision to which H$_0$ can be locally measured.  It provides the largest increase in size we can anticipate in the remaining lifetime of {\it HST} as it now includes all suitable SNe~Ia (of which we are aware) observed between 1980 and 2021 at $z<0.011$ and which slowly accrue at $\sim 1$~yr$^{-1}$.  We have also reprocessed and reanalyzed the NIR observations of Cepheids in the previous 19 hosts reported by R16 for consistency with the new sample.  We make use of an automated pipeline \citep[][hereafter Y22b]{Yuan:2021_SN} to find the Cepheids in 18 new hosts (and reanalyze past hosts) which follow the steps developed manually for the first 19 hosts \citep[][hereafter H16]{Hoffmann:2016}. We benefit from a factor of 3 increase in the sample of Cepheids within NGC$\,$4258 discovered by observing 4 new fields with {\it HST}, fully reanalyzed using the same pipeline by \citet[][hereafter Y22a]{Yuan:2021_N4258}. Extensive details concerning the analysis of SN~Ia data are given by \citet{Scolnic:2021} and \citet{Brout:2022}.

We present the formalism for measuring H$_0$ from the distance ladder in \S\ref{sc:2}; new Cepheid data in \S\ref{sc:3} and in Y22a+Y22b; ancillary data used to measure H$_0$ in \S\ref{sc:4}; our baseline, local determination of H$_0$ and a simultaneous measurement of H$_0$ and the expansion history with high-redshift SNe~Ia in \S\ref{sc:5}; extensions to the baseline and variants of the local measurement of H$_0$ in \S\ref{sc:6}; discussion in \S\ref{sc:7}; and conclusions in \S\ref{sc:8}. Appendices provide further details on characterizing the spectral and photometric properties of the 42 calibrator SNe~Ia (\ref{sc:appa}),  independent tests of the accuracy of Cepheid photometry (\ref{sc:appb}), the Cepheid metallicity scale (\ref{sc:appc}), and alternative applications of ``Wesenheit'' magnitudes (\ref{sc:appd}).  

\section{Measuring the Hubble Constant\label{sc:2}}
\subsection{Distance-Ladder Formalism}

The Hubble constant is the present relation between redshift and distance, $cz=$ H$_0 D$, measured at cosmological distances where expansion is the dominant source of redshift.  Here it is measured via a three-step (or three-rung) distance ladder employing a single, simultaneous fit between (1) geometric distance measurements to standardized Cepheid variables, (2) standardized Cepheids and colocated SNe~Ia in nearby galaxies, and (3) SNe~Ia in the Hubble flow. The fit is accomplished simultaneously by optimizing a $\chi^2$ statistic to determine the most likely values of the parameters in the relevant relations.   The data include measurements, their uncertainties, and their covariances as described in the next section.  The parameters are the distances of all hosts and five additional parameters: the fiducial luminosity of SNe~Ia and Cepheids, two parameters standardizing Cepheid luminosities (their dependence on period and metallicity), and H$_0$. This parameterization of the distance ladder can be expressed as a simple system of linear equations in a compact set of matrices as given below, useful for transmission,  and for which the maximum-likelihood solution is easily found.  The distance-ladder data and scheme are displayed in approximate form in Fig.~\ref{fg:datasrc}. 

To briefly summarize the relevant relations, the distance modulus of a source is $\mu_0=m_0-M_0=5\, {\rm log}\,D + 25$, with $D$ the luminosity distance in Mpc, $m$ the apparent magnitude (flux), $M$ the absolute magnitude (luminosity), and the subscript 0 denoting a magnitude free of (or corrected for) intervening absorption by interstellar dust. The form of the dependence of Cepheid or SN~Ia luminosity on observed characteristics (i.e., ``standardization'') has been well-determined by prior work and is only briefly reviewed here.  The dereddened Cepheid apparent magnitudes \citep[also called ``Wesenheit'' magnitudes;][]{madore82} at mean-phase will be described in \S\ref{sc:3.3} and are identified here as $m_H^W$  (in a specific host like NGC$\,$4258, those Cepheids would have magnitudes given as $m^W_{H,N4258}$).  For the $j$-th such Cepheid magnitude in the $i$-th host given the period $P_{i,j}$ in days and metallicity ${\rm [O/H]}_{i,j}$ relative to the Sun, we have
\bq m_{H,i,j}^W=\mu_{0,i}+M_{H,1}^W+b_W\, (\log\,P_{i,j}-1) +Z_W {\rm [O/H]}_{i,j}, \label{eq:cephmagalt} \eq
\noindent where $M_{H,1}^W$ is the fiducial absolute magnitude (in the Wesenheit magnitude system of \ref{sc:3.4}), of a Cepheid with $\log\,P\!=\!1$ ($P\!=\!10$~days) and solar metallicity, and the parameters $b_W$ and $Z_W$ (sometimes called $\gamma$ in the literature) define the empirical relation between Cepheid period, metallicity, and luminosity.  The ${\rm [O/H]}_{i,j}$ is inferred at its galactocentric radial position as described in \S\ref{sc:3.5}.  

Any number of Cepheid hosts may have an independent geometric distance which contributes to calibrating the Cepheid luminosity; e.g., for variables observed in the maser host NGC$\,$4258 we adopt a distance modulus $\mu_{0,\rm N4258}$, the best estimate of the distance,  with formal uncertainty $\sigma(\mu_{0,{\rm N4258}})$ from \citet{Reid:2019}.  In this case the individual host distance parameter $\mu_{0,i}$ above is replaced with the external constraint, converting the apparent magnitudes to absolute,
\bq M_{H,j}^W=m_{H,N4258,j}^W\!-\!\mu_{0,\rm N4258}+\Delta \mu_{\rm N4258},\label{eq:bkg} \eq

\noindent and introducing a new parameter $\Delta \mu_{\rm N4258}$ as the difference from the measured and true distance with the additional, simultaneous constraint equation $0=\Delta \mu_{\rm N4258} \pm \sigma(\mu_{ 0,{\rm N4258}})$.  This definition allows the simultaneous use of multiple geometric ``anchors'' to calibrate the distance ladder; as we later show, it also allows the use of additional distance indicators for the same anchor, such as the tip of the red giant branch (TRGB) or Mira variable stars, while keeping track of their mutual dependence on the same geometric constraint.

A set of hosts of both SNe~Ia and Cepheids connects the two distance indicators. Thus, for an SN~Ia in the $i$-th Cepheid host, 
\bq m_{B,i}^0=\mu_{0,i}\!+\!M_B^0, \label{eq:snmagalt} \eq
\noindent where $m_{B,i}^0$ is its maximum-light apparent magnitude which has been {\it standardized} \citep[i.e., corrected for variations around the fiducial color, luminosity, and any host dependence; see][]{Scolnic:2021}, $M_B^0$ is the fiducial SN~Ia luminosity, and $\mu_{0,i}$ is the same parameter as in Equation~(\ref{eq:cephmagalt}).  For SNe~Ia, unlike Cepheids, the convention for keeping track of covariance in the standardization, as described by \cite{Scolnic:2021}, is to employ a set of standardized $m_{B,i}^0$, their uncertainties, and the covariance between any pair.  This is an equally mathematically valid approach as keeping track of covariance through the standardizing relation for Cepheids in Equation~(\ref{eq:cephmagalt}).

\ \par

The ladder is completed with a set of SNe~Ia that measure the expansion rate quantified as the intercept, $a_B$, of the distance (or magnitude)--redshift relation.  This is simply $a_B=\log\,cz - 0.2m_B^0$ in the low-redshift limit ($z \approx 0$) but given for an arbitrary expansion history and for $z>0$ as \bq a_B=\log\,cz \left\{ 1 + {\frac{1}{2}}\left[1-q_0\right] {z} -{\frac{1}{6}}\left[1-q_0-3q_0^2+j_0 \right] z^2 + O(z^3) \right\} - 0.2m_B^0, \label{eq:aB} \eq

\noindent measured from a set of SNe~Ia ($z, m_B^0$) where $z$ is the redshift due to expansion, $q_0$ is the deceleration parameter, and $j_0$ is the jerk \citep[see][for definitions]{Visser:2004}. The determination of H$_0$ follows from

\bq \log\,{\rm H}_0={0.2 M_B^0\!+\!a_B\!+\!5}. \label{eq:h0alt} \eq

\ \par

If the set of standardized SN~Ia magnitudes in the hosts of Cepheids which serve to calibrate $M_B^0$ (hereafter ``calibrators'' or CC~SNe~Ia) and those in the Hubble flow used to measure $a_B$ (hereafter HF SNe~Ia) have no common sources of uncertainty (i.e., no covariance), then Equations~(\ref{eq:snmagalt}) and (\ref{eq:aB}) and the ladder parameters they provide ($M_B^0$ and $a_B$) can be determined independently. This was the approach taken by R16. However, an increasingly thorough quantification of systematic uncertainties in SN~Ia measurements, and the standardization undertaken as common practice for the determination of $w$ \citep{Scolnic:2018}, have demonstrated nontrivial covariance of SN~Ia data, quantified following the approach of \cite{Conley:2011} and \cite{Dhawan:2020}. We therefore undertake the optimization of Equations~(\ref{eq:snmagalt}) and {\bf (\ref{eq:h0alt})} {\it simultaneously}. 

\ \par

It is useful to expand the same set of Equations~(\ref{eq:cephmagalt})--(\ref{eq:h0alt}) in the form of matrices that organize the data into a vector of magnitude measurements ${\bf y}$, covariance matrix of standard errors of the magnitude measurements ${\bf C}$, equation matrix ${\bf L}$, and vector of free parameters ${\bf q}$ (hence the model, ${\bf Lq}$) as follows:

\clearpage

\bq
y =\begin{array}{ll}
\left( \begin{array}[c]{c}
  m^W_{H,1} \\
  \td \\
  m^W_{H,\textrm{nh}} \\
  \hline
  m^W_{H,\textrm{N4258}}-\mu_{0,\textrm{N4258}} \\
  m^W_{H,\textrm{M31}}  \\
  m^W_{H,\textrm{LMC}}-\mu_{0,\textrm{LMC}}  \\
  \hline
  m_{B,1}^0 \\
  \td \\
  m_{B,\textrm{ncc}}^0 \\
  \hline
  M_{H,1,\textit{HST}}^W \\
  M_{H,1,{\textit Gaia}}^W \\
  0 \\
  0 \\
  0 \\
  \hline
  m_{B,1}^0 - 5 \log cz_1\{\} -25  \\
  \td\\
  m_{B,\textrm{nhf}}-5\log cz_{\textrm{nhf}}\{\}-25
\end{array} \right)
&
\begin{array}[c]{@{}l@{\,}l}
\left. \begin{array}{c} \vphantom{m^W_{H,\textrm{nh},j}} \\ \vphantom{\td} \\ \vphantom{\textrm{\LARGE HELLO}} \end{array} \right\} & \text{Cepheids in SN Ia hosts} \\
\left. \begin{array}{c} \vphantom{m^W_{H,\textrm{nh},j}} \\ \vphantom{m^W_{H,\textrm{nh},j}} \\ \vphantom{\textrm{\LARGE HELLO}} \end{array} \right\} & \text{Cepheids in anchors or non-SN~Ia hosts\hspace{1.5in}} \\
\left. \begin{array}{c} \vphantom{m^W_{H,\textrm{nh},j}} \\ \vphantom{\td} \\ \vphantom{\textrm{\LARGE HELLO}} \end{array} \right\} & \text{SNe~Ia in Cepheid hosts} \\
\left. \begin{array}{c} \vphantom{m^W_{H,\textrm{nh},j}} \\ \vphantom{\textrm{\LARGE HELLO}} \\ \vphantom{0} \\ \vphantom{m^W_{H,\textrm{nh},j}} \\ \vphantom{\textrm{\LARGE HELLO}} \end{array} \right\} & \text{External constraints} \\
\left. \begin{array}{c} \vphantom{m^W_{H,\textrm{nh},j}} \\ \vphantom{\td} \\ \vphantom{\textrm{\LARGE HELLO}} \end{array} \right\} & \text{SNe~Ia in the Hubble flow} \\
\end{array}
\end{array} 
\nonumber
\eq

\bq
C = \begin{pmatrix}
\sy{\sigma_{{\rm tot},1}^2}\!\!\!\! &\td & Z_{\textrm{cov}} & Z_{\textrm{cov}} & \sy{0} & \sy{0} & \sy{0} & \td &\sy{0} & \sy{0} & \sy{0} & \sy{0} & \sy{0} & \sy{0} & \sy{0} & \td & \sy{0} \\
\td &\td & \td &\td &\td &\td &\td &\td &\td &\td &\td &\td &\td & \td & \td & \td & \td \\
Z_{\textrm{cov}} &\td & \sy{\sigma_{{\rm tot},{\rm nh}}^2}\!\!\!\! & Z_{\textrm{cov}} & \sy{0} & \sy{0} & \sy{0} & \td &\sy{0} & \sy{0} & \sy{0} & \sy{0} & \sy{0} & \sy{0}  & \sy{0} & \td & \sy{0} \\
\hline
Z_{\textrm{cov}} &\td & Z_{\textrm{cov}}& \sy{\sigma_{{\rm tot},{\rm N4258}}^2}\!\!\!\! & \sy{0} & \sy{0} & \sy{0} & \td &\sy{0} & \sy{0} & \sy{0} & \sy{0} & \sy{0} & \sy{0}  & \sy{0} & \td & \sy{0} \\
\sy{0} &\td & \sy{0} & \sy{0} & \sy{\sigma_{{\rm tot},{\rm M31}}^2}\!\!\!\! & \sy{0} & \sy{0} & \td &\sy{0} & \sy{0} & \sy{0} & \sy{0} & \sy{0} & \sy{0}  & \sy{0} & \td & \sy{0} \\
\sy{0} &\td & \sy{0} & \sy{0} & \sy{0} &  \sy{\sigma_{{\rm tot},{\rm LMC}}^2}\!\!\!\! & \sy{0} & \td & \sy{0} & \sy{0} & \sy{0} & \sy{0} & \sy{0} & \sy{0}  & \sy{0} & \td & \sy{0} \\
\hline
\sy{0} &\td & \sy{0} & \sy{0} &  \sy{0} & \sy{0} & \sy{\sigma^2_{m_{\textsc{b}},1}}\!\!\!\! &\td & {\rm SN}_{\textrm{cov}} & \sy{0} & \sy{0} & \sy{0} & \sy{0} & \sy{0}  & {\rm SN}_{\textrm{cov}} & \td & {\rm SN}_{\textrm{cov}} \\
\td    &\td & \td    & \td    &\td     & \td    & \td    & \td    &\td & \td    &\td     &\td &\td & \td  & \td & \td & \td \\
\sy{0} &\td & \sy{0} & \sy{0} & \sy{0} & \sy{0} & {\rm SN}_{\textrm{cov}} & \td & \sy{\sigma^2_{m_{\textsc{b},\textrm{ncc}}}}\!\!\!\! & \sy{0} & \sy{0} & \sy{0} & \sy{0} & \sy{0}  & {\rm SN}_{\textrm{cov}} & \td & {\rm SN}_{\textrm{cov}}\\
\hline
\sy{0} &\td & \sy{0} & \sy{0} & \sy{0} & \sy{0} & \sy{0} &\td & \sy{0} &  \sy{\sigma_{\it HST}^2} & \sy{0}\!\!\!\! & \sy{0} & \sy{0} & \sy{0}  & \sy{0} & \td & \sy{0} \\
\sy{0} &\td & \sy{0} & \sy{0} & \sy{0} & \sy{0} & \sy{0}   & \td &  \sy{0} & \sy{0} & \sy{\sigma_{\it Gaia}^2} & \sy{0}\!\!\!\! & \sy{0} & \sy{0} & \sy{0}  & \td & \sy{0} \\
\sy{0} &\td & \sy{0} & \sy{0} & \sy{0} & \sy{0} & \sy{0}  &\td &   \sy{0} & \sy{0} & \sy{0} & \sy{\sigma_{\rm grnd}^2}\!\!\!\! & \sy{0} & \sy{0} & \sy{0}  & \td & \sy{0}  \\
\sy{0} &\td & \sy{0} & \sy{0} & \sy{0} & \sy{0} & \sy{0} &  \td & \sy{0}  & \sy{0} & \sy{0} & \sy{0} & \sy{\sigma_{\mu,{\rm N4258}}^2}\!\!\!\! & \sy{0} & \sy{0}  & \td & \sy{0} \\
\sy{0} &\td & \sy{0} & \sy{0} & \sy{0} & \sy{0} & \sy{0}  &\td &  \sy{0} & \sy{0}  & \sy{0} & \sy{0} & \sy{0} & \sy{\sigma_{\mu,{\rm LMC}}^2} & \sy{0}  & \td & \sy{0}  \\
\hline
\sy{0} &\td & \sy{0} & \sy{0} & \sy{0} & \sy{0} & {\rm SN}_{\textrm{cov}} &\td &  {\rm SN}_{\textrm{cov}}  & \sy{0} & \sy{0} & \sy{0} & \sy{0} & \sy{0}  & \sy{\sigma^2_{m_{\textsc{b}},z,1}}\!\!\!\!  & \td & {\rm SN}_{\textrm{cov}}  \\
\td    &\td & \td    & \td    &\td     & \td    & \td    & \td    &\td & \td    &\td     &\td &\td & \td  & \td & \td & \td \\
\sy{0} &\td & \sy{0} & \sy{0} & \sy{0} & \sy{0} & {\rm SN}_{\textrm{cov}} &\td &  {\rm SN}_{\textrm{cov}}  & \sy{0} & \sy{0} & \sy{0} & \sy{0} &  \sy{0} & {\rm SN}_{\textrm{cov}} &  \td  & \sy{\sigma^2_{m_{\textsc{b},z,\textrm{nhf}}}}\!  ,  \end{pmatrix}
\nonumber
\eq
where $\sigma_{{\rm tot},1}^2$ representing the $n\times n$ covariance matrix for the Cepheids in the first host is expanded as
\bq
\sigma_{{\rm tot},j}^2=\!\begin{pmatrix}  \sy{\sigma_{{\rm tot},1,1}^2}\!\!\!\! &\td & C_{1,1,n,bkgd} \\
\td &\td & \td \\
C_{1,n,1,bkgd} &\td &  \sy{\sigma_{{\rm tot},1,n}^2}\\
\end{pmatrix} ;
\nonumber
\eq

\bq
L=\begin{array}{ll}
\left( \begin{array}[c]{cccccccccccc}
\sy{1} &\td & \sy{0} & \sy{0} & \sy{1} & \sy{0} & \sy{0} &\lp_{N,1}-1            & \sy{0} & \oh_{N,1}              & \sy{0} & \sy{0} \\  
\td    &\td & \td    & \td    & \td    & \td    & \td    &\td                             & \td    & \td                    & \td    & \sy{0} \\  
\sy{0} &\td & \sy{1} & \sy{0} & \sy{1} & \sy{0} & \sy{0} &\lp_{N,nh}-1           & \sy{0} & \oh_{N,nh}              & \sy{0} & \sy{0} \\  
\hline
\sy{0} &\td & \sy{0} & \sy{1} & \sy{1} & \sy{0} & \sy{0} &\lp_{\textrm{N4258}}-1 & \sy{0} & \oh_{\textrm{N4258}} & \sy{0} & \sy{0} \\  
\sy{0} &\td & \sy{0} & \sy{0} & \sy{1} & \sy{0} & \sy{1} &\lp_{\textrm{M31}}-1   & \sy{0} & \oh_{\textrm{M31}}   & \sy{0} & \sy{0} \\  
\sy{0} &\td & \sy{0} & \sy{0} & \sy{1} & \sy{1} & \sy{0} &\lp_{\textrm{LMC}}-1   & \sy{0} & \oh_{\textrm{LMC}}    & \sy{1} & \sy{0} \\  
\hline
\sy{1} &\td & \sy{0} & \sy{0} & \sy{0} & \sy{0} & \sy{0} &\sy{0}                          & \sy{1} & \sy{0}                 & \sy{0} & \sy{0} \\  
\td    &\td & \td    & \td    & \td    & \td    & \td    &\td                             & \td    & \td                    & \td    & \sy{0} \\  
\sy{0} &\td & \sy{1} & \sy{0} & \sy{0} & \sy{0} & \sy{0} &\sy{0}                          & \sy{1} & \sy{0}                 & \sy{0} & \sy{0} \\  
\hline
\sy{0} &\td & \sy{0} & \sy{0} & \sy{1} & \sy{0} & \sy{0} &\sy{0}                          & \sy{0} & \sy{0}                 & \sy{0} & \sy{0} \\
\sy{0} &\td & \sy{0} & \sy{0} & \sy{1} & \sy{0} & \sy{0} &\sy{0}                          & \sy{0} & \sy{0}                 & \sy{0} & \sy{0} \\
\sy{0} &\td & \sy{0} & \sy{0} & \sy{0} & \sy{0} & \sy{0} &\sy{0}                          & \sy{0} & \sy{0}                 & \sy{1} & \sy{0} \\
\sy{0} &\td & \sy{0} & \sy{1} & \sy{0} & \sy{0} & \sy{0} &\sy{0}                          & \sy{0} & \sy{0}                 & \sy{0} & \sy{0} \\
\sy{0} &\td & \sy{0} & \sy{0} & \sy{0} & \sy{1} & \sy{0} &\sy{0}                          & \sy{0} & \sy{0}                 & \sy{0} & \sy{0} \\
\hline
\sy{0} &\td & \sy{0} & \sy{0} & \sy{0} & \sy{0} & \sy{0} &\sy{0}                          & \sy{1} & \sy{0}                 & \sy{0} & \sy{-1}\\
\td    &\td & \td    & \td    & \td    & \td    & \td    &\td                             & \td    & \td                    & \td    & \td    \\
\sy{0} &\td & \sy{0} & \sy{0} & \sy{0} & \sy{0} & \sy{0} &\sy{0}                          & \sy{1} & \sy{0}                 & \sy{0} & \sy{-1}\\
\end{array} \right)

&
\begin{array}[c]{@{}l@{\,}l}
\left. \begin{array}{c} \vphantom{\lp^l_{N}} \\ \vphantom{\lp^l_{N}} \\ \vphantom{\textrm{\LARGE HELLO}} \end{array} \right\} & \text{Cepheids in SN~Ia hosts} \\
\left. \begin{array}{c} \vphantom{\lp^l_{N}} \\ \vphantom{\lp^l_{N}} \\ \vphantom{\textrm{\LARGE HELLO}} \end{array} \right\} & \text{Cepheids in anchors or non-SN~Ia hosts\hspace{0.28in}} \\
\left. \begin{array}{c} \vphantom{\lp^l_{N}} \\ \vphantom{\lp^l_{N}} \\ \vphantom{\textrm{\LARGE HELLO}} \end{array} \right\} & \text{SNe~Ia in Cepheid hosts} \\
\left. \begin{array}{c} \vphantom{\lp^l_{N}} \\ \vphantom{\lp^l_{N}} \\ \vphantom{\lp^l_{N}} \\ \vphantom{\lp^l_{N}} \\ \vphantom{\textrm{\LARGE HELLO}} \end{array} \right\} & \text{External constraints} \\
\left. \begin{array}{c} \vphantom{\lp^l_{N}} \\ \vphantom{\lp^l_{N}} \\ \vphantom{\textrm{\LARGE HELLO}} \end{array} \right\} & \text{SNe~Ia in the Hubble flow} \\
\end{array}
\end{array}
\nonumber \eq

\bq
q=\begin{pmatrix} \mu_{0,1} \\\td \\  \mu_{0,\textrm{nh}} \\  \Delta \mu_{\rm N4258} \\ M_{H,1}^W\! \\ \Delta \mu_{\rm LMC} \\ \mu_{M31} \\ b_W \\ M_B^0\! \\ Z_W \\ \Delta {\rm zp} \\ 5\,\log\,\textrm{H}_0 \end{pmatrix} .
\nonumber
\eq

The number of Cepheid hosts is \textit{nh}, the number of SNe~Ia in these hosts is \textit{ncc}, and the number of SNe~Ia in the Hubble flow is \textit{nhf}. Period uncertainties are comparatively negligible \citep{Yuan:2021_N4051}. $\sigma_{\it HST}, \ \sigma_{\it Gaia}$ denote the uncertainties in  $M_{H,1,Gaia}^W$ and $M_{H,1,HST}^W$ as derived from parallaxes, respectively, while $\sigma_{\it grnd}$ denotes the uncertainty in ground-based photometry. The term ${\rm SN}_{\textrm{cov}}$ is the covariance between SNe, $Z_{\textrm{cov}}$ is the metallicity covariance given later in Equation~(\ref{eq:covmet}), and $C_{i,j,k,bkgd}$ is the background covariance given later in Equation~(\ref{eq:coverr}). The $\chi^2$ statistic is given as \bq \chi^2=(y-Lq)^TC^{-1}(y-Lq)\label{eq:chisq} \eq and maximum-likelihood parameters are given as $q_{\textit{best}}=(L^TC^{-1}L)^{-1}L^TC^{-1}y$, while the standard errors and the covariance matrix of the parameters come from the matrix $cq=(L^TC^{-1}L)^{-1}$.   The value of H$_0$ is derived from the final entry of ${\bf q}$, $5\,\log\,$H$_0$, and its error from the square root of the corner entry of ${\bf cq}$.  This provides a compact form for storing and transmitting {\it the full dataset} used to determine H$_0$, to edit or augment it, and to enable others to determine its value. The above formalism is the same as used by R16 with the additions of SN and Cepheid covariance.  We will also derive the parameters independently of the analytical solution in {\bf \S\ref{sc:5.1}} by sampling the $\chi^2$ statistic using a Markov Chain Monte Carlo (MCMC) approach to verify the analytical result with a different methodology.

\section{Cepheid Observations in SN~Ia Hosts and the Maser Host NGC$\,$4258\label{sc:3}}
\subsection{Optical Cepheid Discovery\label{sc:3.1}}
   
The SH0ES program has been selecting the SNe~Ia that are most suitable for calibrating their fiducial luminosity (with selection criteria given in \S\ref{sc:1.1}) through observations of Cepheids in their hosts.   The results here include a complete sample of all such SNe~Ia of which we are aware within $z<0.011$ (40 objects), with the addition of two beyond this limit that are useful for testing the range of Cepheid distance measurements, for a total of 42 SN~Ia calibrators.

Fig.~\ref{fg:hstobs} and Table~\ref{tb:wfc3} show the sources of the  {\it HST} data obtained for every SN~Ia and host we measured, gathered from the indicated  {\it HST}  cameras, filters, and time periods. All of these publicly available data are readily obtained from the Mikulski Archive for Space Telescopes (MAST). The imaging data are used for both Cepheid discovery and their flux measurement. For the former, a campaign using a filter with central wavelength in the visual band and $\sim 11$--12 epochs with nonredundant spacings spanning $\sim 60$--100~days is optimal to identify Cepheid variables by their unique light curves and large amplitudes ($\sim 1$~mag peak-to-trough), and accurately measure their periods \citep{madore91,saha96,Stetson:1996}. Image subtraction may find additional Cepheids \citep{Bonanos:2003}, but these objects will be subject to greater photometric biases owing to blends which suppress their amplitudes and chances of discovery in time-series data \citep{Ferrarese:2000}. The data were collected from $\sim 150$ {\it HST} orbits with WFC3-NIR and $\sim 700$ orbits obtained for the optical identification (350 from WFC3, 170 from ACS, and 180 from WFPC2), including $\sim 200$ orbits from NICMOS superseded by WFC3-NIR, for a total of $\sim 1050$ {\it HST}  orbits.  Additional observations of Cepheids in the MW and LMC anchors utilized $\sim 200$ orbits or snapshots.
    
Earlier efforts contribute $\sim 200$~orbits of imaging of these hosts: 35 orbits from the {\it HST} Key Project, 105 from the SNe~Ia Luminosity Program, 36 from \citet{Mager:2013}, and 50 from \citet{macri06}, with the remaining $\sim 800$ from SH0ES.

The procedure for identifying Cepheids from time-series optical data in visual or white light bands has been described extensively \citep{saha96,Stetson:1996,riess05,macri06,Hoffmann:2016}; details of the procedures followed for this sample are presented by \citet{Yuan:2021_SN}.  These procedures utilize the DAO suite of software tools \citep{Stetson:1987,Stetson:1994} for crowded-field PSF photometry, and are similar to those used previously by the SH0ES team \citep{Hoffmann:2016} and to a large extent by the first generation of {\it HST}-based H$_0$ measurements.  Past work has demonstrated that the use of different photometry algorithms yields a largely overlapping list of Cepheids with similar periods and photometry \citep{Ferrarese:1998}.
   
\begin{figure}[ht]
\includegraphics[width=\textwidth]{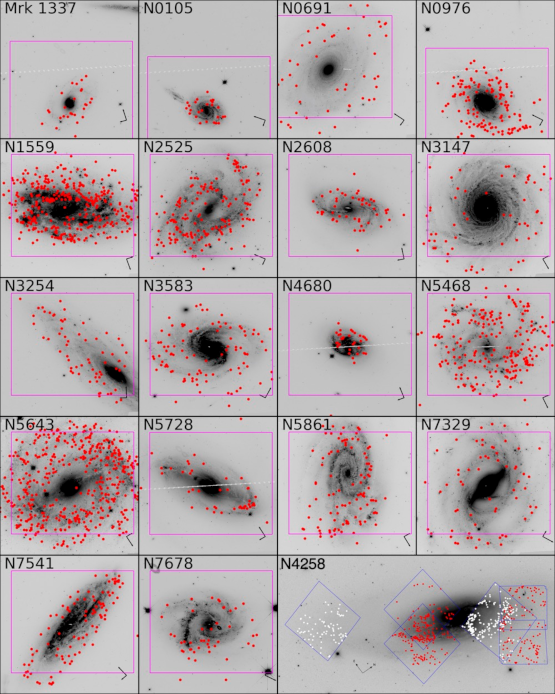}
\caption{\label{fg:hstfovs} Images of 18 newly-observed hosts of 23 SNe~Ia and NGC$\,$4258. Each image is of the Cepheid host indicated.  The magenta outline shows the $2.7' \times 2.7'$ field of the WFC3/NIR observations.  Red dots indicate the positions of the Cepheids.  Compass indicates north (long axis) and east (short axis). The NGC$\,$4258 image shows Cepheids from \citet{macri06} in white and new ones from \citet{Yuan:2021_N4258} in red.} 
\end{figure}

The end result is a set of high-confidence Cepheids which have passed selection and quality cuts described by Y22b, with periods and mean colors ({\it F555W--F814W}) measured in the {\it HST} WFC3 photometric system. For each Cepheid, we estimate a precise position in the WFC3/NIR {\it F160W} images using a geometric transformation derived from the optical images using bright and isolated stars, with resulting mean position uncertainties for the variables $<0.03$~pix.  The positions of the Cepheids are indicated in Fig.~\ref{fg:hstfovs}.

We present composite light curves of all Cepheids with $10 < P < 80$~d in Fig.~\ref{fg:cmplc} based on $\sim 60,000$ individual {\it F555W} or {\it F350LP} photometric measurements, which identify these as bona-fide classical Cepheids with the characteristic sawtooth light-curve shape of fundamental-mode pulsators. There are more subtle Cepheid light-curve features which are not apparent in individual examples at these distances.  However, we can leverage the statistical power of the sample to look for these features as a strong validation test of the universality of Cepheids, as presented below.

\subsection{Cepheid Validation: The Hertzsprung Progression\label{sc:3.2}}
   
The {\it shapes} of Cepheid light curves change in subtle but characteristic ways \citep[often referred to as the ``Hertzsprung Progression'';][]{Hertzsprung:1926} as a function of period, visible only with high signal-to-noise-ratio (SNR) observations. \citet{Hertzsprung:1926}  noted the characteristic ``quick rise and slow decrease'' for most periods but also a small range of periods (9--13~d) where the light curves are quite symmetric, and the presence of a ``bump'' or small local maximum whose phase occurs earlier with increasing period, appearing on the descending phase at $P \approx 6$~d, merging with the main peak at $P \approx 10$~d, and appearing on the rising phase at $P \approx 10$--15~d before disappearing (variables where the bump is apparent are sometimes referred to as ``bump'' Cepheids). While the general fast rise and slow decline is a characteristic of a star pulsating in the fundamental mode, the ``bump'' is understood through modeling to arise from a 2:1 resonance between the fundamental and second overtone \citep{Cox:1993,Bono:2002} and is a unique feature of classical Cepheids.  Cepheids with $P>40$~d occur at the high-mass end of the luminosity function and are thus rare, with the MW hosting only a few examples which show a broadening of the light curve and a decrease in amplitude.  

\begin{figure}[hb]
\includegraphics[width=\textwidth]{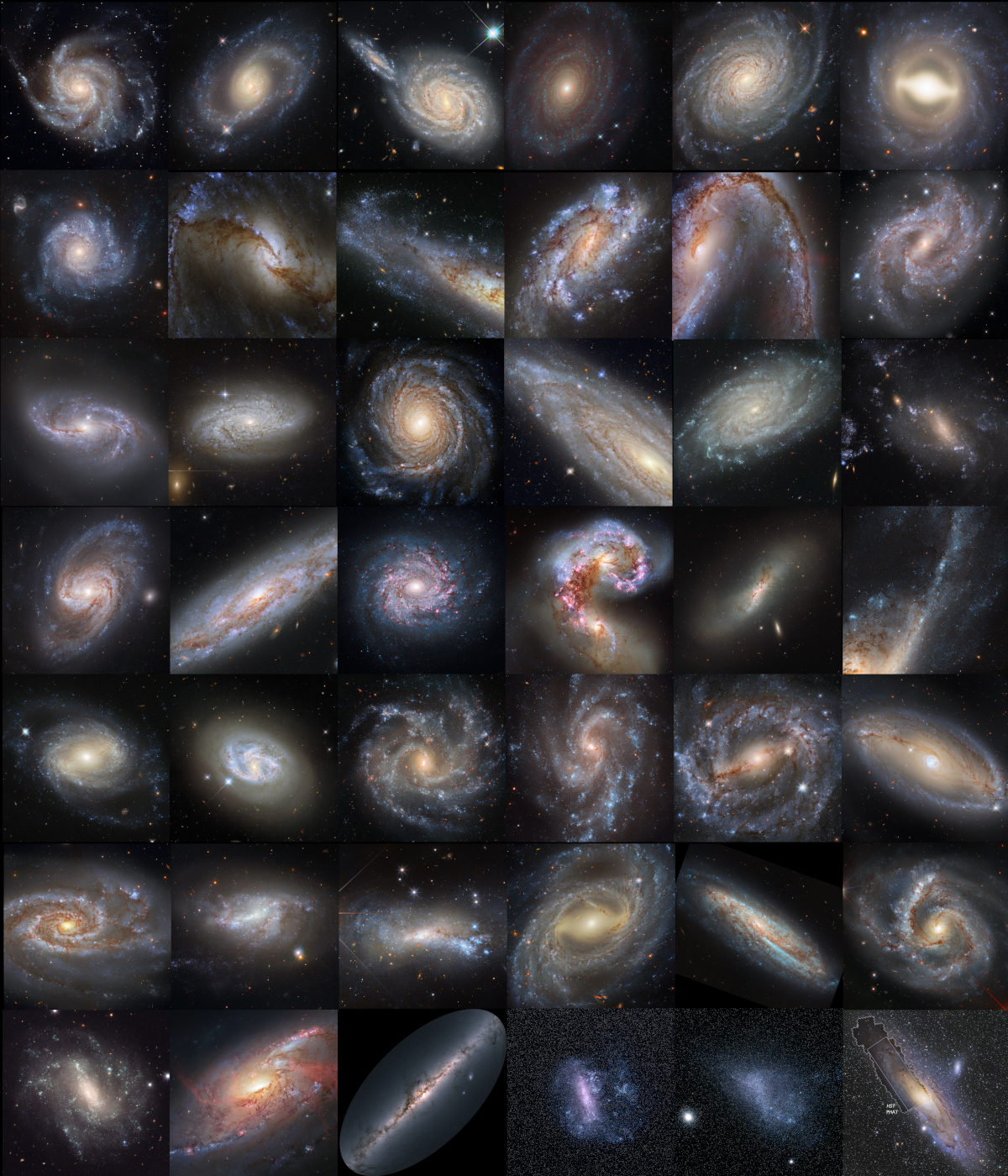}
\caption{\label{fg:hstcol} \bf Pseudocolor images of all Cepheid-bearing galaxies analyzed in this work. From top left, 37 hosts of 42 SNe~Ia presented in the same order as Table~\ref{tb:wfc3}. The last row includes our three anchors (NGC$\,$4258, MW, LMC) and two supporting galaxies (SMC, M31). Galaxies are presented at arbitrary plate scales, though in most cases the panels encompass the entire ACS or WFC3/UVIS field of view. Credits: SN hosts and NGC$\,$4258 -- ESA {\it Hubble} site; MW, LMC, and SMC -- ESA {\it Gaia} site; M31 -- STScI.} 
\end{figure}

\clearpage

\begin{figure}[t]
\begin{center}
\includegraphics[width=0.78\textwidth]{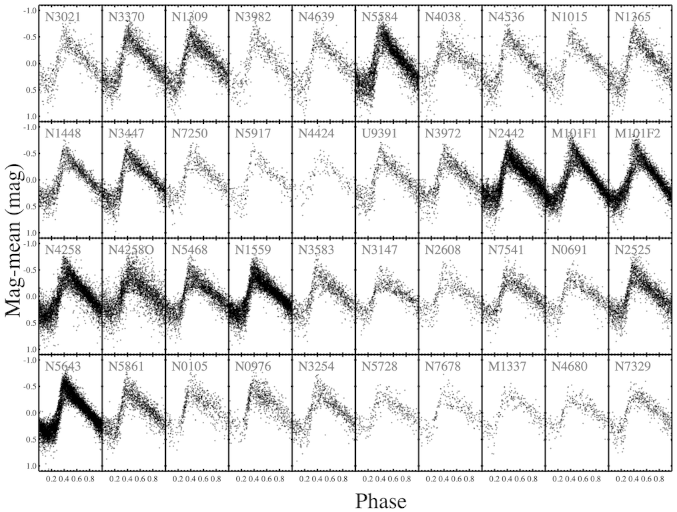}
\end{center}
\caption{\label{fg:cmplc} Composite visual ($F555W$) or white-light ($F350LP$) Cepheid light curves.  Each {\it HST} Cepheid light curve with $10 < P < 80$~d is plotted after subtracting the mean magnitude and determining the phase of the observation.  }
\end{figure}

\begin{figure}[b]
\begin{center}
\includegraphics[width=0.78\textwidth]{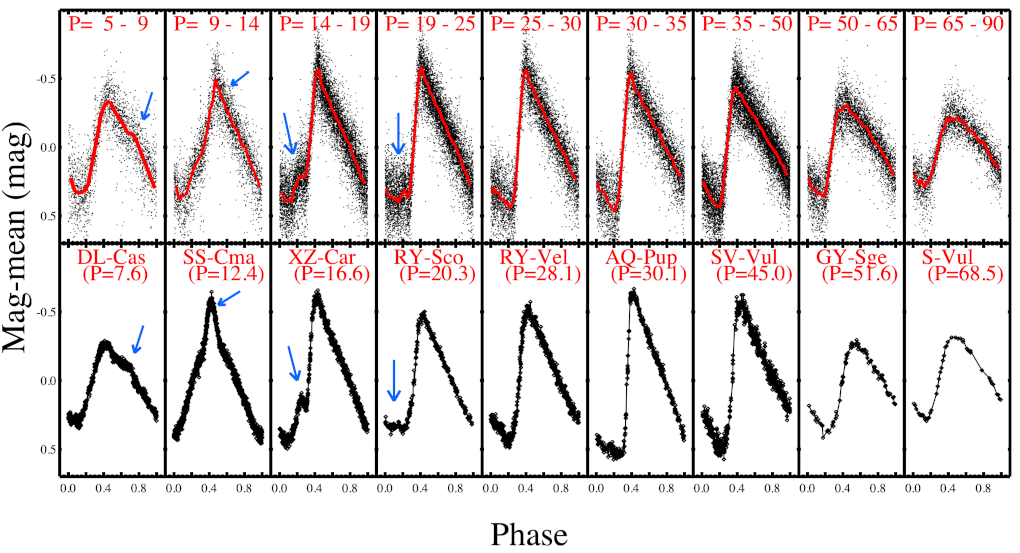}
\end{center}
\caption{\label{fg:hzprg} Composite visual ($F555W$) or white-light ($F350LP$) Cepheid light curves binned by period (upper) and compared with individual MW Cepheids near the middle period of the bin.  The ``Hertzsprung Progression'' (relation between light-curve shape and period) is apparent, including subtle features like the progression in phase of a resonance ``bump'' between the second overtone and fundamental pulsation for $P<20$~d.  The red line is a cubic spline constrained by the averages of bins in phase.}
\end{figure}
\clearpage
   
By binning all of the extragalactic light curves for the 37 SN~Ia hosts and NGC$\,$4258 by period as shown in Fig.~\ref{fg:hzprg}, a striking display of this more subtle light-curve structure emerges.  The resonance-induced bump is indicated on the descending phase at $P = 5$--9~d with the MW Cepheid DL~Cas ($P=7.6$~d) shown for comparison, the narrow-peaked, symmetric curve at $P=9$--14~d as in SS~CMa ($P=12.4$~d), the resonance bump on the rising phase in the $P=14$--19~d as in XZ~Car ($P=16.6$~d), before transitioning to sawtoothed curves in the next 3 bins ($P=19-35$~d). Beyond this range, we see the gradual transition to flatter and more sinusoidal curves matching the MW sequence of SV~Vul ($P=45$~d), GY~Sge ($P=52$~d), and S~Vul ($P=69$~d). 

While our pipeline selection requires candidate Cepheids to have amplitudes in the characteristic range of 0.2--1~mag, we impose no requirement regarding the presence of these more subtle light-curve features, which are not detectable at the data quality typical of single extragalactic Cepheids \citep{Hoffmann:2016}.  The presence of these features in aggregate light curves at the appropriate periods offers an additional level of scrutiny and validation of these sources as classical Cepheids, as well as the means for a more detailed comparison to those in the MW.  The impressive similarity of these light-curve sequences demonstrates the universality of the physics producing these pulsating standard candles, and provides an important test of their consistency along the distance ladder.  The appearance of the Hertzsprung Progression among the extragalactic sample demonstrates these objects are bona-fide Cepheid variables.

\begin{figure}[b]
\includegraphics[width=\textwidth]{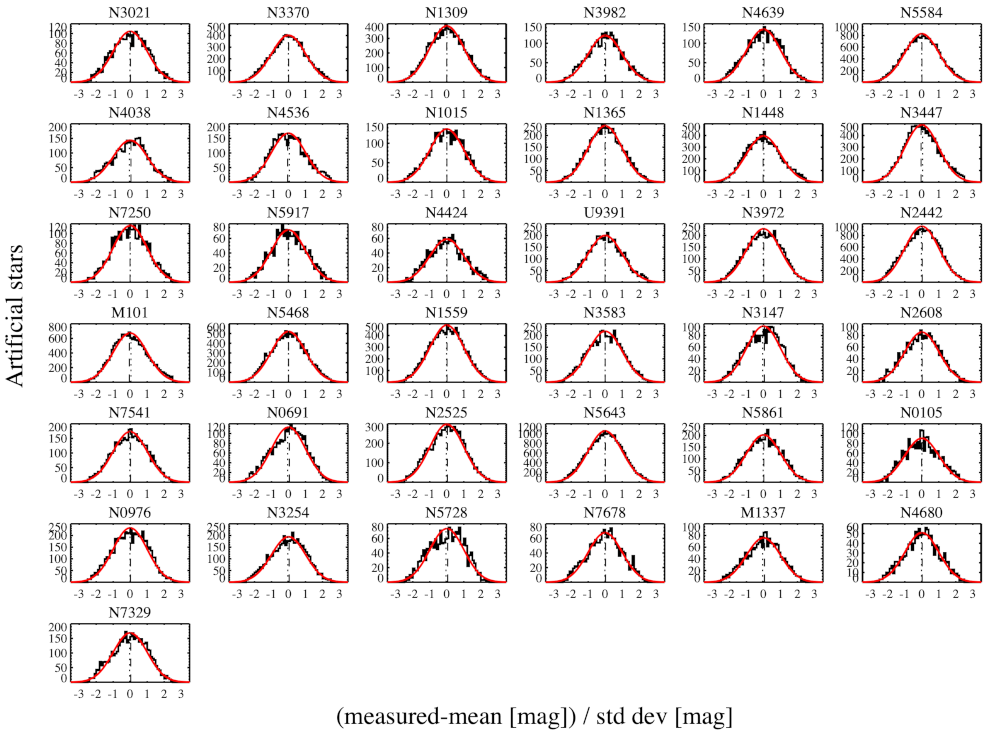}
\caption{\label{fg:artstar} Distributions of recovered artificial stars in each host.  For each real Cepheid, 100 artificial stars of the expected Cepheid magnitude (based  on its period and the \PLs relation) are added to the host images within an annulus with an outer radius of $2.4''$ and measured following the same procedures as for the Cepheids to produce a distribution in magnitudes (log flux).  The distributions for each host are combined after subtracting their mean and dividing by their standard deviation.  They compare well to a Gaussian with a difference in median and mean (dashed and dotted lines) of $\sim 0.03\sigma$.  }
\end{figure}

\subsection{NIR Photometry and Validation in NGC$\,$4258\label{sc:3.3}}
   
Following the same procedures described by R16, we perform NIR photometry of the Cepheids using their positions derived and fixed from the higher-resolution optical data.  The NIR images of all Cepheids were reprocessed for this analysis using revised calibrations by the STScI pipeline (including new flatfields and geometric-distortion tables), and we were able to predict the expected NIR positions of the Cepheids found in the optical data with enhanced precision compared to our previous reduction \citep{Hoffmann:2016}. Additional sources found in the NIR images are fit simultaneously with the Cepheids using an empirical model of the PSF derived from the solar-analogue star P330E.  The background calibration\footnote{Sometimes called a ``crowding correction'' because it adds the mean level of randomly superimposed sources to the initial background estimate derived from the unresolved or constant sky level.} and photometric uncertainties are determined from retrieval of 100 locally-placed artificial stars per Cepheid as described by R11 and R16.

In Fig.~\ref{fg:artstar} we show the distributions of the artificial-star measurements divided by their standard deviation, which demonstrates that they are well approximated by Gaussian distributions {\it in magnitude space} around their mean out to $3\sigma$, a consequence of the log-normal distribution of underlying surface brightness fluctuations (primarily red giants, with larger and brighter collections increasingly rare) in the NIR.  The mean difference between the mean and median of the artificial-star distributions, which vanishes for a true Gaussian, is $0.03\sigma$.  Averaged across all SN hosts, the difference between mean and median of the magnitude distribution drops to $0.01\sigma$ or 4~millimag, showing no apparent correlation between its third moment (skewness) and the host distance.  The distribution for the geometric calibrator, NGC$\,$4258, appears similar to that of the others, with a difference between mean and median of 0.01~mag.  These measurements justify the use of Gaussian statistics in magnitude space in the calculations to follow.

The accuracy of the background estimates owes to Cepheids being randomly superimposed on scenes, a consequence of our perspective whose local levels can be measured {\it statistically}.  A caveat to this approach would be the presence of associated flux (colocated with the Cepheid) which becomes important {\it only} if it is then resolved for nearby Cepheids but not for distant ones.  The level of such ``associated flux'' has been measured statistically from hundreds of Cepheids in M31 by \citet{Anderson:2018} to be $\sim 7$~millimag at distances beyond a few Mpc, and is due to associated open clusters that would not be resolved at those distances.  This term is explicitly included here in the background estimates, in order to compensate for this effect.  An additional and direct consequence of a potential miscalibration of the background, independent of Cepheid mean flux, would be a change in apparent light-curve amplitude.  \citet{Riess:2020} determined that the NIR amplitudes of Cepheids in SN hosts are fully consistent with those in the Milky Way, yielding an independent upper limit of 0.03~mag for the possible misestimation of the background.  The sensitivity of H$_0$ to the unresolved background can be further mitigated by the use of a distant anchor with similar background as the SN~Ia hosts such as NGC$\,$4258; this will be addressed in \S\ref{sc:4}.  

\begin{figure}[t]   
\begin{center}
\includegraphics[width=0.56\textwidth]{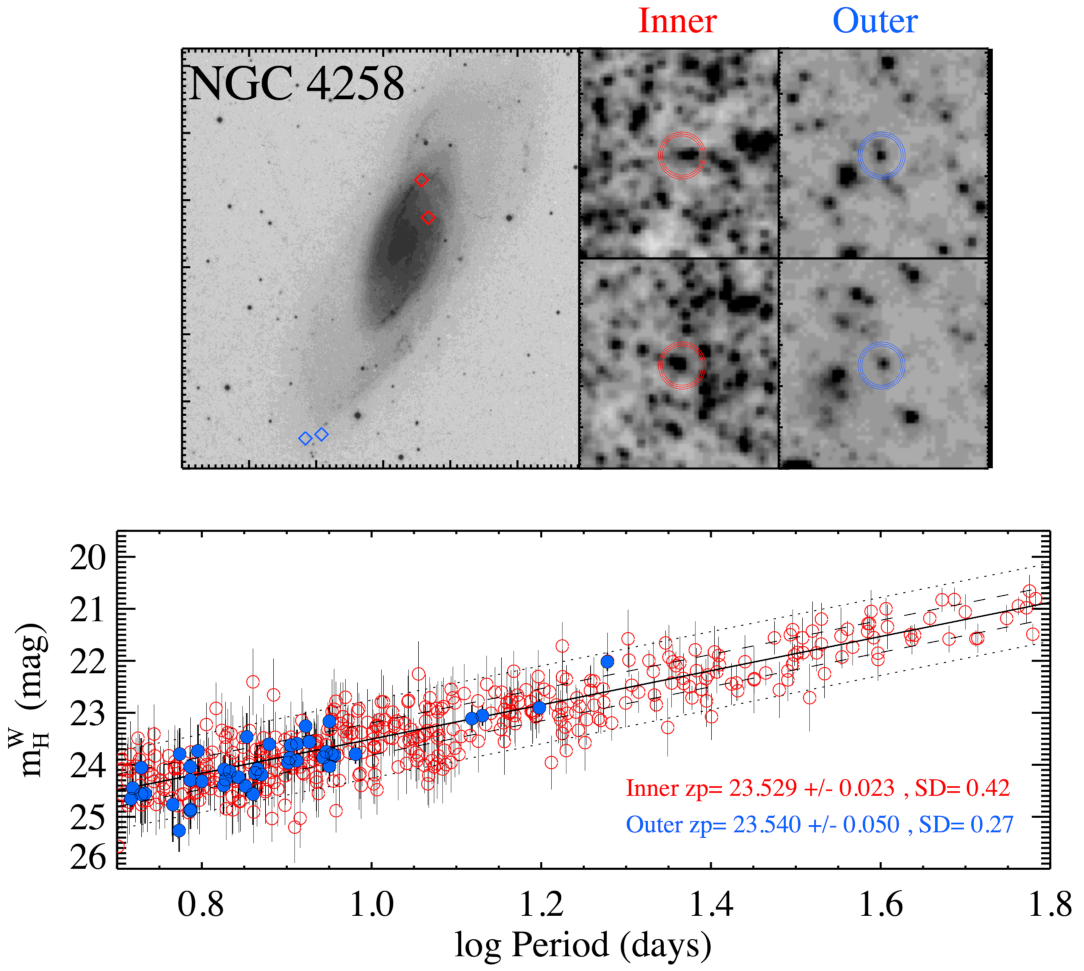}
\end{center}
\caption{\label{fg:crowd} Comparison of Cepheids measured in a dense (inner) field (in red) and sparse (outer) field (in blue) of NGC$\,$4258.  Because these Cepheids are at the same distance, the comparison shows the accuracy of the background estimates which differ in the mean over the same sampled range, $ 0.7 < \log (P/{\rm day}) < 1.2 $, by 0.45~mag (relative to the Cepheids), yet yield a consistent intercept to $\sigma \approx 0.05$~mag.   The difference in metallicity between the samples, $\Delta$[O/H] $= 0.08$~dex, corresponds to a difference of 0.02~mag, smaller than the precision of this comparison.}
\end{figure}

In Fig.~\ref{fg:crowd} we provide a strong test of the background estimation by comparing the Cepheid photometry in dense (inner) and sparse (outer) regions of NGC$\,$4258.  Because the Cepheids in both fields are at the same distance from us and the metallicity gradient in NGC$\,$4258 is small \citep{Bresolin:2016}, an apparent difference of the dereddened magnitudes would be a consequence of misestimating the background in the dense field. The difference in intercept (i.e., distance) is 0.01~mag and well within the indicated errors of the means, demonstrating that Cepheid PSF photometry is accurate in the presence of the same level of crowded backgrounds seen in the SN hosts..

In Appendix~\ref{sc:appb} we provide an independent validation of the PSF photometry using aperture photometry, a method which is accurate when the background is measured from the {\it mean} of pixels in concentric annuli and uncomplicated to apply, albeit less precise than the standard approach of using PSFs to model photometry.  This test validates the mean PSF photometry in SN hosts to $\sigma \approx 0.02$~mag. In that Appendix we further perform an additional ``null test'' of the background estimates by regressing them against the distance ladder fit residuals, finding a dependence of $0.010 \pm 0.014$~mag per magnitude of source background (in the sense of overestimating the background but with no significance). 

In \S\ref{sc:7.1} we review multiple tests of Cepheid PSF photometry in addition to six strong tests of background estimates in the presence of crowded backgrounds, all of which indicate that the Cepheid measurements are accurate.

\clearpage

\subsection{Dereddened Magnitudes\label{sc:3.4}}

The SH0ES program uses observations in  {\it HST}  filters at known Cepheid phases in optical ({\it F555W}, {\it F814W}) and NIR ({\it F160W}) bands to correct for the effects of interstellar dust and the finite width in temperature of the Cepheid instability strip. We employ NIR ``Wesenheit'' magnitudes \citep{madore82} to deredden Cepheids throughout, defined as
\bq m_H^W = m_{H} - R \, (m_{V}\!-\!m_{I}), \label{eq:wh} \eq 
\noindent where $m_H=$ {\it F160W}, $m_V=$ {\it F555W}, and $m_I=$ {\it F814W} and $V-I=m_V-m_I$ in the  {\it HST} system, and $R\equiv A_H/(A_V\!-\!A_I)$.   Wesenheit magnitudes are {\bf not} conventional magnitudes, which compare the brightness of one star to another; rather, they are used to compare the brightness of one standardized candle to another through the removal of  their unequal extinction, as reviewed in Appendix~\ref{sc:appd}. While the value of $R$ obtained from well-characterized extinction laws for these bands is $\sim 0.4$, we note that the correlation between Cepheid intrinsic color and luminosity at a fixed period has the same sense as extinction (cooler is fainter), and is similar in size with an intrinsic value of $\sim 0.6$ (as discussed in \S\ref{sc:6.2}).  Therefore, the value of $R$ derived for extinction effectively also reduces the intrinsic scatter caused by the breadth of the instability strip.  We analyze the sensitivity of H$_0$ to values of $R$ derived from different extinction laws in \S\ref{sc:6.3}. In Appendix~\ref{sc:appd}, we discuss pitfalls associated with varying $R$ in Equation~(\ref{eq:wh}) {\it between} galaxies if the intrinsic color is not first subtracted from the observed color\footnote{This might be pursued to allow  the extinction law to vary in every host, but if the intrinsic color is not first subtracted, it has the unintended consequence of producing a large variation in the luminosity of the standard candle itself which is unrelated to dust, is inconsistent with the premise of a distance ladder where stars (once standardized) have luminosities independent of the rung they live on, and most importantly is not supported by the data as shown in Appendix~\ref{sc:appd}.} \citep{Follin:2017, Mortsell:2021}. 

To avoid a magnitude bias, we include only Cepheids with periods above the completeness limit of detection in our primary fit for each host (Y22b).  The measurements of Cepheids in SN~Ia hosts are provided in Table~\ref{tb:wfc3irceph}, while Table~\ref{tb:nirpl} summarizes the properties of the resulting NIR \PLs relations. We identify a number of improvements realized here since our previous Cepheid measurements in SN hosts $\sim 6$~yr ago (R16, H16), and $\sim 16$~yr ago for NGC$\,$4258 \citep{macri06}.

\begin{deluxetable}{lllrccccccl}[t]
\tablenum{2}
\tablewidth{0pc}
\tabletypesize{\scriptsize}
\tablecaption{WFC3/NIR Cepheids\label{tb:wfc3irceph}}
\tablehead{\colhead{Field} & \colhead{$\alpha$} & \colhead{$\delta$} & \colhead{ID} & \colhead{$P$} & \colhead{Color$^a$} & \colhead{$\sigma_{\rm col}$} & \colhead{{\it F160W}} & \colhead{$\sigma_{\rm tot}$} & \colhead{[O/H]} & \colhead{Note}\\[-0.25cm] 
\colhead{} & \multicolumn{2}{c}{(J2000)} & \colhead{} & \colhead{[d]} & \multicolumn{4}{c}{[mag]} & \colhead{[dex]}}
\startdata
M101 & 210.91148 & 54.357230 & 54672 &  6.853  &  0.91  &  0.11  &  24.04  &  0.44  &  0.05  & {\it HST}\\
M101 & 210.88464 & 54.338980 & 110830 &  6.873  &  1.05  &  0.19  &  23.36  &  0.62  &  0.09  & {\it HST}
\enddata
\tablecomments{(a): {\it F555W--F814W}.}
\end{deluxetable}
\vspace{-36pt}
\begin{enumerate}
\item Sky annuli: The size of the annulus used to estimate the level of the sky value of the region around the Cepheid (after subtracting all detected sources) in $F160W$ was reduced in size to inner and outer radii of 0\farcs24 and 0\farcs8 (from 0\farcs96 and 1\farcs44 in R16) based on extensive simulations.  This determination of the sky level is more precise, although it requires taking into account the contribution to the sky from the wings of the PSF; the resulting offset of  0.008~mag is robustly determined from bright stars and corrected in the final photometry.  The random sky error is propagated to the photometry error through the sampling of sky values in the artificial-star analyses).

\item Determination of the background and covariance: As in R16, artificial stars are added  in the {\it F160W} images in the vicinity of each Cepheid at the apparent magnitude expected from its period and trial fits of the \PLs relation. The difference between the input and recovered magnitudes is used to refine the initial estimate of the sky background and revise the Cepheid photometry.  This process is necessarily iterative as the \PLs relation used to predict the magnitude of a given Cepheid from its period is determined {\it after} correcting the photometry based on the retrieval of the artificial stars. Thanks to greater processing power, we increased the number of iterations since R16, and found that in some cases the iterations in our previous work were not adequate (as they had not fully converged). The new process fully converges with improved determination of the trial intercepts and slopes, and we use the final iteration to estimate a systematic uncertainty of 20\% of the background (in units of the Cepheid magnitude correction) as the covariance ``error floor'' of any pair of Cepheids, $j,k$ in the same $i$-th host, given by 

\bq C_{i,j,k,{\rm bkgd}}=0.2^2(\textit{bkgd}_{i,j})(\textit{bkgd}_{i,k}) . \label{eq:coverr} \eq 

This provides a systematic uncertainty in the range of 0.03--0.06~mag for all Cepheids within each host, with the term {\it bkgd} representing the change in Cepheid magnitude due to the addition of the mean level of the crowded background from unresolved sources derived from the artificial stars.   The artificial star magnitudes are determined from the trial P-L relations of each host independently from every other host so there is no source of background covariance between different hosts.

\item Reference Files: The photometry benefits from the latest STScI data pipelines, including new flatfields (with better ``blob'' mapping), bias frames, long-history dark frames, pixel-based CTE corrections for optical data, and geometric-distortion corrections, yielding improved alignment between the optical and NIR frames.

\item Count Rate Nonlinearity (CRNL): We adopt a new calibration of the WFC3/NIR CRNL \citep{Riess:2019}.
By convention, this is applied to Cepheids in anchors between their flux and the background level of Cepheids in SN hosts.

\item M101 long-period Cepheids: A re-examination of the Cepheids in M101, together with simulations, revealed that the baseline of the original monitoring campaign \citep[carried out in 2006 and reported by][]{Mager:2013} was too short to provide reliable periods for apparent Cepheids with $P > 35$~d (equivalent to $1.2\times$ the time span of the observations). Two additional epochs obtained 7~yr later and separated by a week are insufficient to resolve the issue, because at these periods the two epochs provide effectively only one phase measurement, and the prevailing period uncertainty of $>0.5$~d makes the phasing of the two sets unreliable. For that reason we exclude M101 Cepheids with $P>35$~d (about 10\% of the sample used by R16).  

\item We correct Cepheid periods to rest-frame values owing to $(1+z)$ time dilation \citep{Anderson:2019}, a small  (given our typical $z \approx 0.005$) but one-sided effect.

\end{enumerate}

 We note that the Cepheid color measurements, $V\!-\!I$, employed in Eq.~\ref{eq:wh} to determine the baseline value of H$_0$ here and in R16 (as well as most variants) are relatively insensitive to the previously noted improvements to the calibration of the Cepheid optical measurements realized in the last 6 or 16 years and included in Y22a,b. These include the use of artificial stars to estimate the crowded backgrounds, revised flatfields, archival dark frames, updated geometric distortion maps to rectify frames, and pixel-based CTE corrections, most of which were {\it not implemented} by H16 or by \cite{macri06} from which the N$\,$4258 data in H16 were derived.  These improvements cancel to first order in the difference, $V\!-\!I$, as do changes in CCD sensitivity which can decline on orbit or change abruptly when the electronics are refurbished as occurred for ACS in 2009.  However, the use of optical Wesenheit magnitudes to determine H$_0$ without bias (as we explore in \S\ref{sc:6.13}) requires fully-calibrated optical magnitudes rather than only accurate optical colors (as in H16 and provided in R16). The fully-calibrated optical magnitudes available in Y22a,b are suitable for this purpose.

We do not attempt to quantify the impact of each of these improvements individually; however, in the aggregate, matching Cepheids within 1\arcsec\ of those in R16, the net change to Cepheid photometry in {\it F160W} is that 63\% (37\%) of Cepheids are fainter (brighter).  We provide additional details of this comparison in Appendix~\ref{sc:appb}.

\begin{deluxetable}{lrrrrr} [t]
\tablenum{3}
\tablewidth{0pc}
\tabletypesize{\normalsize}
\tablecaption{Properties of NIR $P$--$L$ Relations\label{tb:nirpl}}
\tablehead{\multicolumn{1}{l}{Galaxy} & \multicolumn{3}{c}{Number} & \colhead{$\langle P \rangle$} & \multicolumn{1}{r}{$\langle$[O/H]$\rangle^a$} \\[-0.18cm]
\colhead{} & \colhead{FoV} & \colhead{meas.$^b$} & \colhead{fit$^c$} & \colhead{[day]} & \colhead{[dex]}}
\startdata
M101 &           311 &          260 &          259 &  15.8  &  0.10   \\[-0.1cm]
Mrk1337 &         21 &           20 &           15 &  52.9  &  -0.18   \\[-0.1cm]
N0105 &           32 &            8 &            8 &  41.5  &  -0.13   \\[-0.1cm]
N0691 &           31 &           28 &           28 &  46.5  &  0.09   \\[-0.1cm]
N0976 &           57 &           35 &           33 &  40.2  &  0.02   \\[-0.1cm]
N1015 &           26 &           20 &           18 &  52.5  &  -0.03   \\[-0.1cm]
N1309 &           57 &           53 &           53 &  54.1  &  -0.08   \\[-0.1cm]
N1365 &           66 &           47 &           45 &  29.0  &  -0.14   \\[-0.1cm]
N1448 &           90 &           77 &           73 &  35.2  &  -0.11   \\[-0.1cm]
N1559 &          136 &          110 &          110 &  34.4  &  0.00   \\[-0.1cm]
N2442 &          238 &          177 &          177 &  36.0  &  0.00   \\[-0.1cm]
N2525 &           85 &           73 &           73 &  40.4  &  0.10   \\[-0.1cm]
N2608 &           25 &           22 &           22 &  45.4  &  0.11   \\[-0.1cm]
N3021 &           26 &           16 &           16 &  32.2  &  0.06   \\[-0.1cm]
N3147 &           29 &           28 &           27 &  52.3  &  0.17   \\[-0.1cm]
N3254 &           54 &           48 &           48 &  41.5  &  -0.18   \\[-0.1cm]
N3370 &           82 &           73 &           73 &  42.5  &  -0.12   \\[-0.1cm]
N3447 &          116 &          102 &          101 &  36.0  &  -0.16   \\[-0.1cm]
N3583 &           62 &           54 &           54 &  41.6  &  -0.06   \\[-0.1cm]
N3972 &           66 &           54 &           52 &  32.3  &  0.03   \\[-0.1cm]
N3982 &           31 &           27 &           27 &  29.9  &  -0.14   \\[-0.1cm]
N4038 &           38 &           29 &           29 &  53.6  &  0.03   \\[-0.1cm]
N4424 &           17 &           10 &            9 &  31.1  &  0.06   \\[-0.1cm]
N4536 &           45 &           41 &           40 &  36.0  &  -0.15   \\[-0.1cm]
N4639 &           36 &           30 &           30 &  38.7  &  -0.01   \\[-0.1cm]
N4680 &           18 &           11 &           11 &  55.1  &  -0.06   \\[-0.1cm]
N5468 &          118 &           93 &           93 &  54.9  &  -0.10   \\[-0.1cm]
N5584 &          196 &          167 &          165 &  36.8  &  -0.10   \\[-0.1cm]
N5643 &          294 &          251 &          251 &  31.8  &  0.13   \\[-0.1cm]
N5728 &           25 &           20 &           20 &  44.3  &  0.15   \\[-0.1cm]
N5861 &           60 &           41 &           41 &  43.8  &  0.06   \\[-0.1cm]
N5917 &           17 &           14 &           14 &  37.9  &  -0.30   \\[-0.1cm]
N7250 &           30 &           21 &           21 &  37.3  &  -0.28   \\[-0.1cm]
N7329 &           38 &           31 &           31 &  54.6  &  0.17   \\[-0.1cm]
N7541 &           50 &           33 &           33 &  49.1  &  -0.12   \\[-0.1cm]
N7678 &           21 &           16 &           16 &  42.8  &  0.02   \\[-0.1cm]
U9391 &           36 &           33 &           33 &  39.6  &  -0.22   \\[-0.07cm]
\hline
SN Total &         2680  &         2173 &         2150& 36.5  & -0.01  \\[-0.07cm]
\hline
N4258 &          555 &          451 &          443 &  14.4  &  -0.10   \\[-0.1cm]
M31 &        -- &           55 &           55 &  19.1  &  -0.11   \\[-0.1cm]
LMC$^d$ &        -- &          342 &          339 &  13.3  &  -0.29   \\[-0.1cm]
SMC &        -- &          145 &          143 &  13.3  & -0.72  \\[-0.07cm]
\hline
Total All& --- &         3165 &         3129 & --  & --
\enddata
\tablecomments{(a) Solar value: 12 + log\,[O/H] $= 8.69$, \citet{asplund09}. (b) Good-quality measurement, within allowed color range, period above completeness limit. (c) After 3.3$\sigma$ outlier rejection (1.2\% of sample).  (d) 69 of these are from {\it HST} and 270 from the ground.}
\end{deluxetable}

\subsection{Cepheid Metallicities\label{sc:3.5}}

As in R16 and H16, we measured radial gradients of the strong-line abundance ratios ($R_{23}$) of oxygen to hydrogen in H~II regions in the Cepheid hosts. The optical spectra were obtained with the Low-Resolution Imaging Spectrometer \citep[LRIS;][]{Oke:1995}) on the Keck-I 10~m telescope on Maunakea, Hawaii. See R16 and H16 for details regarding the observations and data reduction.

We define the metallicity of each Cepheid to be the value of this linear function at its galactocentric radius. We have revised the calibration of these strong-line abundance measurements relative to those from \citet[][hereafter Z94]{zaritsky94} used by R16, taking advantage of more recent calibrations between $R_{23}$ and the metallicity 12 + log\,[O/H]. We adopt the {\it average of nine} recent literature calibrations, with the transformations between the Z94 system and the newer ones given by \citet{Teimoorinia:2021}. Further details about the metallicity measurements and their uncertainties are given in Appendix~\ref{sc:appc}.

We also use direct abundance measurements derived from high-resolution spectra of Cepheids in the MW, LMC, and Small Magellanic Cloud (SMC).  In Appendix~\ref{sc:appc}, we evaluate the consistency of the direct and radial (strong-line) abundance measures by comparing H~II regions and Cepheid spectral abundances in the MW.  While we use the average of nine strong-line abundance calibrations for our baseline results, we select the \citet[][hereafter PP04]{Pettini:2004} calibration for comparison to the mean, as it is known to provide a good match to measured extragalactic stellar abundances \citep[][and F.~Bresolin, priv.~comm.]{Bresolin:2016}, and to estimate the uncertainty of the strong-line metallicity estimates.  Specifically, we propagate a systematic uncertainty in the strong-line abundance scale as well as its covariance by including in our fit covariance matrix (given in \S\ref{sc:3}) the product of the difference between the mean calibration and the PP04 calibration for the $i$-th and $j$-th Cepheid,
\bq  C_{i,j,{\rm syst}} =Z_W^2(\textrm{[O/H]}_{i,\textrm{avg}}-\textrm{[O/H]}_{i,\textrm{PP04}})(\textrm{[O/H]}_{j,\textrm{avg}}-\textrm{[O/H]}_{j,\textrm{PP04}}). \label{eq:covmet} \eq 

This approach requires an iteration to use the same value of $Z_W$ in Equation~(\ref{eq:covmet}) that is determined from optimization of the global $\chi^2$. The mean difference of $\sim 0.05$~dex between the average and PP04 scale represents a systematic uncertainty in the abundance scale, propagated here in the covariance matrix, which is consistent with the empirical assessment in Appendix~\ref{sc:appc} 

\section{Anchor Constraints and Ancillary Data\label{sc:4}}

The strongest constraints on Cepheid \PLs and {\it P--L--C} relations come, not surprisingly, from the nearest star-forming galaxies whose samples of Cepheids have better temporal sampling, higher resolution, wider wavelength coverage, and far greater SNR than we can expect to achieve from distant Cepheids presented above that occupy the second rung of the distance ladder.  An accurate and precise determination of H$_0$ {\it requires} leveraging such data to empirically constrain Cepheid properties. Neglecting such data naturally reduces the precision in H$_0$ and as a consequence the significance of any Hubble tension, but this is not a reasonable approach to determine the source of any tension.   We adopt uncertainty estimates from the indicated, external sources as provided and without alteration, and will test their internal consistency within the distance ladder in the following sections.

Here we describe the data we use, in addition to those presented in \S\ref{sc:3}.  Some of these data, namely those described in \S\ref{sc:4.1} through~\ref{sc:4.2}, are {\it anchor constraints} --- they provide direct information on the zeropoint of the Cepheid \PLs relation.  For this purpose, we require (a) that Cepheids be observed directly in our three-filter {\it HST} photometric system, most importantly in the NIR with the WFC3 {\it F160W} filter, and (b) that the distance determination, to individual Cepheids or to their host, be purely geometric.  Additionally, we make use of several datasets, to which we refer as {\it ancillary data}, which do not directly constrain the zeropoint of the \PLs relation, but provide useful information on other characteristics of Cepheids; these include Cepheid measurements in nearby hosts with high-precision photometry (in similar filters but not in our standard {\it HST} system), Cepheids in hosts without a precision geometric distance, and information on SNe~Ia in the Hubble flow.  

\begin{deluxetable}{llrrclll}[t]
\tablenum{4}
\tablewidth{0pc}
\tabletypesize{\scriptsize}
\tablecaption{Ancillary Cepheid Data\label{tb:anccep}}
\tablehead{\colhead{Sample} & \colhead{Reference} & \colhead{N} & \colhead{$\langle P \rangle$} &  \colhead{$\langle$[O/H]$\rangle$} & \colhead{Photometry} & \colhead{Selection} & \colhead{Notes} \\[-0.25cm]
\colhead{} & \colhead{} & \colhead{} & \colhead{[d]} & \colhead{[dex]} & \colhead{} & \colhead{} & \colhead{}}
\startdata
MW {\it Gaia} EDR3& \citet{Riess:2021} & 66 & 12.5 & 0.13 & {\it HST $m_H,m_V,m_I$} & see ref. & $M^W_{H,1,Gaia}=-5.903,$ $\sigma_{Gaia}=0.024^a$,\\
+{\it HST}&&&&&&&$Z_W=-0.20\pm0.12$\\
MW WFC3 SS & \citet{Riess:2018b} & 8 & 22.6 & 0.05 & {\it HST $m_H,m_V,m_I$} & see ref. & $M^W_{H,1,HST}=-5.810,$ $\sigma_{HST}=0.054^a$  \\
LMC {\it HST} & \citet{Riess:2019} & 70 &16.0 & $-0.29^b$ &  {\it HST $m_H,m_V,m_I$}  & see ref. &  \\
LMC ground & \citet{Macri:2015} & 272 & 12.6 & $-0.29^b$ &  ground {\it $m_H,m_V,m_I$} & $P>5$~d & $m_H$ transformed to 2MASS$^5$ \\
SMC ground & \citet{Kato:2007} & 145 & 9.9 & $-0.72^b$ & ground {\it $m_H,m_V,m_I$}  & $P>5$~d, & $m_H$ transformed to 2MASS$^5$ \\
&&&&&&$r<0.6^{\circ}$&\\
M31 SH0ES & \citet{Li:2021} & 55 & 19.1 & $-0.11$ & {\it HST $m_H,m_V,m_I$} & $P>4$~d&  \\
M31 PHAT & \citet{Kodric:2018} & 463 & 10.5 & 0.12 & {\it HST $m_H,m_J$} & $P>4$~d & {\it $m_H,m_J$} transformed  to {\it $m_H,m_V,m_I$}
\enddata
\tablecomments{(a) measured following \citet{Riess:2021} with global fit \PL parameters (b) From \citet{Romaniello:2021} and \citet{Romaniello:2008}.} 
\end{deluxetable}

\subsection{Milky Way Cepheids\label{sc:4.1}}
    
Trigonometric parallaxes to MW Cepheids offer a direct source of geometric calibration of their luminosities. We employ two samples with precise parallaxes and fluxes measured on the same  {\it HST}  system as the extragalactic Cepheids and with direct spectroscopic metallicity measurements.  The first is a sample of 8 MW Cepheids from \cite{Riess:2018b} with parallaxes measured with  {\it HST}  WFC3 spatial scanning.   The other contains 75 MW Cepheids with parallaxes from {\it Gaia} EDR3 as given by \citet{Riess:2021}. We do not use the sample from \citet{benedict07} that was part of R16, because their fluxes are too bright ($0\!<\!m_H\!<\!3$~mag) to be measured {\it directly} with {\it HST} nor have they been measured with good accuracy from the ground; the parallax sample from {\it Gaia} EDR3 provides superior information on MW Cepheids.

Since the distance uncertainties are measured (and Gaussian) in parallax space jointly and are not insignificant ($>5$\%), the direct transformation of parallax to distance modulus (i.e., distance to magnitudes) would yield a bias  \citep[often called the ``Lutz-Kelker'' bias, following][]{lutz73}. One can compensate for it by estimating approximate statistical corrections between parallax and distance moduli for individual Cepheids, assuming the form of their spatial distribution in the MW using Bayesian inference \citep{Bailer-Jones:2021}. However, it is simpler and more reliable to analyze the MW Cepheid data directly in parallax space, in order to retain the Gaussian parallax errors, and to derive their joint constraint on the Cepheid absolute-magnitude zeropoint ($M^W_{H,1}$) following  \citet{Riess:2021}, or similarly through the use of the ``astrometric-based luminosity''  \citep{Arenou:1999}. 

Given that the constraints from the parallaxes across a Cepheid sample are related by the \PLs relation, the combined constraint is evaluated in parallax space and the resulting error in the mean is greatly reduced (for our two samples to 1\% and 3\%), so that the resulting mean constraint on the zeropoint in magnitudes can then be approximated as Gaussian to better than 0.1\%.  We note that the 1\% calibration from {\it Gaia} EDR3 {\it includes marginalization} over the {\it Gaia} parallax offset term as described in \citet{Riess:2021} and that the measured $\sigma=6$ micro-arcsecond uncertainty in the offset is the source of 0.9\% uncertainty for the MW Cepheid sample due to its mean 650  micro-arcsecond parallax. These constraints are given in Table~\ref{tb:anccep} for the \PLs relation determined here. The constraints are not identical to those provided in \citet{Riess:2021} because they employ the same \PLs parameters, $b$ and $Z_W$, that optimizes the global $\chi^2$ for all Cepheids, not just MW Cepheids.  \cite{Riess:2021} showed that the global \PLs parameters, including slope and metallicity dependence, are good fits for the MW Cepheids on their own.

\subsection{Cepheids in NGC$\,$4258\label{sc:4.4}}

The R16 analysis used 143 Cepheids observed with  {\it HST} in the maser host NGC$\,$4258 with a geometric distance from \cite{Reid:2019}.  With new {\it HST} imaging campaigns in four fields shown in Fig.~\ref{fg:hstfovs}, we have more than tripled that sample to 443 Cepheids; details of their identification are given by \citet{Yuan:2021_N4258}. These Cepheids are included in Table~\ref{tb:wfc3irceph}.

\subsection{Cepheids in the LMC\label{sc:4.2}}
       
We include in the joint constraint the sample of 70 Cepheids in the LMC measured on the same {\it HST}  three-band photometric system as given by R19. These have also been corrected with the best-fit planar geometry model of the LMC to be at a single distance as described by R19.  The distance to the center of the LMC is given by \cite{Pietrzynski:2019} from a sample of 20 DEBs as $\mu_0=18.477\pm0.0263$~mag. \citet{Romaniello:2021} has obtained spectra for 68 of these variables, demonstrating that they are consistent within the errors of having a single, common abundance of [Fe/H] $=-0.40$~dex. 60\% of the variables have sufficiently-measured lines to further provide a mean spectroscopic abundance [O/H] $=-0.29 \pm 0.02$~dex which we adopt for the full LMC sample.   LMC Cepheids are corrected for the WFC3 CRNL across the 5~dex between the Cepheid flux and the SN host background \citep{Riess:2019b}.  The LMC Cepheids from {\it HST} also set the intrinsic scatter in $m^W_H$ owing to the finite width of the instability strip to be 0.07~mag, which we include for all Cepheids.
       
The LMC also provides ancillary data, as defined above, that can be used independent of their inferred zeropoint to refine the characterization of the \PLs relation. For this purpose we use the ground-based sample of 785 LMC Cepheids from \citet{Macri:2015} with {\it $m_H,m_V,m_I$} photometry limited to 270 with $P>$ 5~d.  This photometry has been transformed to the {\it HST} system as described by Equations 10--12 in R16; however, we assign a common, systematic uncertainty of $\sigma_{\textit grnd}\!=\!0.10$~mag to the transformed magnitudes, hence the simultaneous constraint takes the form 
$0=\Delta zp \pm \sigma{\textit grnd}\!$, where $\Delta zp$ is a parameter describing the difference between the ground and HST zeropoints, to account for possible systematics associated with ground-based observations.   Because the assigned systematic uncertainty is a factor of $\sim 10$ larger than the mean of the smaller {\it HST} LMC sample (which has no such photometric system difference), the consequence is that the ground-based sample has negligible ($100\times$~less) weight in the tie to the LMC distance; therefore, the ground sample only helps constrain the {\it slope} of the \PLs relation, still an invaluable contribution.
           
\subsection{Cepheids in the SMC\label{sc:4.3}}
   
An important development in support of the distance ladder is the recent measurement of a geometric distance to the SMC from 10 DEBs \citep{Graczyk:2020} with a precision of better than 2\%.  While this precision and method would make the SMC suitable as an additional anchor, two issues present limitations.  The first is the considerable line-of-sight depth of the SMC, which can cause a 10\% dispersion in star distances across the full SMC structure and an offset between the DEBs and the Cepheids in the SMC.  Here we follow the approach of \citet{Breuval:2021} to make use of a sample of SMC Cepheids by (1) correcting for the depth by adopting the same geometry of the SMC as used by \citet{Graczyk:2020} to characterize the DEBs, and (2) limiting the Cepheids to the inner core of the SMC, a radius of $0.6^\circ$. This combination yields a Cepheid sample which, based on the SMC geometric model, is offset in depth by 2~millimag from the mean DEB distance (with an uncertainty of a small fraction of this) and has a modeled dispersion in depth of $\sigma=0.024$~mag, while still providing a sample of 145 Cepheids (with $P>5$~d) that can yield valuable constraints on the Cepheid metallicity term. Earlier studies \citep{Gieren:2018} were unable to employ the DEB-based distance from \citet{Graczyk:2020} and did not focus on the core of the SMC \citep{Wielgorski:2017}.  

The other limitation is the lack of {\it HST} photometry for SMC Cepheids.  Elsewhere we have been able to negate zeropoint uncertainties in the distance scale by using Cepheids observed with the same instruments. However, the SMC Cepheids can still provide a powerful constraint on the Cepheid metallicity term which is well-constrained using {\it only} differential, cross-calibrated ground-based photometry and the differential DEB distance measurement between the SMC and LMC, the latter of which is known {\it better} than the simple difference in the individual DEB cloud distances.
   
As discussed by \citet{Graczyk:2020}, most of the uncertainty in the DEB distance estimates to the LMC and SMC is systematic and propagates from the uncertainty in the surface brightness vs.~color calibration of red giants, the zeropoint of the $V$-band and $K$-band photometry, and the uncertainty in the extinction law. Since the aforementioned SMC and LMC DEB measurements utilized the same relations, observational setup, and reduction methodologies, their {\it differential} distance from DEBs as given by \citet{Graczyk:2020} is far better constrained to $0.500\pm 0.017$~mag, {\it independent of the absolute calibrations and their uncertainties in the DEB method}. Use of consistently-calibrated LMC and SMC ground photometry and this differential distance, even without reference to the absolute DEB distances, helps constrain the metallicity term as we will show in \S\ref{sc:6.2}.  Here we use the 2MASS Point Source Catalog \citep{Cutri:2003} to produce a consistent calibration\footnote{\citet{Macri:2015}: LMC, $H_{\textsc{2mass}}\!=\!m_H\!+\!0.0116\!-\!0.0054(J\!-\!K\!-\!0.4)\!-\!0.0189(J\!-\!K\!-\!0.4)^2$, root-mean square (rms) = 0.038~mag ($N\approx 34,000$).\\\citet{Kato:2007}:SMC, \hspace{0.09in}$H_{\textsc{2mass}}\!=\!m_H\!-\!0.0246\!-\!0.0228(J\!-\!K\!-\!0.4)\!+\!0.0106(J\!-\!K\!-\!0.4)^2$, rms = 0.033~mag ($N\approx 14,000$).} of the $H$-band LMC Cepheid data from \citet{Macri:2015}  and the $H$-band SMC data from \citet{Kato:2007}, and $m_V$ and $m_I$ data from OGLE~III \citep{Soszynski:2008}, with data provided in Table~\ref{tb:wfc3irceph}.  The precision of these transformations allows us to fully leverage the constraint on the distance difference from  \citet{Graczyk:2020}.

We take the mean SMC Cepheid metallicity to be $\Delta$[Fe/H] $=-0.43$~dex relative to the LMC or [O/H] $=-0.72$~dex \citep{Romaniello:2021,Romaniello:2008}. We add to this a systematic uncertainty (common covariance) in the mean LMC and SMC metallicity of 0.05~dex, as in Equation~(\ref{eq:covmet}), each, separately, and relative to other hosts.

\subsection{Cepheids in M31\label{sc:4.5}}
  
In \citet{Li:2021} we presented measurements of 55 Cepheids in M31 using the same three-filter {\it HST} system adopted elsewhere. Owing to their low dispersion and broad range in period, these Cepheids provide additional constraints on the slope of the \PLs relation, independent of any prior knowledge of the distance to M31; this is how we employ these data for our baseline analysis. Given the M31 metallicity gradient of $-0.023$~dex~kpc$^{-1}$ \citep{Zurita:2012}, these Cepheids span only a narrow range of abundances with a mean [O/H] $=-0.1$~dex and a standard deviation of 0.05~dex.
   
A much larger sample of M31 Cepheids is available from the  {\it HST}  PHAT Treasury program \citep{Dalcanton:2012}, but the filters used to observe it do not correspond to the three used here, limiting its utility.  The PHAT program observed these Cepheids with {\it WFC3} {\it F160W} and a ``wide-$J$'' filter ({\it F110W}); \citet{Riess:2012} defined a transformation to the $m_H^W$ system. In R16 we included measurements for 375 PHAT Cepheids before the availability of those from \citet{Li:2021}. An expanded compilation from \cite{Kodric:2018} includes 522 Cepheids from the PHAT program with $3\!<\!P\!<\!78$~d. We use the latter sample as an alternative to \citet{Li:2021} in some variants of our baseline analysis in \S\ref{sc:6.5} because of its powerful leverage to examine evidence of a possible break in the \PLs relation near $P \approx 10$~d.
     
\begin{figure}[t]
\includegraphics[width=\textwidth]{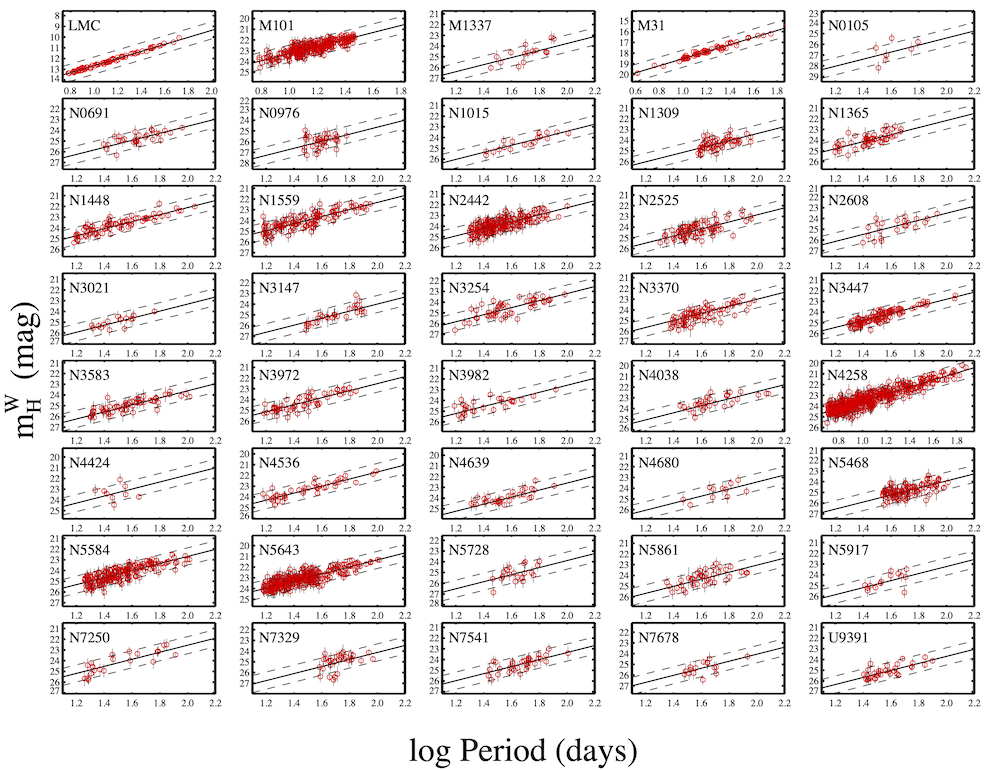}
\caption{\label{fg:plr} {\it HST} NIR Wesenheit Cepheid \PLs relations. The Cepheid magnitudes are shown for 37 SN~Ia hosts, M31, and 2 of the 3 possible distance-scale anchors (LMC and NGC$\,$4258). The uniformity of the photometry and metallicity reduces systematic errors along the distance ladder.  A single slope is shown and used for the baseline, but we also allow for a break (two slopes) as well as limited period ranges in some analysis variants. }
\end{figure}

\subsection{Period--Luminosity Relations\label{sc:4.6}}
   
Fig.~\ref{fg:plr} shows the 40 individual Cepheid host $m^W_H$ \PLs relations (not including the MW as explained above) for a period range of $5\!<\!P\!<\!120$~d fitted with a common slope. Before combining the results of many hosts in a global analysis,  we examine in Fig.~\ref{fg:slope} the independently-fitted slopes of the $m_H^W$ \PLs relations across all Cepheid hosts.

The slopes are all consistent (at the $\sim 2\sigma$ level) with a mean in the range of $-3.27$ to $-3.30$~mag~dex$^{-1}$. The most tightly constrained slope comes from the LMC with $-3.284\pm0.017$~mag~dex$^{-1}$, mostly from the ground sample. Overall, we see no evidence to reject the null hypothesis of a single slope. There have been claims in the past of a break at $P\approx 10$~d which we will consider as a variant of the primary analysis in the next section.  However, the mean slopes from these data below and above $P=10$~d are $-3.33\pm0.02$ and $-3.21\pm0.06$~mag~dex$^{-1}$ (respectively), a difference of $1.8\sigma$; individual hosts with the strongest constraint (LMC and the M31 PHAT sample) show slope changes at $P=10$~d in {\it opposite} directions.  Furthermore, formal uncertainties in these slopes are somewhat {\it underestimated} because of the uneven sampling of periods between hosts.  A Monte Carlo analysis (bootstrap resampling with replacement) from hosts with the largest samples of Cepheids shows variations owing to uneven period sampling increases the formal slope uncertainty typically by 10\% and up to 35\% for the LMC. We find that the metallicity dependence has a negligible effect on the mean slope.  Therefore, in the following we will consider a single slope for $5<P<120$~d in our baseline analysis, but we analyze the impact of a break or limited period range on the determination of H$_0$ as variants of the baseline analysis.  
  
\subsection{Geometric Distance Priors\label{sc:4.7}}
  
We make use of the geometric distances to NGC$\,$4258 from masers \citep{Reid:2019} of $\mu_\textsc{n4258}=29.398\pm0.032$~mag, the LMC from DEBs \citep{Pietrzynski:2019} of $\mu_\textsc{lmc}=18.477\pm0.0263$~mag, and the SMC from DEBs \citep{Graczyk:2020} in the form of $\Delta\mu_\textsc{lmc-smc}=0.500\pm0.017$~mag, as well as the MW {\it Gaia} EDR3 and {\it HST} spatial-scan parallaxes discussed in \S\ref{sc:4.1}. These are all given in Table~\ref{tb:anccep}.

\subsection{SN Magnitudes\label{sc:4.8}}
  
We adopt standardized SN~Ia magnitudes $m_{B}^0$ from the Pantheon+ analysis \citep{Scolnic:2021,Brout:2022}, where the value $m_{B,i}^0$ is a measure of the maximum-light apparent $B$-band brightness of a SN~Ia in the {\it i}-th host at the time of $B$-band peak, corrected to the fiducial color and luminosity determined for each SN~Ia from its multiband light curves and a light-curve-fitting algorithm.  We use the uncertainties and covariance of $m_{B}^0$ as given by the Pantheon+ analysis.  The SN~Ia covariance matrix has substantial off-diagonal terms and is displayed in Fig.~\ref{fg:covarlo}.  An improvement in the current analysis of calibrator SNe~Ia over R16 is our use of multiple SN light-curve datasets for most calibrators, 77 sets in all for 42 SNe~Ia, a mean of $\sim 2$ independent sets per SN, reducing measurement errors (not intrinsic scatter, which is covariant among multiple samples of the same SN) by a mean factor of 1.4.  These datasets are given in \citet{Scolnic:2021}.  

\clearpage

\begin{figure}[t]   
\begin{center}
\includegraphics[height=0.55\textheight,angle=90]{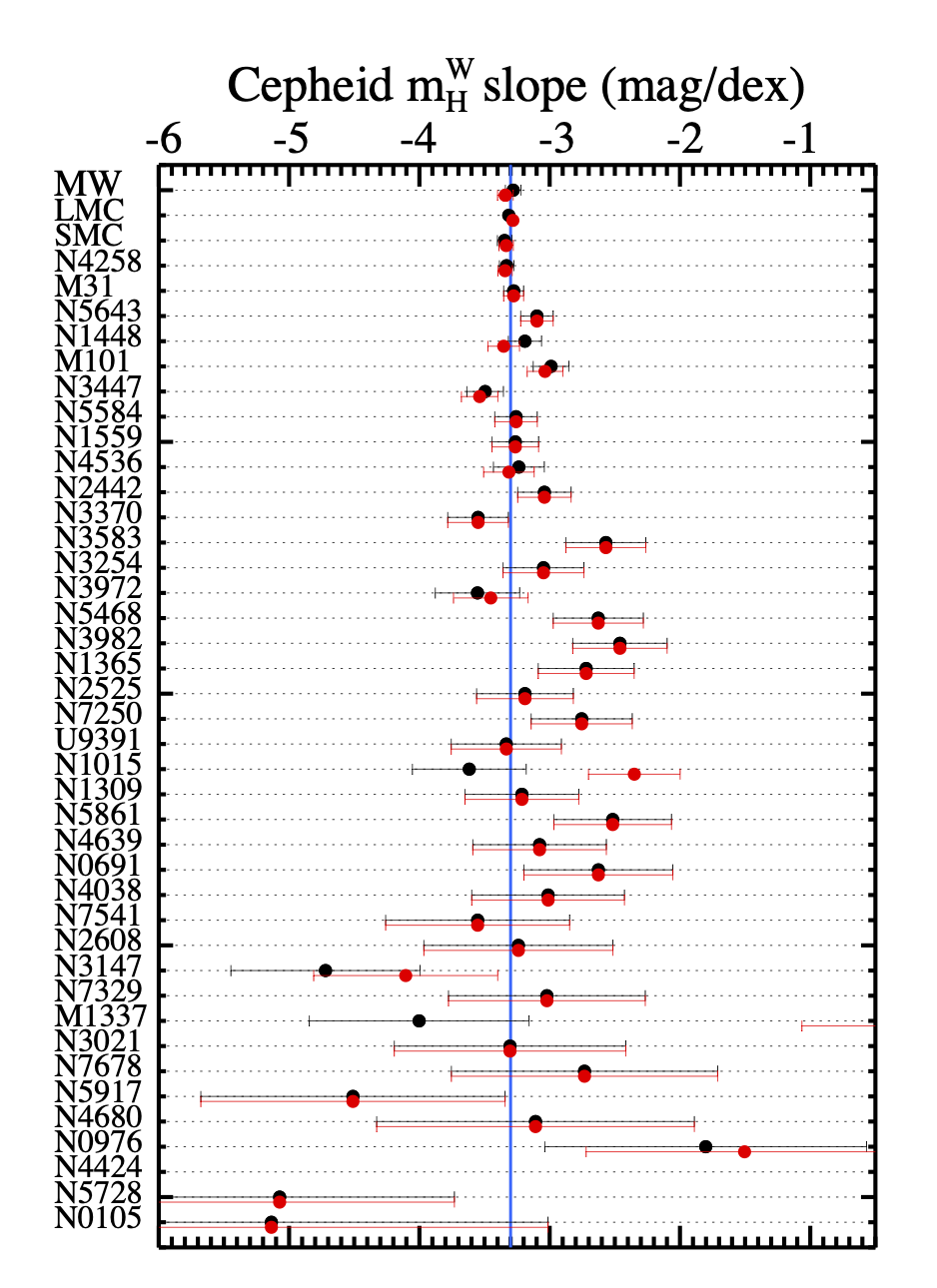}
\end{center}
\caption{\label{fg:slope} Independent determinations of the slope of the \PLs relation in each host plotted in order of precision.  Black (red) points are after (before) outlier rejection. We find that the slopes are consistent with a single value of $-3.27$ to $-3.30$ as indicated by the blue line.  The most metal-poor (SMC) and metal-rich (MW) also have consistent slopes. Error bars are for the fit error only and do not include uneven period sampling or background covariance.  The large change in slope for Mrk 1337 was due to the relatively small sample size and rejection of a few Cepheids far from its center.}
\end{figure}

\begin{figure}[b]   
\begin{center}
\includegraphics[height=0.35\textheight]{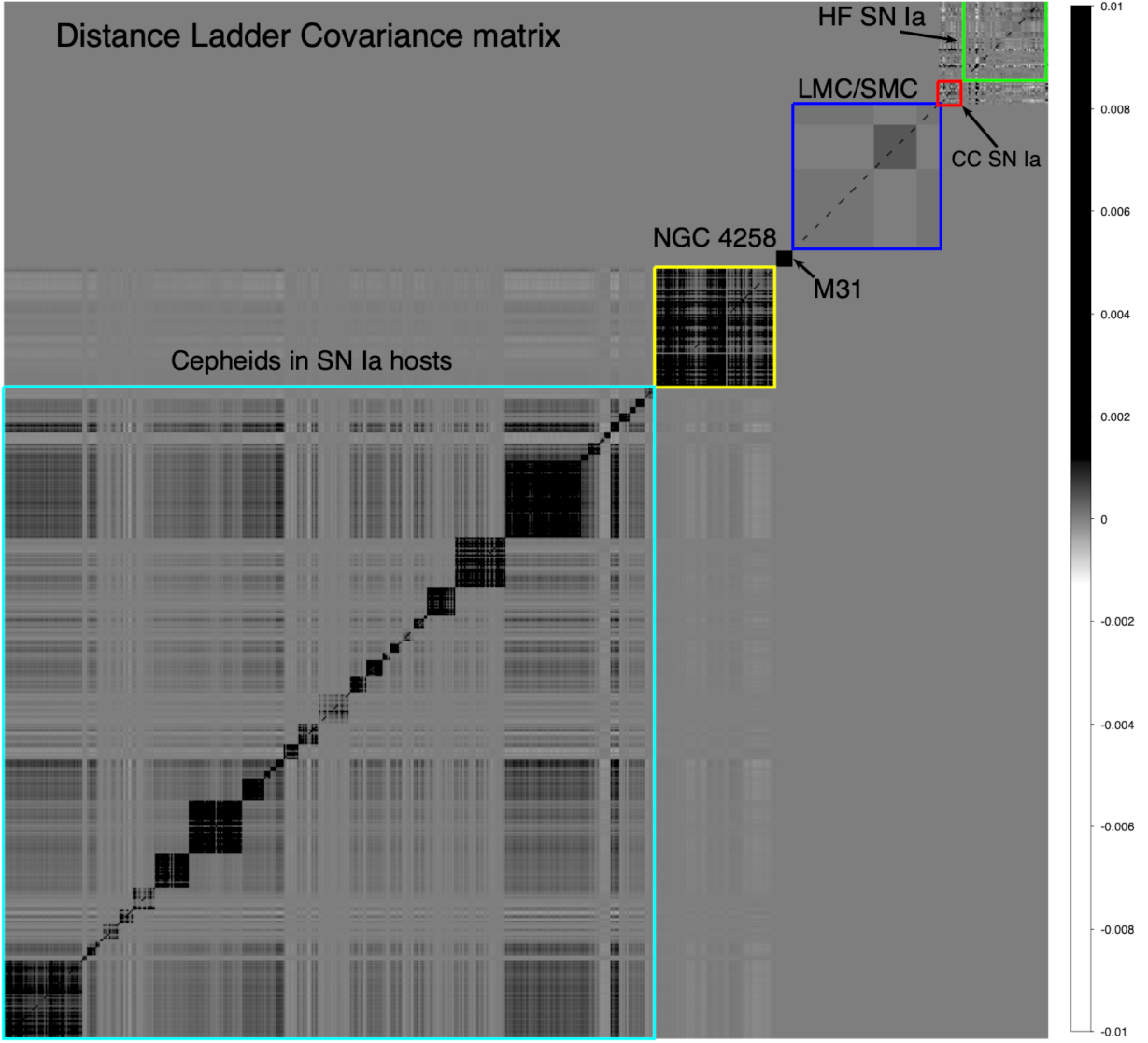}
\end{center}
\caption{\label{fg:covarlo} Distance ladder covariance matrix.  Graphical representation of the covariance matrix of measurements with values applicable to different sources indicated.  The covariance matrix contains past sources of systematic error that propagate here as error in the fit.  Among Cepheid hosts, square blocks along the diagonal indicate the background covariance within a host (Eq.~\ref{eq:coverr}) while the covariance pattern between hosts farther off the diagonals comes from the covariance of the metallicity scale (Eq.~\ref{eq:covmet}).  The SN-SN covariance is characterized in the Pantheon+ analysis \citep{Scolnic:2021,Brout:2022}.}
\end{figure}

\clearpage

\startlongtable
\begin{deluxetable}{clcccccccc}
\tabletypesize{\footnotesize}
\tablewidth{0pc}
\setlength{\tabcolsep}{0.33em}
\tablenum{5}
\tablecaption{Fits for H$_0$\label{tb:h0var}}
\tablehead{\colhead{Fit}&\colhead{Variant}&\colhead{$\chi^2_{dof}$}&\colhead{N}& \colhead{H$_0$}&\colhead{b}&\colhead{$\gamma$}&\colhead{$M_W^0$}&\colhead{$M_B^0$}&\colhead{$a_b$}}
\startdata
\hline
{\bf       1 } & {\bf Baseline } & {\bf  1.03  } & {\bf         3445 } & {\bf 73.04 1.01  } & {\bf -3.299 0.015  } & {\bf -0.217 0.046  } & {\bf  -5.894  } & {\bf  -19.253  } & {\bf  0.714158} \\[-0.015cm]
\hline
\hline
\multicolumn{10}{c}{Cepheid Clipping Variants \S 6.1} \\[-0.015cm]
\hline
       2 & global $\sigma_\textrm{clip}=3.3$ &  1.03  &         3446 & 73.19 1.01  & -3.298 0.015  & -0.216 0.046  &  -5.891  &  -19.249  &  0.714174\\[-0.015cm]
       3 & individual PL $\sigma_\textrm{clip}=3.3$ &  0.99  &         3370 & 73.25 1.02  & -3.296 0.015  & -0.201 0.045  &  -5.893  &  -19.248  &  0.714315\\[-0.015cm]
       4 & tight:one-by-one MAD $\sigma_\textrm{clip}=3.0$ &  0.99  &         3429 & 73.12 1.01  & -3.299 0.015  & -0.220 0.045  &  -5.893  &  -19.251  &  0.714175\\[-0.015cm]
       5 & tight:global $\sigma_\textrm{clip}=3.0$ &  0.99  &         3432 & 73.22 1.01  & -3.298 0.015  & -0.215 0.045  &  -5.891  &  -19.248  &  0.714194\\[-0.015cm]
       6 & loose:global $\sigma_\textrm{clip}=5.0$ &  1.16  &         3475 & 73.34 1.01  & -3.294 0.016  & -0.221 0.049  &  -5.888  &  -19.244  &  0.714183\\[-0.015cm]
       7 & loose:one-by-one MAD $\sigma_\textrm{clip}=5.0$ &  1.15  &         3474 & 73.42 1.01  & -3.295 0.016  & -0.222 0.048  &  -5.888  &  -19.242  &  0.714178\\[-0.015cm]
       8 & loose:individual PL $\sigma_\textrm{clip}=5.0$ &  1.05  &         3397 & 73.35 1.01  & -3.296 0.015  & -0.202 0.046  &  -5.892  &  -19.244  &  0.714257\\[-0.015cm]
       9 & none &  1.23  &         3481 & 73.41 1.01  & -3.290 0.016  & -0.206 0.050  &  -5.885  &  -19.242  &  0.714248\\[-0.015cm]
\hline
\multicolumn{10}{c}{Geometric Anchors Variants \S 6.2} \\[-0.015cm]
\hline
      10 & N4258 &  1.06  &         3454 & 72.51 1.54  & -3.294 0.015  & -0.204 0.051  &  -5.905  &  -19.269  &  0.714174\\[-0.015cm]
      11 & Milky Way &  1.03  &         3446 & 73.02 1.19  & -3.298 0.015  & -0.208 0.051  &  -5.895  &  -19.254  &  0.714179\\[-0.015cm]
      12 & LMC &  1.03  &         3446 & 73.59 1.36  & -3.298 0.015  & -0.208 0.051  &  -5.878  &  -19.237  &  0.714178\\[-0.015cm]
      13 & N4258+MW &  1.03  &         3446 & 73.00 1.09  & -3.298 0.015  & -0.207 0.050  &  -5.895  &  -19.254  &  0.714171\\[-0.015cm]
      14 & N4258+LMC &  1.03  &         3446 & 73.35 1.17  & -3.299 0.015  & -0.211 0.050  &  -5.885  &  -19.244  &  0.714179\\[-0.015cm]
      15 & MW+LMC &  1.03  &         3446 & 73.25 1.05  & -3.298 0.015  & -0.216 0.046  &  -5.889  &  -19.247  &  0.714169\\[-0.015cm]
\hline
\multicolumn{10}{c}{Cepheid Dust-Color Treatment Variants \S 6.3} \\[-0.015cm]
\hline
      16 & Fitzpatrick 99 law $R_V=2.5$ &  1.03  &         3446 & 73.24 1.00  & -3.291 0.015  & -0.209 0.046  &  -5.850  &  -19.247  &  0.714171\\[-0.015cm]
      17 & CCM law $R_V=3.1$ &  1.04  &         3445 & 73.09 1.00  & -3.310 0.015  & -0.226 0.046  &  -5.957  &  -19.252  &  0.714183\\[-0.015cm]
      18 & Nataf law $R_V=3.3$ &  1.03  &         3445 & 73.32 0.99  & -3.276 0.015  & -0.204 0.046  &  -5.804  &  -19.245  &  0.714141\\[-0.015cm]
      19 & $R_W$ free global &  1.03  &         3446 & 73.40 1.02  & -3.286 0.016  & -0.204 0.046  &  -5.835  &  -19.242  &  0.714167\\[-0.015cm]
      20 & intrin. col. subtr. F99 $R_V=3.3$ &  1.03  &         3446 & 73.13 1.01  & -3.201 0.015  & -0.218 0.046  &  -5.604  &  -19.250  &  0.714170\\[-0.015cm]
      21 & intrin. col. subtr. F99 $R_V=$ free &  1.03  &         3446 & 73.34 1.01  & -3.201 0.015  & -0.206 0.046  &  -5.582  &  -19.244  &  0.714170\\[-0.015cm]
      22 & intrin. col. subtr. $R_V$(host mass-SFR) &  1.05  &         3446 & 73.85 1.02  & -3.202 0.015  & -0.228 0.046  &  -5.605  &  -19.229  &  0.714230\\[-0.015cm]
      23 & None($A_H$ values assumed to cancel) &  1.11  &         3437 & 74.78 1.03  & -3.188 0.016  & -0.115 0.048  &  -5.434  &  -19.201  &  0.714086\\[-0.015cm]
\hline
\multicolumn{10}{c}{PL Break/span Variants \S 6.4} \\[-0.015cm]
\hline
      24 & Break at $P=10$~d &  1.03  &         3444 & 72.68 1.04  & -0.10 0.05 & -0.222 0.046  &  -5.887  &  -19.264  &  0.714151\\[-0.015cm]
      25 & Only use $P>10$~d &  1.06  &         3004 & 73.15 1.11  & -3.337 0.023  & -0.169 0.051  &  -5.861  &  -19.250  &  0.714158\\[-0.015cm]
      26 & Only use $P<60$~d &  0.93  &         3699 & 73.99 1.04  & -3.261 0.011  & -0.236 0.044  &  -5.898  &  -19.226  &  0.714276\\[-0.015cm]
\hline
\multicolumn{10}{c}{M31 Cepheid Sampl Variants \S 6.5} \\[-0.015cm]
\hline
      27 & M31 PHAT sample &  1.02  &         3854 & 73.21 1.00  & -3.297 0.013  & -0.234 0.044  &  -5.893  &  -19.248  &  0.714184\\[-0.015cm]
      28 & M31 PHAT sample +Break at $P=10$~d &  1.02  &         3852 & 72.74 1.03  & -0.07 0.04 & -0.240 0.044  &  -5.891  &  -19.262  &  0.714156\\[-0.015cm]
\hline
\multicolumn{10}{c}{Metallicity Variants \S 6.6} \\[-0.015cm]
\hline
      29 & no metallity dependence &  1.04  &         3446 & 73.52 1.01  & -3.296 0.015  & \nd &  -5.860  &  -19.239  &  0.714194\\[-0.015cm]
      30 & PP04 metallicity scale &  1.04  &         3446 & 72.84 1.00  & -3.297 0.015  & -0.166 0.042  &  -5.883  &  -19.259  &  0.714162\\[-0.015cm]
\hline
\multicolumn{10}{c}{TRGB Inclusion Variants \S 6.7} \\[-0.015cm]
\hline
      31 & adds EDD TRGB+N4258tip &  1.03  &         3457 & 72.76 0.95  & -3.301 0.015  & -0.197 0.045  &  -5.891  &  -19.262  &  0.714253\\[-0.015cm]
      32 & adds CCHP TRGB+N4258tip &  1.03  &         3457 & 72.29 0.94  & -3.304 0.015  & -0.208 0.046  &  -5.894  &  -19.276  &  0.714254\\[-0.015cm]
\hline
\multicolumn{10}{c}{Hubble Flow Sample Variants \S 6.8} \\[-0.015cm]
\hline
      33 & all host types $0.0233<z<0.15$ &  1.03  &         3652 & 73.32 0.99  & -3.298 0.015  & -0.216 0.046  &  -5.891  &  -19.246  &  0.714479\\[-0.015cm]
      34 & highz:all host types $0.0233<z<0.80$ &  1.00  &         4483 & 73.68 0.98  & -3.298 0.015  & -0.216 0.045  &  -5.891  &  -19.244  &  0.716225\\[-0.015cm]
      35 & skip local alltypes $0.06<z<0.15$ &  1.04  &         3318 & 73.35 1.06  & -3.298 0.015  & -0.217 0.046  &  -5.891  &  -19.245  &  0.714311\\[-0.015cm]
      36 & highz:skip local alltypes $0.06<z<0.8$ &  1.00  &         4149 & 73.90 1.01  & -3.298 0.015  & -0.217 0.045  &  -5.891  &  -19.242  &  0.716991\\[-0.015cm]
      37 & highmass:hubble flow host logmass$>10$ &  1.04  &         3304 & 72.97 1.04  & -3.298 0.015  & -0.217 0.046  &  -5.891  &  -19.251  &  0.713297\\[-0.015cm]
\hline
\multicolumn{10}{c}{Calibrator Sample  Variants \S 6.9} \\[-0.015cm]
\hline
      38 & complete calibrator sample $z<0.011$ &  1.03  &         3446 & 73.30 1.02  & -3.298 0.015  & -0.217 0.046  &  -5.891  &  -19.245  &  0.714191\\[-0.015cm]
      39 & complete sample $z<0.011$+TRGB &  1.04  &         3458 & 72.88 0.95  & -3.301 0.015  & -0.196 0.045  &  -5.888  &  -19.258  &  0.714297\\[-0.015cm]
      40 & highmass:calibrator host logmass$>10$ &  1.03  &         3445 & 73.54 1.10  & -3.296 0.015  & -0.214 0.046  &  -5.891  &  -19.238  &  0.714201\\[-0.015cm]
      41 & Use least crowded half &  1.02  &         3446 & 73.34 1.16  & -3.297 0.015  & -0.223 0.046  &  -5.892  &  -19.246  &  0.714597\\[-0.015cm]
      42 & Use most crowded half hosts &  1.02  &         3445 & 73.35 1.37  & -3.293 0.015  & -0.215 0.046  &  -5.891  &  -19.243  &  0.714063\\[-0.015cm]
      43 & only 19 hosts from R16 &  1.02  &         3446 & 73.47 1.17  & -3.297 0.015  & -0.229 0.046  &  -5.893  &  -19.242  &  0.714505\\[-0.015cm]
      44 & only hosts since R16 &  1.02  &         3445 & 73.07 1.31  & -3.293 0.015  & -0.210 0.046  &  -5.890  &  -19.252  &  0.714178\\[-0.015cm]
      45 & closer half hosts $m_b<13$ &  1.03  &         3446 & 73.07 1.16  & -3.295 0.015  & -0.219 0.046  &  -5.891  &  -19.252  &  0.714191\\[-0.015cm]
\hline
\multicolumn{10}{c}{Excluded SN Surveys Variants \S 6.10} \\[-0.015cm]
\hline
      46 & No SDSS SNe &  1.03  &         3446 & 72.90 1.02  & -3.298 0.015  & -0.217 0.046  &  -5.891  &  -19.249  &  0.712428\\[-0.015cm]
      47 & No CSP SNe &  1.02  &         3445 & 73.43 1.06  & -3.298 0.015  & -0.217 0.046  &  -5.891  &  -19.246  &  0.715130\\[-0.015cm]
      48 & No literature SNe &  1.03  &         3446 & 73.47 1.05  & -3.298 0.015  & -0.213 0.046  &  -5.890  &  -19.240  &  0.714072\\[-0.015cm]
      49 & No LOSS SNe &  1.02  &         3447 & 73.26 1.04  & -3.297 0.015  & -0.213 0.046  &  -5.891  &  -19.251  &  0.715022\\[-0.015cm]
      50 & No {\it Swift} SNe &  1.03  &         3445 & 73.09 1.02  & -3.297 0.015  & -0.217 0.046  &  -5.891  &  -19.248  &  0.713551\\[-0.015cm]
      51 & No CfA1/2 SNe &  1.03  &         3446 & 73.03 1.03  & -3.298 0.015  & -0.218 0.046  &  -5.891  &  -19.250  &  0.713464\\[-0.015cm]
      52 & No CfA3/4 SNe &  1.03  &         3447 & 73.31 1.02  & -3.298 0.015  & -0.214 0.046  &  -5.891  &  -19.245  &  0.714257\\[-0.015cm]
      53 & No Foundation SNe &  1.01  &         3446 & 73.46 1.03  & -3.298 0.015  & -0.217 0.045  &  -5.891  &  -19.241  &  0.714196\\[-0.015cm]
      54 & No pre-2000 SNe &  1.02  &         3446 & 73.20 1.09  & -3.297 0.015  & -0.218 0.046  &  -5.891  &  -19.245  &  0.713530\\[-0.015cm]
\hline
\multicolumn{10}{c}{SN Fitting Variants \S 6.11} \\[-0.015cm]
\hline
      55 & SN scatter monochromatic &  1.00  &         3444 & 73.54 1.08  & -3.296 0.015  & -0.223 0.045  &  -5.892  &  -19.238  &  0.714067\\[-0.015cm]
\hline
\multicolumn{10}{c}{Peculiar Velocity Variants \S 6.12} \\[-0.015cm]
\hline
      56 & 2MRS &  1.03  &         3446 & 73.12 1.01  & -3.298 0.015  & -0.216 0.046  &  -5.891  &  -19.249  &  0.713775\\[-0.015cm]
      57 & CMB frame z &  1.04  &         3446 & 72.70 1.00  & -3.298 0.015  & -0.216 0.046  &  -5.891  &  -19.249  &  0.711260\\[-0.015cm]
      58 & $q_0=-0.52$ &  1.03  &         3446 & 73.19 1.01  & -3.298 0.015  & -0.216 0.046  &  -5.891  &  -19.249  &  0.714149\\[-0.015cm]
      59 & $q_0=-0.52$ all highz &  1.00  &         4483 & 73.65 0.98  & -3.298 0.015  & -0.216 0.045  &  -5.891  &  -19.244  &  0.715950\\[-0.015cm]
\hline
\multicolumn{10}{c}{Optical Wesenheit Variants \S 6.13} \\[-0.015cm]
\hline
      60 & optical Wesenheit clipping=one-by-one &  0.94  &         3626 & 72.70 1.03  & -3.299 0.010  & -0.248 0.041  &  -5.858  &  -19.264  &  0.714413\\[-0.015cm]
      61 & optical Wesenheit clipping=global &  0.94  &         3626 & 72.90 1.03  & -3.291 0.010  & -0.247 0.041  &  -5.857  &  -19.259  &  0.714417\\[-0.015cm]
      62 & optical Wesenheit F99 $R_V=2.5$ &  0.92  &         3618 & 73.20 1.03  & -3.230 0.010  & -0.202 0.041  &  -5.623  &  -19.249  &  0.714359\\[-0.015cm]
      63 & optical Wesenheit CCM $R_V=3.1$ &  0.97  &         3626 & 72.46 1.03  & -3.335 0.010  & -0.270 0.042  &  -6.020  &  -19.272  &  0.714469\\[-0.015cm]
      64 & optical Wesenheit N4258 only &  0.92  &         3623 & 74.85 2.31  & -3.291 0.010  & -0.211 0.045  &  -5.797  &  -19.201  &  0.714429\\[-0.015cm]
      65 & optical Wesenheit MW only &  0.93  &         3623 & 71.93 1.15  & -3.291 0.010  & -0.211 0.045  &  -5.883  &  -19.288  &  0.714421\\[-0.015cm]
      66 & optical Wesenheit LMC only &  0.92  &         3623 & 74.26 1.39  & -3.291 0.010  & -0.211 0.045  &  -5.814  &  -19.218  &  0.714416\\[-0.015cm]
      67 & optical Wesenheit+TRGB &  0.94  &         3638 & 72.15 0.94  & -3.301 0.010  & -0.239 0.041  &  -5.858  &  -19.281  &  0.714385\\[-0.015cm]
\enddata
\end{deluxetable}

\begin{deluxetable}{lllrrrrrrrrr}[b]
\tabletypesize{\footnotesize}
\tablewidth{0pc}
\tablenum{6}
\tablecaption{Approximations for Distance Parameters\label{tb:distpar}}
\tablehead{\colhead{N} & \colhead{Host} & \colhead{SN} & \colhead{$m_{B,i}^0$} & \colhead{$\sigma$} & \colhead{$\mu_{\rm Ceph}^a$} & \colhead{$\sigma$} & \colhead{$M_{B,i}^0$} & \colhead{$\sigma$} & \colhead{$\mu_{\rm host}^b$} & \colhead{$\sigma$} & \colhead{$R^c$} \\[-0.15cm]
\colhead{} & \colhead{}  & \colhead{} & \multicolumn{7}{c}{[mag]}}
\startdata
1 & M101 &  2011fe &  9.7800  &  0.115  &  29.193  &  0.039  &  -19.413  &  0.122  &  29.178  &  0.041  &  0.44  \\[-0.12cm]
2 & M1337 &  2006D &  13.655  &  0.106  &  32.922  &  0.124  &  -19.267  &  0.163  &  32.920  &  0.123  &  0.43  \\[-0.12cm]
3 & N0105 &  2007A &  15.250  &  0.133  &  34.531  &  0.253  &  -19.281  &  0.286  &  34.527  &  0.250  &  0.37  \\[-0.12cm]
4 & N0691 &  2005W &  13.602  &  0.139  &  32.846  &  0.109  &  -19.244  &  0.177  &  32.830  &  0.109  &  0.42  \\[-0.12cm]
5 & N0976 &  1999dq &  14.250  &  0.103  &  33.719  &  0.151  &  -19.468  &  0.183  &  33.709  &  0.149  &  0.30  \\[-0.12cm]
6 & N1015 &  2009ig &  13.350  &  0.094  &  32.570  &  0.075  &  -19.220  &  0.120  &  32.563  &  0.074  &  0.46  \\[-0.12cm]
7 & N1309 &  2002fk &  13.209  &  0.082  &  32.546  &  0.060  &  -19.337  &  0.102  &  32.541  &  0.059  &  0.38  \\[-0.12cm]
8 & N1365 &  2012fr &  11.900  &  0.092  &  31.379  &  0.057  &  -19.479  &  0.108  &  31.378  &  0.056  &  0.46  \\[-0.12cm]
9 & N1448 &  2001el &  12.254  &  0.136  &  31.290  &  0.037  &  -19.036  &  0.141  &  31.287  &  0.037  &  0.41  \\[-0.12cm]
10 & N1448 &  2021pit &  11.752  &  0.200  &  31.290  &  0.037  &  -19.538  &  0.203  &  31.287  &  0.037  &  0.41  \\[-0.12cm]
11 & N1559 &  2005df &  12.141  &  0.086  &  31.501  &  0.062  &  -19.360  &  0.106  &  31.491  &  0.061  &  0.35  \\[-0.12cm]
12 & N2442 &  2015F &  12.234  &  0.082  &  31.457  &  0.065  &  -19.223  &  0.105  &  31.450  &  0.064  &  0.36  \\[-0.12cm]
13 & N2525 &  2018gv &  12.728  &  0.074  &  32.067  &  0.100  &  -19.339  &  0.124  &  32.051  &  0.099  &  0.39  \\[-0.12cm]
14 & N2608 &  2001bg &  13.443  &  0.166  &  32.629  &  0.155  &  -19.186  &  0.227  &  32.612  &  0.154  &  0.39  \\[-0.12cm]
15 & N3021 &  1995al &  13.114  &  0.116  &  32.474  &  0.160  &  -19.360  &  0.198  &  32.464  &  0.158  &  0.34  \\[-0.12cm]
16 & N3147 &  2021hpr &  13.843  &  0.159  &  33.044  &  0.165  &  -19.201  &  0.229  &  33.014  &  0.165  &  0.45  \\[-0.12cm]
17 & N3147 &  1997bq &  13.821  &  0.141  &  33.044  &  0.165  &  -19.223  &  0.217  &  33.014  &  0.165  &  0.45  \\[-0.12cm]
18 & N3147 &  2008fv &  13.936  &  0.200  &  33.044  &  0.165  &  -19.108  &  0.260  &  33.014  &  0.165  &  0.45  \\[-0.12cm]
19 & N3254 &  2019np &  13.201  &  0.074  &  32.332  &  0.077  &  -19.130  &  0.107  &  32.331  &  0.076  &  0.52  \\[-0.12cm]
20 & N3370 &  1994ae &  12.937  &  0.082  &  32.123  &  0.052  &  -19.186  &  0.097  &  32.120  &  0.051  &  0.33  \\[-0.12cm]
21 & N3447 &  2012ht &  12.736  &  0.089  &  31.939  &  0.035  &  -19.203  &  0.096  &  31.936  &  0.034  &  0.40  \\[-0.12cm]
22 & N3583 &  ASASSN-15so &  13.509  &  0.093  &  32.808  &  0.081  &  -19.299  &  0.123  &  32.804  &  0.080  &  0.34  \\[-0.12cm]
23 & N3972 &  2011by &  12.548  &  0.094  &  31.644  &  0.090  &  -19.096  &  0.130  &  31.635  &  0.089  &  0.46  \\[-0.12cm]
24 & N3982 &  1998aq &  12.252  &  0.078  &  31.724  &  0.072  &  -19.472  &  0.106  &  31.722  &  0.071  &  0.36  \\[-0.12cm]
25 & N4038 &  2007sr &  12.409  &  0.106  &  31.615  &  0.117  &  -19.206  &  0.158  &  31.603  &  0.116  &  0.57  \\[-0.12cm]
26 & N4424 &  2012cg &  11.487  &  0.192  &  30.856  &  0.130  &  -19.369  &  0.232  &  30.844  &  0.128  &  0.87  \\[-0.12cm]
27 & N4536 &  1981B &  11.551  &  0.133  &  30.838  &  0.051  &  -19.287  &  0.142  &  30.835  &  0.050  &  0.49  \\[-0.12cm]
28 & N4639 &  1990N &  12.454  &  0.124  &  31.818  &  0.085  &  -19.364  &  0.150  &  31.812  &  0.084  &  0.48  \\[-0.12cm]
29 & N4680 &  1997bp &  13.173  &  0.205  &  32.606  &  0.208  &  -19.433  &  0.292  &  32.599  &  0.205  &  0.48  \\[-0.12cm]
30 & N5468 &  1999cp &  13.880  &  0.080  &  33.120  &  0.075  &  -19.240  &  0.110  &  33.116  &  0.074  &  0.39  \\[-0.12cm]
31 & N5468 &  2002cr &  13.993  &  0.072  &  33.120  &  0.075  &  -19.127  &  0.104  &  33.116  &  0.074  &  0.39  \\[-0.12cm]
32 & N5584 &  2007af &  12.804  &  0.079  &  31.775  &  0.053  &  -18.971  &  0.095  &  31.772  &  0.052  &  0.36  \\[-0.12cm]
33 & N5643 &  2013aa &  11.252  &  0.079  &  30.569  &  0.050  &  -19.317  &  0.093  &  30.546  &  0.052  &  0.39  \\[-0.12cm]
34 & N5643 &  2017cbv &  11.208  &  0.074  &  30.569  &  0.050  &  -19.362  &  0.089  &  30.546  &  0.052  &  0.39  \\[-0.12cm]
35 & N5728 &  2009Y &  13.514  &  0.115  &  33.116  &  0.207  &  -19.602  &  0.236  &  33.094  &  0.205  &  0.58  \\[-0.12cm]
36 & N5861 &  2017erp &  12.945  &  0.107  &  32.232  &  0.101  &  -19.287  &  0.147  &  32.223  &  0.099  &  0.33  \\[-0.12cm]
37 & N5917 &  2005cf &  13.079  &  0.095  &  32.363  &  0.122  &  -19.284  &  0.154  &  32.363  &  0.120  &  0.47  \\[-0.12cm]
38 & N7250 &  2013dy &  12.283  &  0.178  &  31.628  &  0.126  &  -19.345  &  0.218  &  31.628  &  0.125  &  0.49  \\[-0.12cm]
39 & N7329 &  2006bh &  14.030  &  0.079  &  33.274  &  0.116  &  -19.244  &  0.140  &  33.246  &  0.117  &  0.39  \\[-0.12cm]
40 & N7541 &  1998dh &  13.418  &  0.128  &  32.504  &  0.121  &  -19.086  &  0.176  &  32.500  &  0.119  &  0.30  \\[-0.12cm]
41 & N7678 &  2002dp &  14.090  &  0.093  &  33.196  &  0.155  &  -19.106  &  0.181  &  33.187  &  0.153  &  0.39  \\[-0.12cm]
42 & U9391 &  2003du &  13.525  &  0.084  &  32.849  &  0.068  &  -19.324  &  0.108  &  32.848  &  0.067  &  0.44  
\enddata
\tablecomments{(a) Approximate, Cepheid-based distance derived without inclusion of the SNe in a given host. (b) Cepheid-based distance derived without inclusion of any SN in any host. (c) Empirical host reddening ratio derived from  mass and star-formation rate (SFR) using EDA (\S6.3). (d) {$m_{B,i}^0$ and error given are an approximation, the simple average from one or more SN surveys}}
\end{deluxetable}

\ \par
\begin{figure}[b]
\begin{center}
\includegraphics[width=\textwidth]{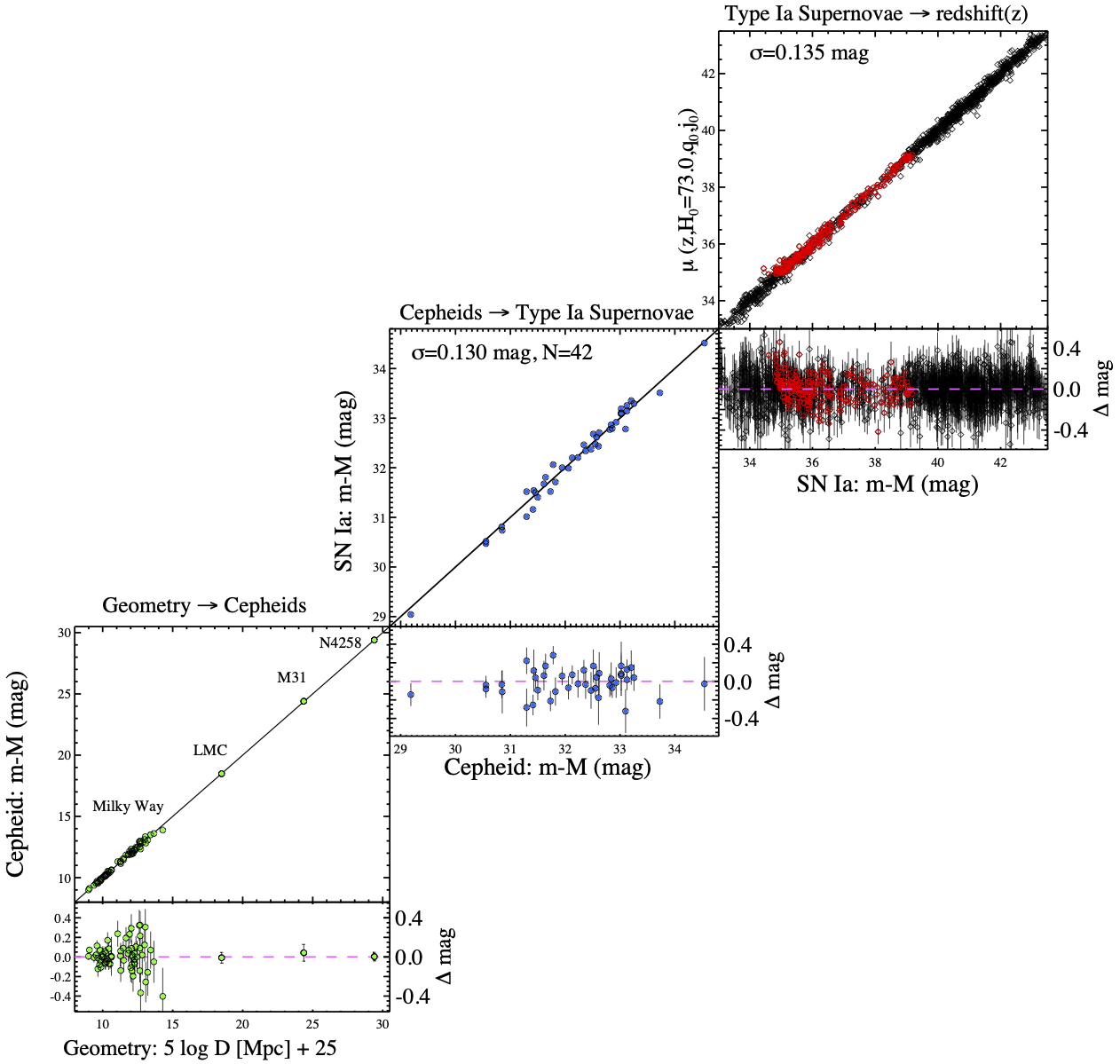}
\end{center}
\caption{\label{fg:ladder} Complete distance ladder.  The simultaneous agreement of distance pairs: geometric and Cepheid-based (lower left), Cepheid- and SN-based (middle), and SN- and redshift-based (top right) provides the measurement of H$_0$.  For each step, geometric or calibrated distances on the abscissa serve to calibrate a relative distance indicator on the ordinate through the determination of $M_B$ or H$_0$.  Results shown are an approximation to the global fit as discussed in the text.  Red SN points are at $0.0233 < z < 0.15$, with the lower-redshift bound producing the {\it appearance} of asymmetric residuals when plotted against distance.}
\end{figure}

\clearpage

\begin{figure}[t]
\begin{center}
\includegraphics[width=1.15\textwidth,angle=90]{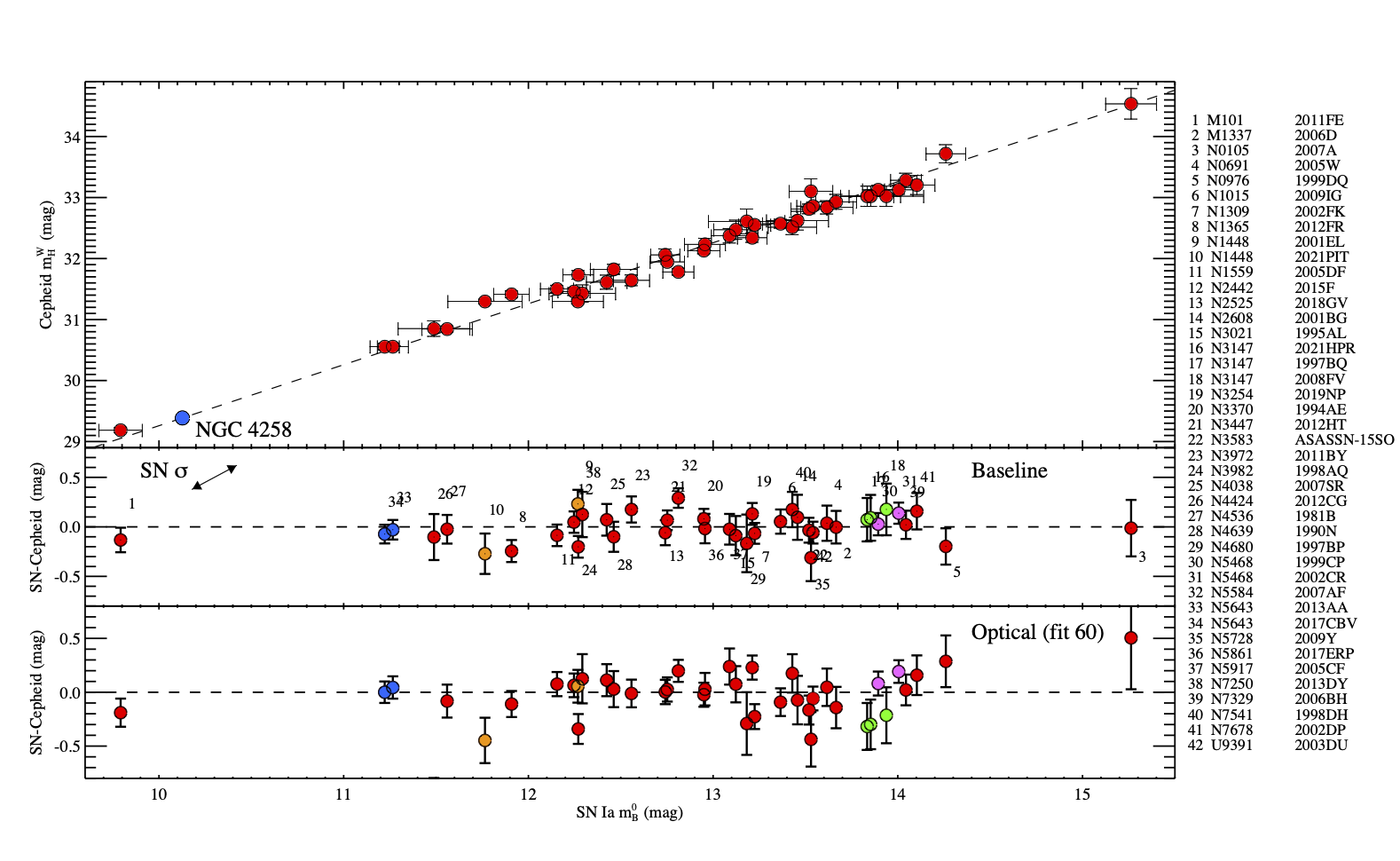}
\end{center}
\caption{\label{fg:comp_ceph} Comparison of SN and Cepheid distance measures on the second rung of the distance ladder, and residuals (middle).  Because the residual panel includes SN measures on both axes, residuals are convolved by the SN error arrow as indicated.  An alternate plot with Cepheids on the abscissa is shown in Fig.~\ref{fg:ladder}.  SNe that share a common host are identified with various colors.  We find no significant trend ($<1.5\sigma$) in this space.  Results shown are an approximation to the global fit as discussed in the text.  The lower panel replaces the NIR-based Cepheid magnitudes with the optical only from Fit~60.}
\end{figure}

\clearpage

\section{The Local Value of H$_0$, the Baseline\label{sc:5}}

Our calibrator sample contains 42 SNe~Ia in the 37 Cepheid hosts presented in the previous section and 277 SNe~Ia in the Hubble flow, all the objects at $0.0233\!<\!z\!<\!0.15$ from the Pantheon+ sample which pass the same quality cuts and are in late-type hosts like the Cepheid calibrators. Criteria for inclusion in the sample used to measure the Hubble flow and variations are further considered in \S\ref{sc:6.7}.  

Our baseline (best) analysis includes the full array of constraints on our model and what we assess to be the optimal choices (considering a wide range of alternatives in \S6) in the treatment of the available data.  Its results, derived from the optimization of $\chi^2$ in Equation~(\ref{eq:chisq}), are given in the first line of Table~\ref{tb:h0var}, which also provides the best-fit parameters.  This fit gives a $\chi^2_\textrm{dof}=$ \chisqbase\, with $N=3445$ degrees of freedom, with Cepheid slope, metallicity, and luminosity parameters $b_W=$ \bslopebase, $Z_W=$ \gammabase, and $M_{H,1}^W=$ \mwbase. These parameters are similar to those found by R16 and updated by R19, with small increases in the absolute values of the slope of the \PLs relation (from $-3.26 \pm 0.03$ in R16) and the metallicity dependence (from $-0.17 \pm 0.06$ in R19). 
 Additional ``nuisance" parameters in the baseline fit are $\Delta \mu_{\rm N4258}=-0.013 \pm 0.022$ mag and $\Delta \mu_{\rm LMC}=0.010 \pm 0.019$ mag. 
Table~\ref{tb:distpar} provides individual host distances and SN parameters. Fig.~\ref{fg:ladder} displays the baseline data, fit, and residuals, while Fig.~\ref{fg:comp_ceph} provides more details about the central panel of the preceding figure.  

The fiducial SN absolute magnitude parameter applicable to the Pantheon+ standardization \citep{Scolnic:2021,Brout:2022} is $M_B^0=$ \mbbase.
The value of the Hubble constant derived from the baseline fit is H$_0$=\hbase.  
Including an additional systematic uncertainty from the analysis variants as discussed in \S\ref{sc:6.14} yields

\bq {\textbf H}\bm{_0}\bm{=}\hsbaseq\ \textbf{(baseline with systematics)}.\eq

\ \par

The determination of H$_0$ in the measured space is $5\,\log$\,H$_0$ $= 9.318 \pm 0.031$ (with systematics) where the errors are symmetric in log space and slightly asymmetric for H$_0$ with errors of ($-0.98$ and $+1.03$ for the fit).  The full difference from the Planck+$\Lambda$CDM prediction of H$_0=67.4 \pm 0.5$ \citep{Planck:2018} in units of $\Delta 5\,\log$\,H$_0$ is $0.176 \pm 0.035$~mag (errors in quadrature), a difference of 5.0$\sigma$ (one in 3.5 million). 

The dispersion (weighted by the measurement errors) between the 42 calibrator SNe~Ia and Cepheids is $\sigma=0.130$~mag which is equivalent to (albeit slightly lower than) the 0.135~mag dispersion between the SN magnitudes and redshifts of the Hubble-flow sample.  We would expect these dispersions to be comparable (as they are) because the additional sources of scatter independent of SNe and applicable to the two comparisons (Cepheid distance errors and peculiar velocities) are both subdominant to SN scatter.  From these dispersions and the good global $\chi^2$ we conclude that there is no unexplained variance in the baseline fit beyond the intrinsic scatter of SNe~Ia and the intrinsic width of the Cepheid instability strip.  

\subsection{Markov Chain Monte Carlo (MCMC) Sampling\label{sc:5.1}}

To check the preceding calculation using a different methodology, we performed MCMC sampling. Our likelihood function is
\begin{equation}\label{equ_likelihood}
    p(q\mid L,C,y) \propto p(q)\ p(L,C,y\mid q)\ ,
\end{equation}
where the $p(L,C,y\mid q)$ is a Gaussian likelihood based on the covariance matrix $ C $; if $ C $ is diagonal, $ C_{ij} = \sigma_i^2 \delta_{ij} $, it has the form
\begin{equation}
    p(L,C,y\mid q) = \prod_i \frac{1}{\sqrt{2\pi}\sigma_y}
    \exp\left[\frac{-(y_i-y_{\rm{fit},i})^2}{2\sigma_y^2}\right]\ .
\end{equation}

Since the covariance matrix $ C $ is independent of $q$ in our setup, Equation~\ref{equ_likelihood} can be simplified to
\begin{equation}
    p(q|L,C,y)\propto p(q) \exp\left(-\frac{1}{2}\sum_i \chi_i^2\right)\ .
\end{equation}

For the calculation of $\chi^2$, we followed the method described in Equation~(\ref{eq:chisq}). For numerical accuracy, we substituted $C^{-1}$ with an inverse of Cholesky-decomposed lower triangular matrix $\tilde C_L^{-1}$ (which satisfies $\tilde C_L \tilde C_L^{-1}=I$ and $\tilde C_L \tilde C_L^T=C$):
\begin{equation}
    \chi^2 = \left(y-L\cdot q\right)\tilde C_L^{-1}\left(y-L\cdot q\right)\ . \label{eq:chimc}
\end{equation}

\clearpage

\ \par
\begin{figure}[b]
\includegraphics[width=\textwidth]{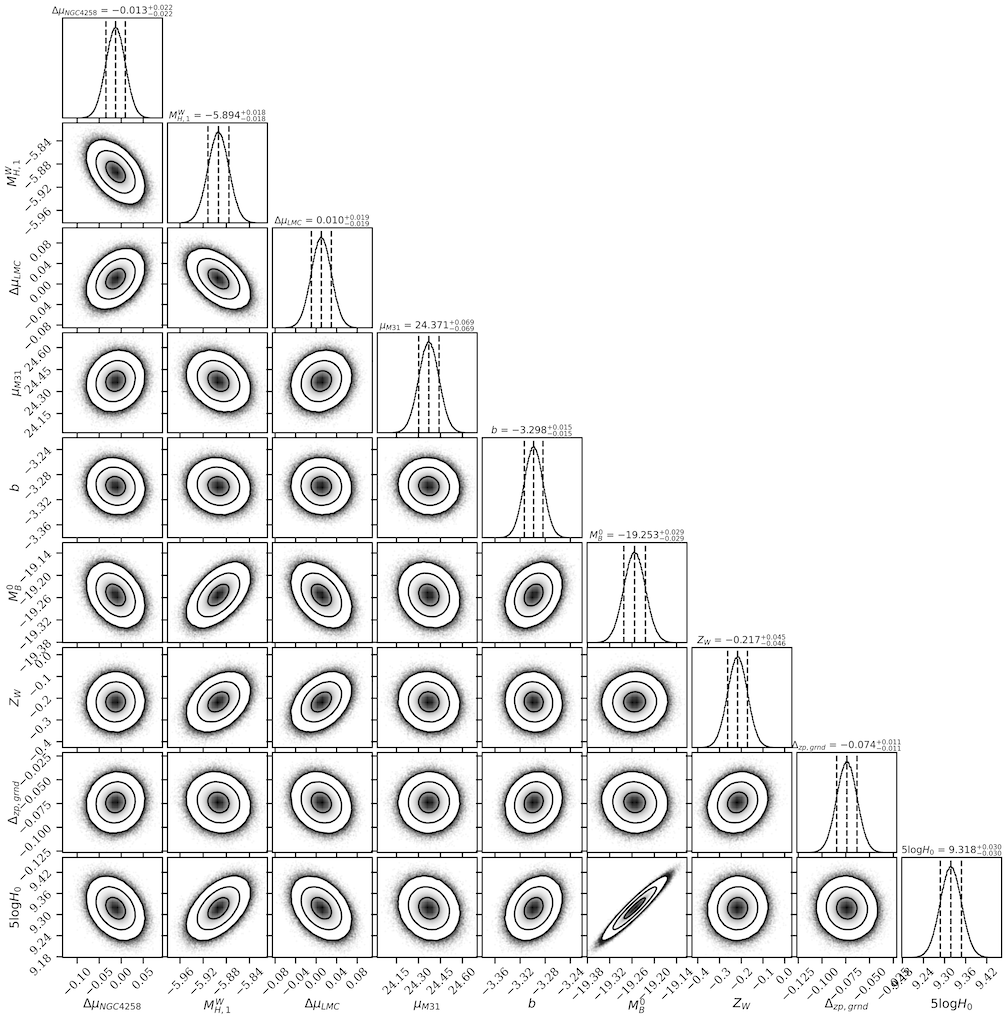}
\caption{\label{fg:corner} MCMC sampling of the $\chi^2$ statistic of the global fit showing all free parameters except the individual host distances.  Contours are 1$\sigma$, 2$\sigma$, and 3$\sigma$ confidence regions.  We find that the means and uncertainties agree very well with the analytical minimization of $\chi^2$.}
\end{figure}

\clearpage

\begin{figure}
\begin{center}
\includegraphics[width=0.75\textwidth]{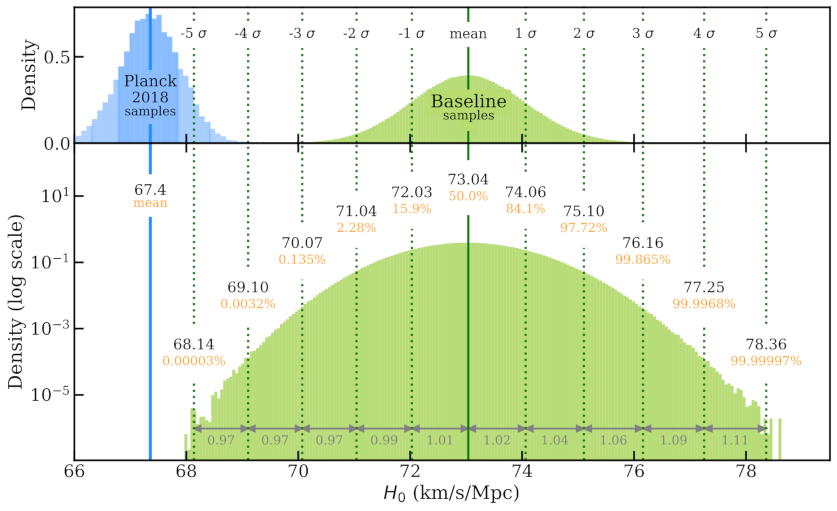}
\end{center}
\caption{\label{fg:pct} Extended MCMC sampling of the posterior for H$_0$ to measure out to the $\pm 5\sigma$ confidence level.  The upper panel shows the probability density for the baseline from SH0ES and from the \citet{Planck:2018} chains.  The bottom panel shows the log of the probability density to improve the ability to see the tails.  We note some asymmetry to the distribution, with intervals on the low-H$_0$ side a little smaller than on the high side, as the measurements are Gaussian in magnitudes and in $5\,\log$\,H$_0$, hence slightly skewed in H$_0$.}
\end{figure}

\begin{figure}   
\begin{center}
\includegraphics[width=0.55\textwidth]{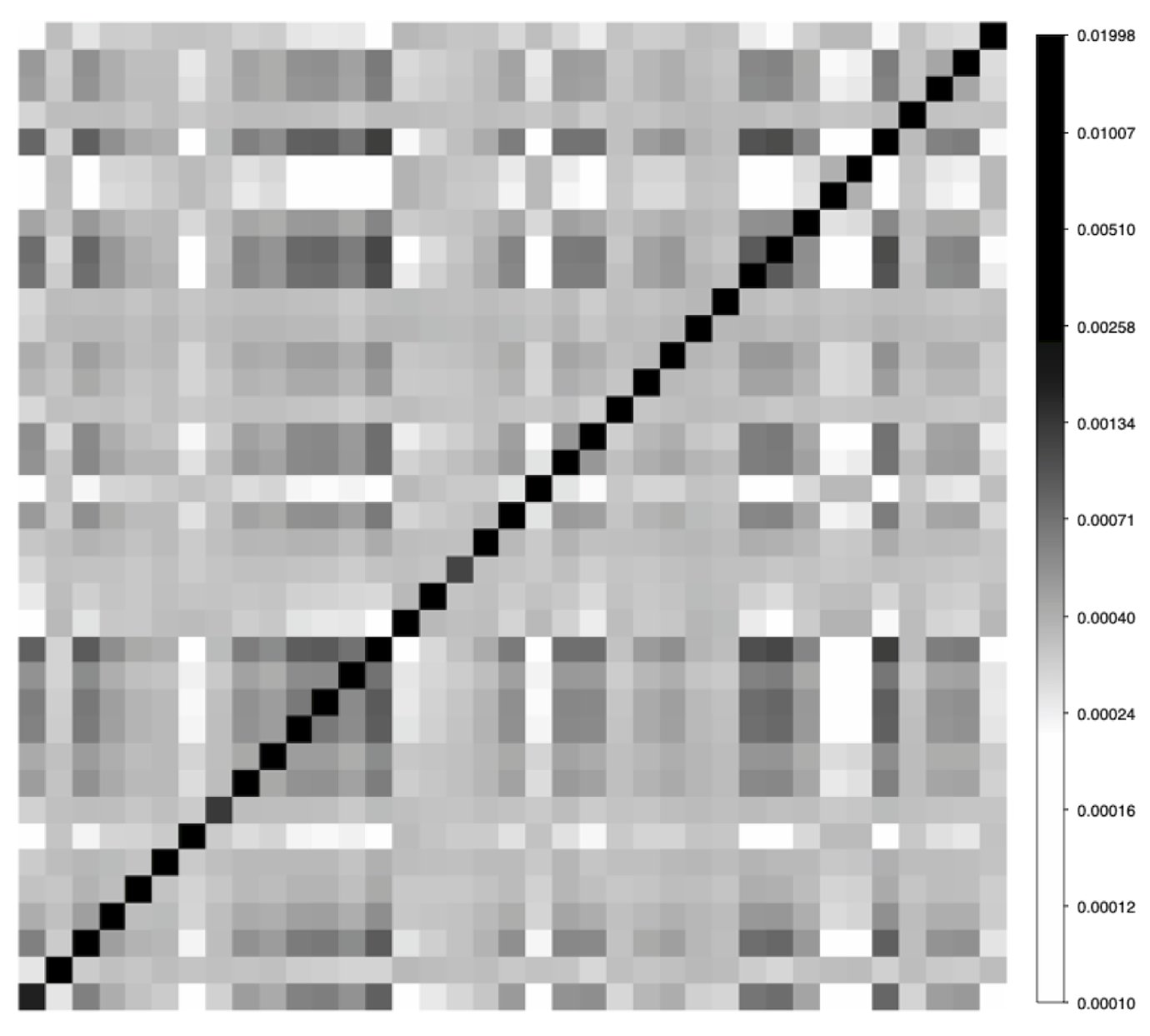}
\end{center}
\caption{\label{fg:covar} Host distance ($\mu_{\rm host}$) covariance matrix.  This is the covariance for the baseline set of 37 hosts derived only from the first two rungs and without the use of any SN~Ia data. The nonzero off-diagonal terms result from common anchors, common parameters of the Cepheid \PL\ relation, and covariance of the metallicity scale. The systematic uncertainty in all host distances may be measured from the square root of the level of the off-diagonal elements and is 0.019~mag, or 0.9\% in distance.}
\end{figure}

\clearpage

As a ``neutral'' prior, we chose the uniform distribution
\begin{equation}
    q_i \sim \rm{Uniform}(\mu_i-10\sigma_i,\mu_i+10\sigma_i),
\end{equation}
which is centered at the results from the analytical result. Its width is set to be sufficiently ($20\times$) larger than the estimated standard deviation from linear regression $\sigma$. We find that this range is more than enough to allow determination of median and standard deviation of the resulting population from $\mu$ and $\sigma$.

We used the \texttt{Python} package \texttt{emcee} \citep[][]{emcee_2013} to perform sampling under conditions mentioned above. The initial states are uniformly distributed in the prior range (Eq.~\ref{eq:chimc}) for a total of 100 walkers.  The convergence is monitored using \texttt{emcee}'s recommended method to estimate the autocorrelation time $\tau$, which employs a method originally proposed by \cite{Goodman_2010_autocorr}. The burn-in time is set to be $5\tau$ to allow chains to fully converge\footnote{We used the final estimation of $\tau(N=N_{\rm max})$, where the relative change of estimated value becomes small enough: $\mathrm d\tau/\mathrm dN/\tau \ll 0.001\%$}.

The convergence of the sampled distribution is checked visually and by the estimated autocorrelation time value. We required the total number of chains after burn-in $N$ to satisfy $N>100\tau$ before any visual inspection for convergence is performed. The samples used in the final analysis are contained well within the inner region of the initial states for all parameters, indicating that our choice of prior size did not affect the final result.

The results for selected parameters are shown in Fig.~\ref{fg:corner}. All parameters exhibit Gaussian-like posterior probability density functions (PDFs), which indicates that the results from the linear-regression method are a good representation of the results from this project.  The locations of Gaussian-equivalent percentiles (i.e., the locations at which the same relative volume is met) in our sample suggest that the samples in H$_0$ are more broadly distributed toward larger values (right-hand side; RHS) than smaller ones (left-hand side; LHS), a consequence of Gaussian errors in magnitudes of $5\,\log$\,H$_0$, with the mean distances between each line $\bar\sigma_R=1.06$ on the RHS and $\bar\sigma_L=0.98$ on the LHS (toward lower H$_0$), similar to the analytical results.

More than 120 million samples were used for the baseline fit to delineate the position of the 5$\sigma$ confidence interval for H$_0$, as shown in Fig.~\ref{fg:pct}. The calculated values of $\sigma$, whose mean is $\bar\sigma_{\rm{all}}=1.02$, contains the estimated uncertainty from the linear-regression method $\sigma_{\rm{reg}}=1.01$, hence showing a full consistency between two methods. Using these three values ($\bar\sigma_L$, $\bar\sigma_{\rm{all}}$, and $\sigma_{\rm{reg}}$), and regarding the $50^{\rm{th}}$ percentile value as the most probable value for the first two cases, our results are
H$_{0,\rm{MCMC-L}}=73.04\pm0.98$, H$_{0,\rm{MCMC-all}}=73.42\pm1.02$, and H$_{0,\rm{reg}}=73.04\pm1.01$~\kms.

The significance of the discrepancy between our results and those of \cite{Planck:2018}, H$_{0,\rm{Planck}}=67.4\pm0.5$, are $5.1\sigma$, $4.9\sigma$, and $5.0\sigma$, respectively. 

\subsection{Simultaneous Constraints on H$_0$ and the Expansion History\label{sc:5.2}}
  
This section illustrates an approach to simultaneously measure H$_0$ and the expansion history for an arbitrary form of H$(z)$, using the full covariance of the datasets.
  
The ``near-field'' determination of H$_0$ in the preceding section is quite insensitive  to specific knowledge of the recent expansion history because the mean redshift at which the measurement is made, $\langle z \rangle=0.055$, is low.  Yet to make the measurement more precise at even these small redshifts, in the prior section we accounted for the derivative of the expansion history in Equation~(\ref{eq:aB}) through the empirical derivative of H$(z)$ measured from higher-redshift SNe~Ia, $q_0$.  We set $q_0=-0.55$ as this is historically a good fit to high-redshift SNe~Ia\footnote{It is also the expectation for a consensus $\Lambda$CDM with $\Omega_M=0.3$ and $\Omega_\Lambda = 0.7$.}. This empirical approach is fully independent of the CMB, so an independent comparison of H$_0$ to the CMB with $\Lambda$CDM is appropriate.
  
However, to consider less-conventional expansion histories such as a rapid change in H$(z)$ together with the measured value of H$_0$, as may be undertaken in the effort to find a resolution of the Hubble tension, it might be necessary and it is certainly more reliable to explicitly account for the dependence of H$_0$ on the form of the expansion history, H$(z)$, at low redshifts \citep{Efstathiou:2021,Camarena:2021}. Furthermore, there is covariance in the measurements of SNe~Ia, whether in Cepheid hosts or at moderate redshifts \citep{Scolnic:2018,Dhawan:2020}, and it is necessary to account for this when SNe~Ia are used {\it simultaneously} to measure H$_0$ and the recent expansion history.

It is important to recognize that standardized SN~Ia data can provide only {\it relative distance measurements} between all SNe measured within the same standardization scheme, with SN parameters, uncertainties, and covariance with values relevant within the context of the algorithm used to standardize the SNe.    Therefore, to avoid inconsistencies between SN standardization schemes or loss of knowledge of measurement covariance, one would ideally make full use of all relevant SN data {\it simultaneously} to determine absolute quantities. A straightforward and reliable path is to use the set of {\it absolute} distances to SN hosts (their uncertainties and covariance) derived from only the first two rungs without the use of any SN data, together with a consistently standardized set of SNe (in these hosts and in the Hubble flow), to determine H$_0$ and H$(z)$ simultaneously, the so-called ``forward'' direction.  Or, one could alternatively follow an ``inverse'' approach, starting with CMB data and using the best-fit $\Lambda$CDM model to calibrate the distance--redshift relation of SNe including the predicted distances of nearby hosts of SNe~Ia and Cepheids (or TRGB)\footnote{A variation of the inverse approach is to predict $M_B$ and compare that with local hosts as suggested by \citet{Efstathiou:2021}, an approach which is equivalent in principle but the dependence of $M_B$ on the method of SN standardization makes it less widely applicable.}. Here we follow the forward approach.

For the set of SN hosts with calibrated distances (using only the first two rungs of the distance ladder), the dereddened absolute distance modulus $\mu_0$ is given by 
\bq \mu_{0,\textrm{host}}-m_B=M_B , \label{eq:muhost} \eq 
with terms given in Equation~(\ref{eq:snmagalt}). We then have a second set of SNe with cosmological redshifts $z>0$, 
\bq m_B=5\,\log\,cz \left\{ 1 + {\frac{1}{2}}\left[1-q_0\right] {z} -{\frac{1}{6}}\left[1-q_0-3q_0^2+j_0 \right] z^2 + O(z^3) \right\} -5\,\log\,{\rm H}_0 + M_B + 25 .\label{eq:mbcosmo} \eq.

\noindent  
The two leading terms on the right-hand side of Equation~(\ref{eq:mbcosmo}) can be replaced with any empirical or cosmological model for H$(z)$, such as the example of $\Lambda$CDM with dark energy equation-of-state parameter $w$ and mass density $\Omega_M$,
\bq m_B=5\,\log\,\left [c{\rm H}_0^{-1}(1+z)  \int_0^z { \frac{dz'}{E(z')}} \right ] +25+M_B , \label{eq:mbempz}\eq
where \bq E(z) \equiv \left \{ \om(1+z')^3 +(1-\om) \times \exp \left [+3\int_0^{\ln(1+z)} d\ln(1+z') (1+w(z')) \right ] \right \}^{1/2}.\eq

The left-hand side of Equations (\ref{eq:muhost}) and (\ref{eq:mbcosmo}) (or \ref{eq:muhost} and \ref{eq:mbempz}) are measured quantities, either $\mu_{0,\textrm{host}}$ or $m_B$, and the free parameters $M_B$, H$_0$, and $q_0$ [or $w$ and $\Omega_M$ for Equation~(\ref{eq:mbempz}) instead of Equation~(\ref{eq:mbcosmo})] are determined by simultaneously optimizing these.

\begin{figure}[b]   
\begin{center}
\includegraphics[width=0.5\textwidth]{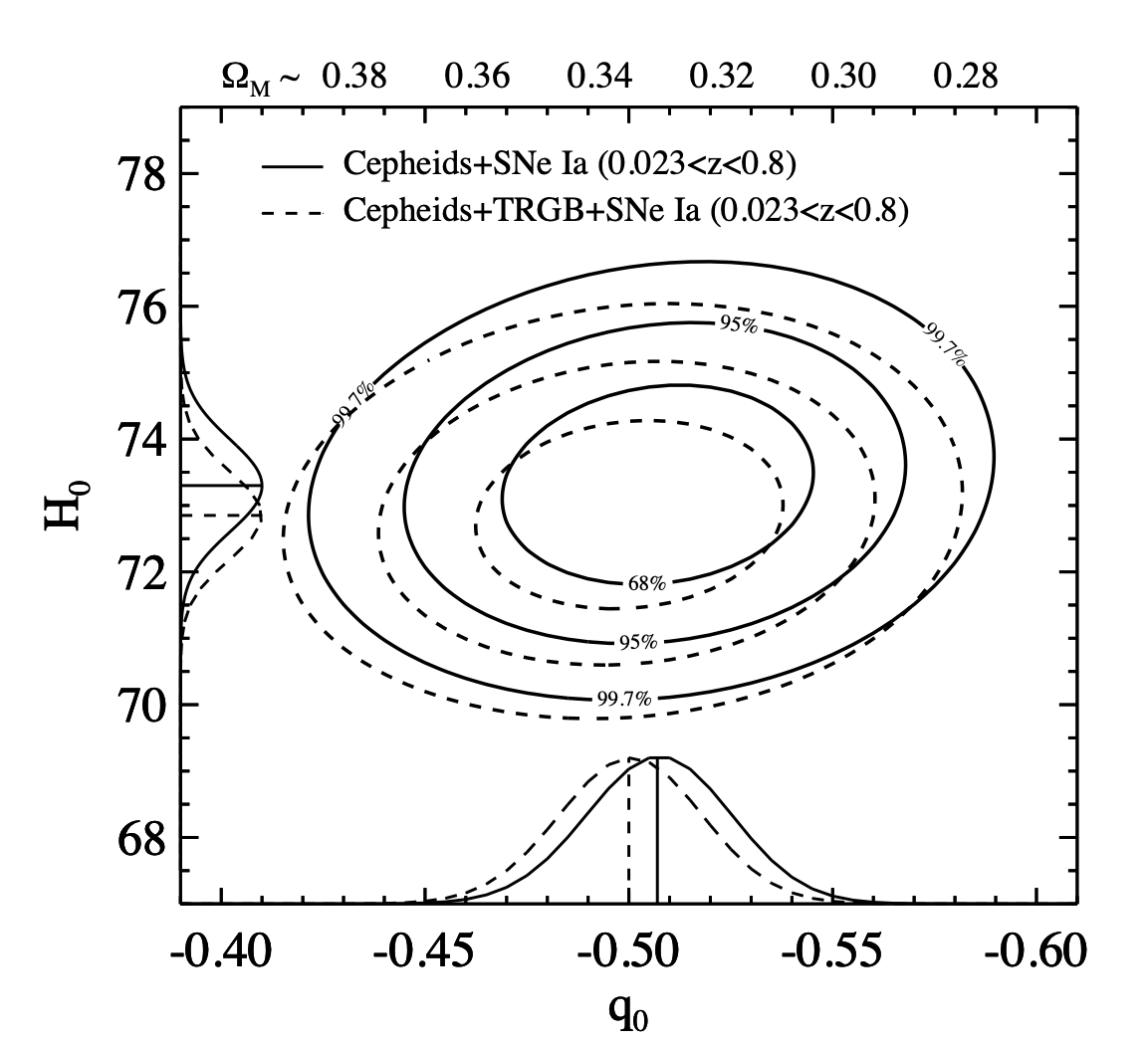}
\end{center}
\caption{\label{fg:h0exp} Simultaneous measurement of H$_0$ and expansion history.  Here the expansion history is fit with a single free parameter, $q_0$, though as discussed in the text, it can be fit with any form for H$(z)$ using the set of $\mu_{\rm host}$ and a set of consistently standardized SNe~Ia in these hosts and in the Hubble flow. }
\end{figure}

Following this approach, the set of 37 values of $\mu_{0,\textrm{host}}$ for 42 SNe~Ia (or 40 values for 46 SNe~Ia including the TRGB data as in \S\ref{sc:6.7}) and their covariance matrix, derived from the first two rungs (i.e., without the use of any SN data), are taken as input.  The covariance matrix of $\mu_{0,\textrm{host}}$ is shown in Fig.~\ref{fg:covar}.  In addition, the values of $m_B$ for the SNe in these hosts and in the Hubble flow (including their redshifts) and the covariance matrix between all SN measurements are taken as additional data and the two constraining relations are solved simultaneously.  

As an example, in Fig.~\ref{fg:h0exp} we follow this approach, marginalizing over the SN standardization parameter, $M_B$, and simultaneously measure H$_0$ and $q_0$.  As expected, H$_0$ will be quite uncorrelated with $q_0$ (or other cosmological parameters) unless a cosmological model produces a much more rapid change in H$(z)$ at $z \ll 1$ than the polynomial in Equation~(\ref{eq:mbcosmo}). For this case we add SNe from Pantheon+ at $0.15<z<0.8$ to the prior analysis, finding $q_0 =$ \qhiz\ and H$_0 =$ \hhiz. The result is very similar to our baseline result and the added freedom in the expansion history has had little impact on the uncertainty. \citet{Brout:2022} give results for a Friedman-Robertson-Walker (FRW) expansion history governed by $w$ and $\Omega_M$.  We provide the SN-independent host distances and their covariance matrix at \url{pantheonplussh0es.github.io} to allow for other forms of H$(z)$.  

\section{Extensions or Variants of the Baseline Analysis\label{sc:6}}

\begin{figure}[b]   
\begin{center}
\includegraphics[height=0.8\textheight]{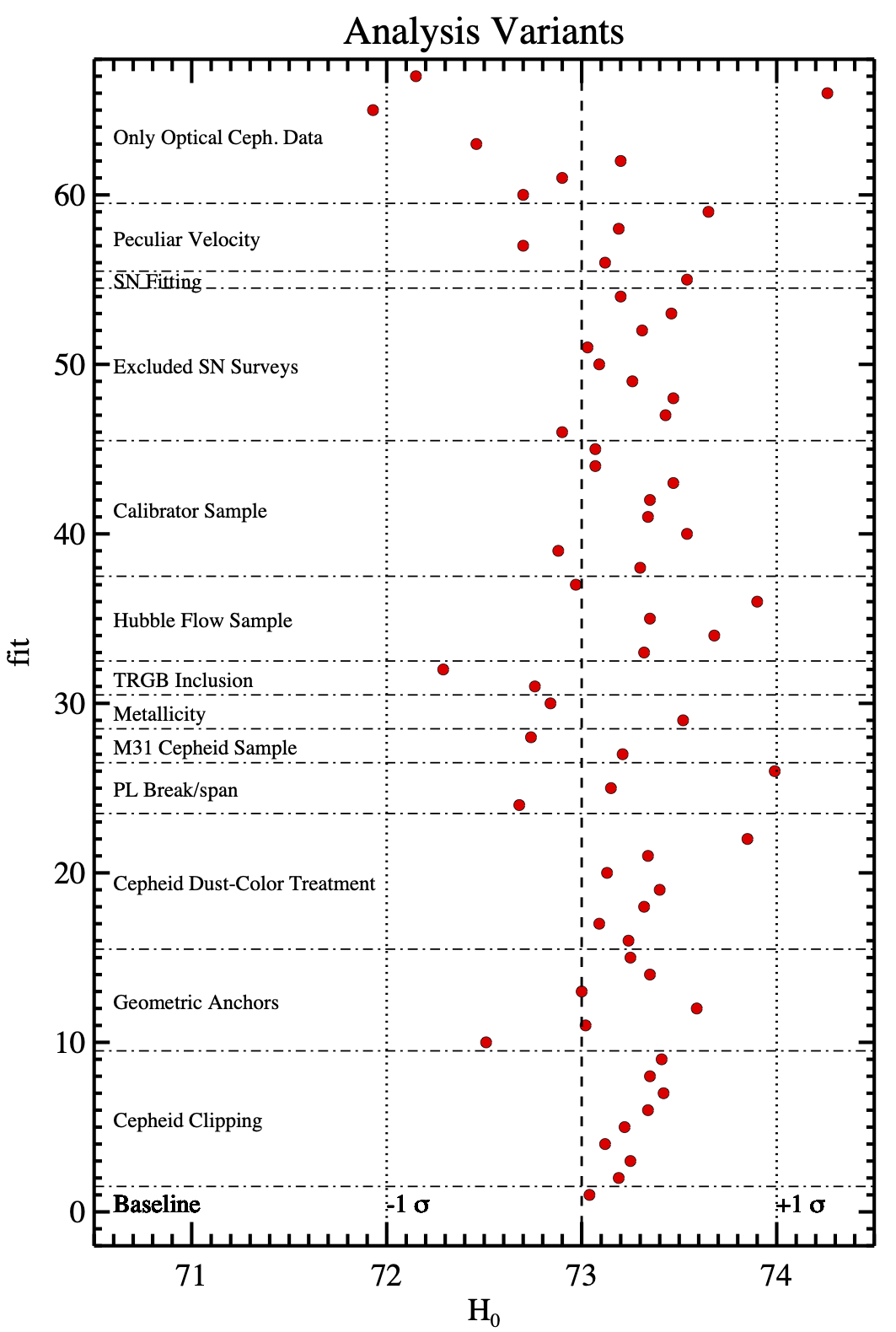}
\end{center}
\caption{\label{fg:vars} Display of 67 fits in 12 categories of alternatives or extensions to the baseline as shown in Table~\ref{tb:h0var}. }
\end{figure}

The baseline analysis was identified as providing the most accurate model of the data that is also the most economical in terms of the number of free parameters. Here we review 12 {\it categories} of alternatives or extensions to the baseline analysis, selected to explore the sensitivity of the results to additional considerations and systematic uncertainties. These are given as 67 fits listed in Table~\ref{tb:h0var}, shown graphically in Fig.~\ref{fg:vars}, and summarized in \S\ref{sc:6.14}.
 
\subsection{Cepheid Clipping\label{sc:6.1}}
  
The optical selection of our Cepheid sample is discussed by H16 and Y22b. As in R16, we include only Cepheids with colors {\it F814W--F160W} within 0.8~mag of the median color in each host to remove blends with unresolved sources of comparable luminosity and different color (e.g., red giants, blue supergiants, unresolved star clusters). This is a useful criterion as it is distance- and period-independent, insensitive to reddening, and anchored to the physical properties of Cepheids (i.e., stars with spectral types F--K). We still may expect a small number of outliers owing to fully-blended yellow supergiants, or the sample may include a small number of objects erroneously identified as Cepheids.
  
Our baseline analysis removes Cepheids that deviate from the global fit at $>3.3\sigma$ (Chauvenet's criterion), iteratively discarding the single largest outlier (i.e., MAD algorithm) until none remain above the threshold.  The fraction of such objects is 1.2\% (these outliers are available upon reasonable request). Fits~2--9 explore other rejection approaches: global (i.e., all objects above threshold are removed as opposed to the most deviant followed by recalculation of the fit), from the individual \PLs relations, and with tighter or looser thresholds ($3\sigma$ or $5\sigma$), as well as no rejection. The median H$_0$ of these alternatives to the baseline clipping is larger by 0.2~\kms. Because the global outlier removal (Fit~2) is faster to calculate and gives results within 0.15~\kmss of the MAD baseline, we use this approach for most other fits unless we explicitly state the use of MAD.  For the number of degrees of freedom here, fits with $\chi^2_{dof}>1.07$ are considered not good (probability to exceed $>3 \sigma$), which applies to Fits 6, 7, and 9 that remove few or no Cepheid outliers.  

\subsection{Geometric Anchors:  Consistency with Metallicity\label{sc:6.2}}
  
\begin{figure}[h]
\begin{center}
\includegraphics[width=\textwidth]{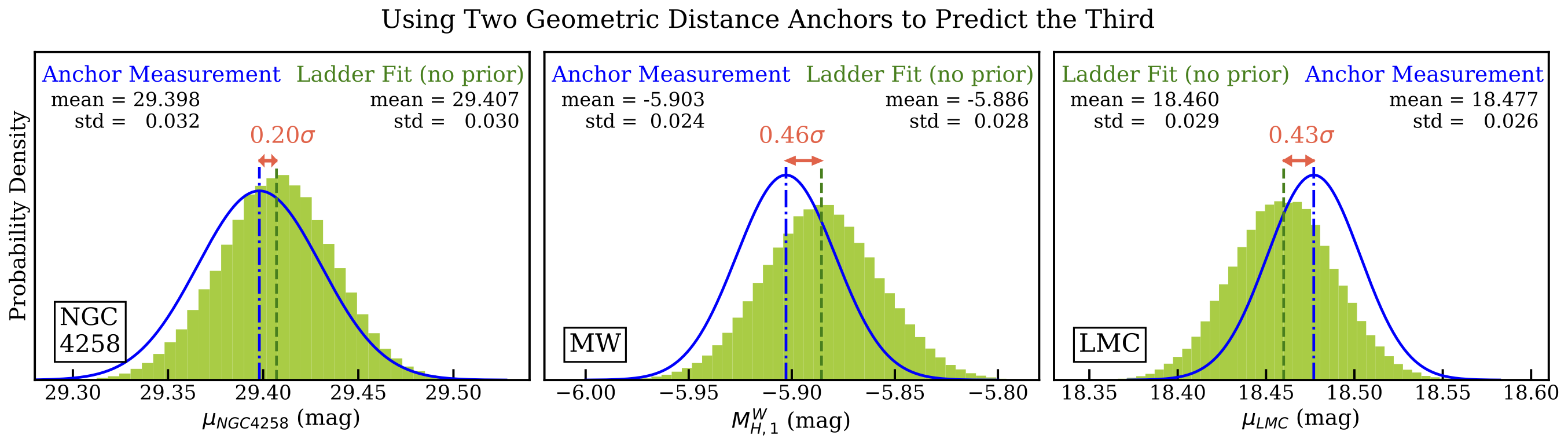}
\end{center}
\caption{\label{fg:anchors} Comparison of each geometric anchor distance with its expected value based only on the Cepheids it hosts and the distance ladder calibrated by the geometric distance of the other two anchors.  The green histograms are MCMC samples from Fits~13, 14, and 15 showing the expected distance of an anchor whose independent distance was excluded from the analysis.  The independent measurement (blue curve) and prediction (green) are in good agreement in all cases to $< 1\sigma$.  }
\end{figure}

Fits~10--15 provide the results of including the geometric distance measurement(s) of only one or at most two anchors,  rather than all three.  The goal here is to explore the possibility of an unexpected error in one of the external geometric distance constraints. In Fig.~\ref{fg:anchors} we compare the geometric distance estimate to each anchor with the value modeled using only its Cepheids and the geometric distance of the other two anchors.  As shown, the external measured distances are consistent with their Cepheids at $-0.2\sigma$ for NGC$\,$4258, $+0.4\sigma$ for the LMC, and $-0.5\sigma$ for the MW.  Their internal consistency (despite having different mean abundances) is a consequence of the Cepheid metallicity dependence.   

\begin{figure}[h]
\begin{center}
\includegraphics[width=0.575\textwidth]{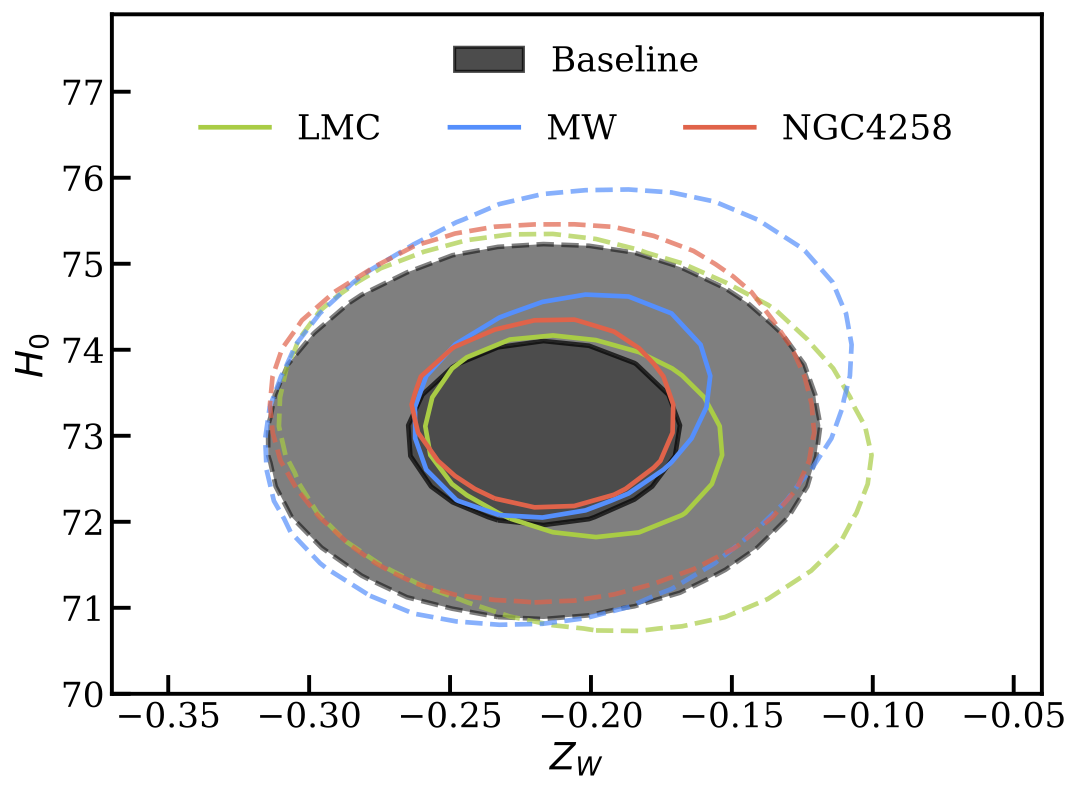}
\end{center}
\caption{\label{fg:h0zw} Marginalized posterior covariance between H$_0$ and the metallicity term, $Z_W$, for the three two-anchor cases and the baseline fit.  The metallicity term is well-constrained, with a substantial tightening owing to the differential DEB distance between the two clouds (LMC and SMC) from \citet{Graczyk:2020}.   The term has little correlation with H$_0$ because the anchor abundances span the range in the SN hosts, but this term provides for the consistency between the anchors.}
\end{figure}

\begin{figure}[h]  
\begin{center}
\includegraphics[width=0.575\textwidth]{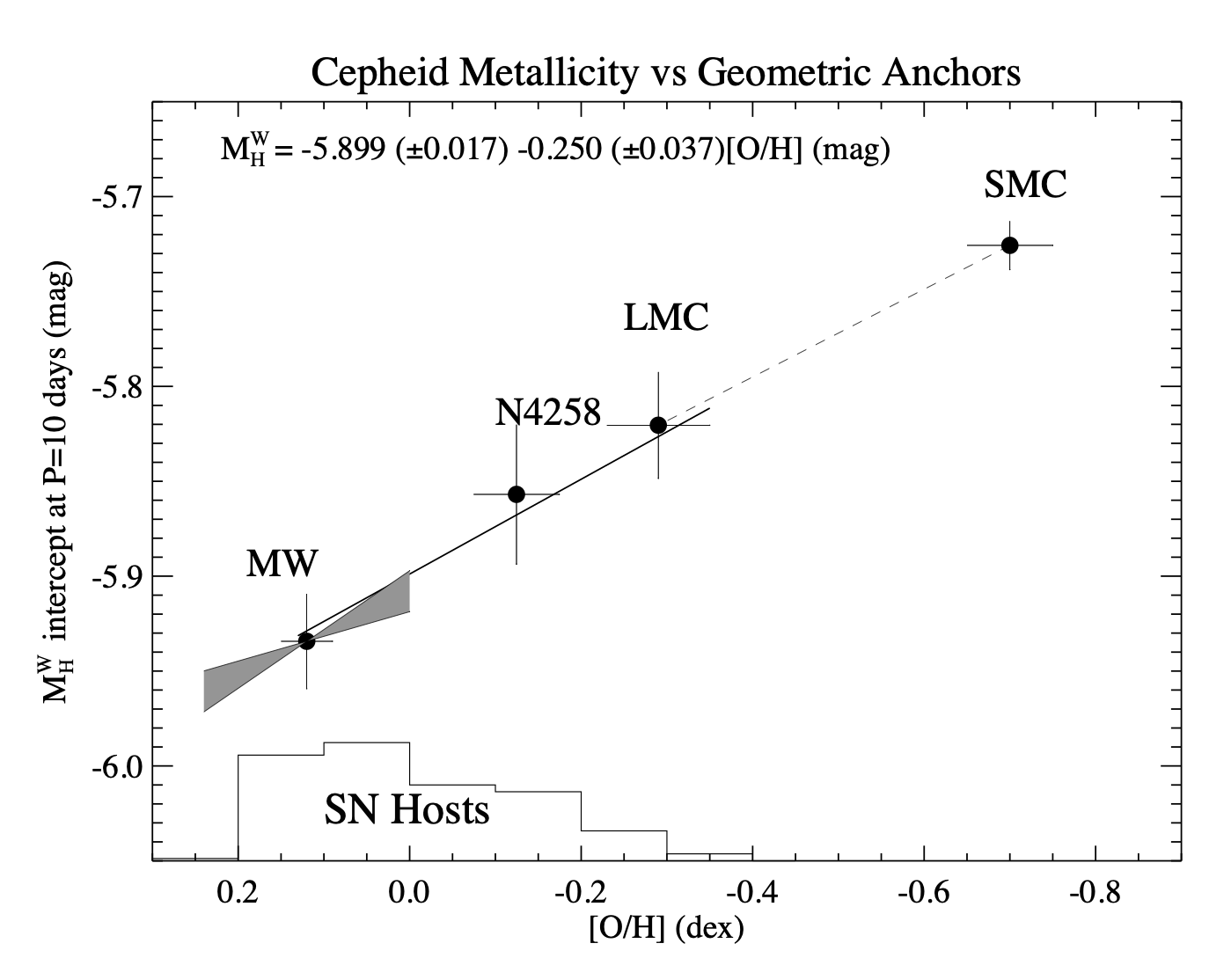}
\end{center}
\caption{\label{fg:mety} The metallicity term and consistency of the geometric anchors.  The mean slope of four hosts with geometric distances is plotted against the intercept of the Wesenheit magnitude \PLs relation.  The DEB distance difference between the SMC and LMC from \citet{Graczyk:2020} as discussed is independent of the calibration uncertainties of the DEB method, making this link, the dashed line, robust and independent of the other anchors (a linear fit to the 4 points is very good, though not unexpectedly so with $P=10$\% to be better). The gray constraint on the MW comes from the breadth of metallicities and individual {\it Gaia} EDR3 parallaxes.  The metallicities of the Cepheids in SN hosts span the range of the anchors making the value of H$_0$ insensitive to the value of the metallicity dependence, with a change in H$_0$ of 0.2~units for a change in $Z_W$ of 0.1~mag~dex$^{-1}$. } 
\end{figure}

The consistency of the anchors is most readily seen by comparing their intercepts and metallicity as shown in Figs.~\ref{fg:h0zw} and \ref{fg:mety}.  Here we determine the intercepts from the Cepheid data table for each host, NGC$\,$4258, LMC, SMC, and the MW using their geometric distances for an absolute measure; see also \citet{Breuval:2021} for a similar analysis.  It is clear that the consistency of the anchors is a direct consequence of the Cepheid metallicity dependence which has been greatly refined since R16.  In our primary analysis we find a metallicity dependence on [O/H] of \gammabase.  In R16 our primary result was $\gamma = -0.13\pm 0.07$ (with variants values ranging from $-0.24$ to $-0.08$) and in R19 it was $-0.17\pm0.06$.   The total $\chi^2$ for the three anchors in \ref{fg:mety} is 0.18, which for a line fit leaves one remaining degree of freedom, and the likelihood to find agreement this good or better is 33\% and thus nominal but not surprising.

A number of recent developments have tightened this constraint considerably while broadening its range.  The DEB distance for the SMC from \citet{Graczyk:2020} as discussed in \S\ref{sc:4.3} provides a differential measurement between the Cepheids in the LMC and SMC which constrains the metallicity dependence. As shown in Fig.~\ref{fg:mety}, this constraint alone gives $\gamma=-0.22\pm0.05$, similar to the values and uncertainties found by \citet{Breuval:2021} for [Fe/H], and falling along the line that joins the other two anchors.  It is one of the strongest constraints available because it comes from the {\it difference} in the DEB distances to each Cloud, a measure which has small uncertainty owing to calibration cancellation and which does not depend on the comparison between the LMC and the other two anchors  \citep{Graczyk:2020}. In addition, the constraint internal to the MW {\it Gaia} EDR3 parallaxes alone ($-0.22\pm0.09$) indicates a similar value\footnote{R21 derived $-0.20\pm0.13$ from 66 Cepheids with {\it HST} photometry.  \citet{Ripepi:2021} used a larger ground-based sample of $N=317$ fundamental and first-overtone pulsators to derive $-0.37\pm0.09$ on the ground system with similar filters, steeper by $\sim 1.5\sigma$ than R21 but less applicable here owing to the presence of overtones and objects with low accuracy. \citet{Groenwegen:2018} used 205 MW Cepheids with {\it Gaia} DR2 parallaxes, somewhat less precise than those in EDR3, to derive a NIR Wesenheit abundance term of $-0.204 \pm 0.14$.  Here we identify 211 Cepheids with basic quality cuts, fundamental-mode only, $V\!-\!I<2$~mag, $m_G>6$~mag, $P>3$~d, and {\it Gaia} GOF $<10$, transforming the ground magnitudes to the {\it HST} filter system to obtain $-0.22\pm0.09$ with a slope of $-3.29$ as indicated in Fig.~\ref{fg:mety}.}.  The global fit also makes use of the internal metallicity gradients in the SN hosts to constrain $Z_W$.  Individually, these are not constraining, with a median uncertainty greater than 1 and a minimum of 0.3. However, combined these are supportive of the results from the local galaxies albeit less constraining, yielding $-0.13 \pm 0.11$~mag~dex$^{-1}$ by combining their independent fits with uncertainties in both axes.  
   
Abundance measurements for 68 of the 70 LMC Cepheids used here \citep{Romaniello:2021} show that they are consistent with a single value, and the lack of any measurable breadth in metallicities negates the ability to measure an abundance dependence internal to the LMC as claimed in prior analyses \citep{Freedman:2011} and further discussed by \citet{Romaniello:2021}. Together, these developments provide a consistent result of a $\sim -0.2$~mag~dex$^{-1}$ metallicity dependence in the NIR that also provides accord amongst the anchors and is little changed by excluding any anchor as shown in Fig.~\ref{fg:h0zw}.

It is important to note that excluding knowledge of the geometric distance to an anchor (e.g., the DEB distance to the LMC) as we do in Fits~10--15 does not exclude the Cepheids in that anchor, which remain extremely valuable (at any distance) for constraining the global properties of Cepheids. Rather, when excluding knowledge of an anchor distance, we allow that distance to become a free parameter which may subsequently be compared to the external geometric estimate.  In this case a parameter such as the slope of the \PLs remains constrained by the excluded anchor Cepheids because these Cepheids are only consistent with a single distance to the host with an accurate value of their slope.  

\subsection{Variants with Color and the Reddening Ratio, $R$\label{sc:6.3}}
  
\begin{figure}[b]  
\begin{center}
\includegraphics[width=0.8\textwidth]{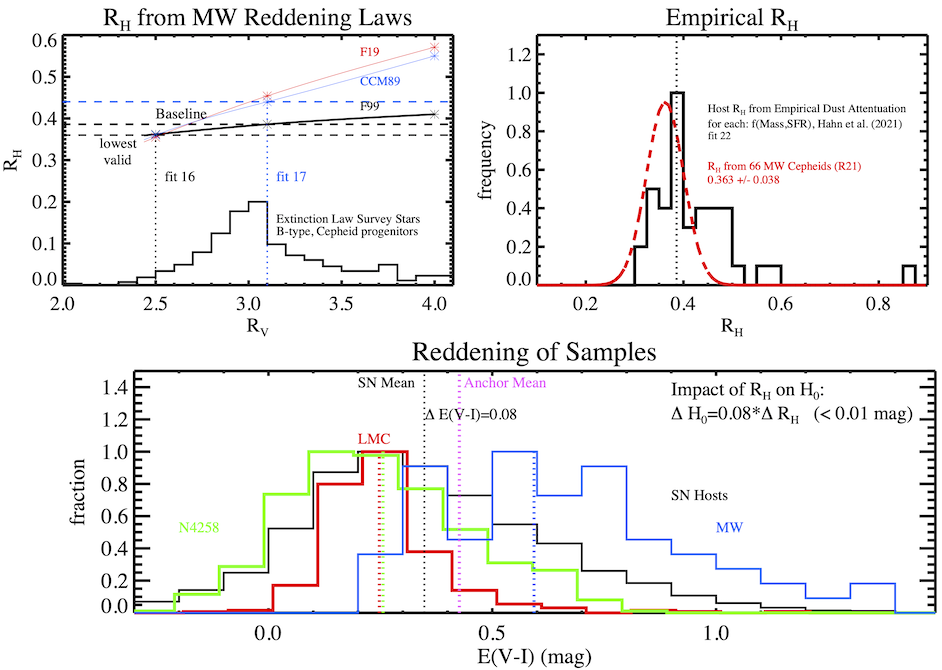}
\end{center}
\caption{\label{fg:rh} Reddening ratios and impact on H$_0$. The upper-left panel shows various reddening laws and their free parameter $R_V$ used to estimate the appropriate ratio $R_H$ for dereddening the Cepheids and the range over which this is applicable. The upper-right panel shows the empirical determination of $R_H$ using the MW Cepheids from R21 and individual estimates parameterized by host, type, mass, and SFR from SDSS galaxies in \citet{Hahn:2021}.  The lower panel shows the color excess, $E(V\!-\!I)$, for Cepheid samples.  The net change to H$_0$ due to differential Cepheid reddening on the ladder is 0.08\,$R_H$ (0.03)~mag, with realistic differences in $R_H$ producing changes $ < 0.01$~mag or  $< 0.5$\%.}
\end{figure}

Here we explore variants in how the Cepheid color, $V\!-\!I$, is used in their distance determination.  Our baseline analysis derives the Wesenheit parameter $R$ (sometimes refereed to as $R_H$ to avoid confusion with the optical reddening law parameter, $R_V$) from the \citet{Fitzpatrick:1999} law with reddening parameter $R_V=3.3$, assuming MW-like reddening for the sample of late-type hosts and thus $R=0.386$.  Fits~16, 17, and 18 (respectively) change the reddening parameter to $R_V=2.5$ or adopt different laws --- \citet[][with $R_V=3.1$]{cardelli89} or \citet[][appropriate for the inner halo]{Nataf:2016}.  The relationship between $R$, $R_V$, and the reddening laws is shown in the upper-left panel of Fig.~\ref{fg:rh}.  Fit~19 allows the value of $R$ for all hosts to be a common but free parameter.  The value derived from MW Cepheids for $R$ is  $0.36\pm0.04$ \citep{Riess:2021} and for the full sample of Cepheids here is $0.34\pm0.02$.  These fits yield a very similar value for H$_0$; including $R$ as a free parameter has little impact on the uncertainty.  
  
If one wishes to allow for differences in the reddening law or the value of $R$ across different hosts, it is necessary to first subtract the intrinsic color of Cepheids using their empirical period--color ($P$--$C$) relation, in order to separate the component of the color that results from dust reddening; see Appendix~\ref{sc:appd} for details.  \citet{sandage04} dereddened MW Cepheids to determine $\langle(V\!-\!I)_0\rangle=0.256\log P+0.497$.  We used the LMC reddening maps from \citet{Skowron:2021} to deredden the Cepheids in the catalog of \citet{Macri:2015} to derive a $P$--$C$ relation of $\langle(V\!-\!I)_0\rangle=0.238\log P+0.513$, very similar to the MW one. From the same data we derived the intrinsic $P$--$C$ relation from the LMC Cepheids as above to be $\Delta m_H=0.635(\pm 0.021)\ \Delta({V\!-\!I})$.

We use a mean relation of $\langle(V\!-\!I)_0\rangle=0.25\log P+0.50$ to first subtract the intrinsic color from all Cepheids using their periods, and then in Fit~20 we substitute the Wesenheit magnitudes in Equation~(\ref{eq:wh}) for $m_H-R\ E({V\!-\!I})$ with $R=0.386$, which reduces H$_0$ by 0.1~\kms. In Fit~21 we allow this reddening ratio to be a free parameter with little change from the result given in Fit~19.
  
We can also consider different values for the reddening ratio, $R$, for different hosts. One might derive these as the value that optimizes the relation between colors and magnitudes within each host.  However, after subtracting the intrinsic $P$--$C$ relation, the small residual color span, coupled with relatively large color measurement uncertainties, does not provide any meaningful constraint on the individual values of $R$ for a given host beyond the few nearest galaxies as demonstrated in Appendix~\ref{sc:appd}.  In addition, Appendix~\ref{sc:appd} shows that determining an unbiased estimate of $R$ requires accounting for uncertainties in both axes \citep[in this case color and brightness with the statistical issue discussed by][]{Tremaine:2002}, and failing to do so leads to large underestimates of $R$ and its uncertainty, as seen (for example) in \citet{Peri:2021} and \citet{Mortsell:2021}.  We find typical uncertainties in individual values of $R$ to be $\sim 1$, and thus uninformative. Likewise, \citet{Follin:2017} also concluded that such color data are ``insufficient to make a completely data-driven inference on [individual] $R$,'' and they used a ``wide-prior'' of $0.39\pm 0.1$, finding distant hosts consistent with that prior and little impact on H$_0$. It is also crucial to recognize that the empirical reddening law is only valid for use at $R_V > 2.5$, as stated by \citet{Fitzpatrick:2019}, because sight lines with $R_V<2$ are not seen among the hundreds of massive MW stars used to determine the law as shown in Fig.~\ref{fg:rh} (indeed, the Rayleigh-scattering limit for absorption corresponds to $R_V \geq 1.5$).   Thus, there is no empirical support for $R<0.3$ from these laws or from the stars (significantly, of the same B-type that later become Cepheids) that inform them.  

An alternative approach with better grounding is to derive individual values of $R$ for each host based on its specific properties and the empirically determined correlation of these properties with dust attenuation.  We use the empirical dust attenuation (EDA) framework from \citet{Hahn:2021}, which derives individual extinction laws for hosts as a function of their mass, type, and star-formation rate (SFR) as determined to best match colors to SDSS galaxy observations.  These individual estimates of $R$ are given in Table~\ref{tb:distpar} and shown in Fig.~\ref{fg:rh}, and have a mean of 0.42 and dispersion of 0.10, well matching the prior used by \citet{Follin:2017}.  The smallest value for any host is 0.30 for NGC$\,$7541, demonstrating that values lower than this become unrealistic as discussed further in Appendix~\ref{sc:appd}.  Fit~22 uses these modeled values of $R$ from \citet{Hahn:2021} and results in an increase in H$_0$ of 0.8~\kms.
  
It is clear why varying the reddening ratio in hosts is not effective at changing H$_0$.  The mean value of $E(V\!-\!I)$ in the SN hosts is 0.35~mag, as shown in Fig.~\ref{fg:rh}.  For the Cepheids in the anchors MW ({\it Gaia} EDR3), LMC, and NGC$\,$4258, it is 0.58, 0.23, and 0.24~mag (respectively), for a mean anchor value (weighted by the precision of the anchor distance) of 0.43~mag.  Thus, the net difference in $E(V\!-\!I)$ between anchors and SN hosts is 0.08~mag, and hence the full impact on H$_0$ of correcting for Cepheid reddening is 0.03~mag or $\sim 1.5\%$, with perturbations to this procedure changing the result by a smaller amount. As shown in Fig.~\ref{fg:rh}, the strongest empirical evidence has the characteristic value of $R$ less than 0.1 from the baseline; hence, e.g., $\Delta$H$_0=0.08\Delta R$~(mag) $< 0.01$~mag.  
In Fit~23 we discard the use of color altogether, representing a reasonable assumption that the Cepheid extinction at 1.6~$\mu$m is modest and to first approximation cancels along the distance ladder, and find that H$_0$ goes up by 2\%.

\subsection{Form of the P--L Relation\label{sc:6.4}}

In \S\ref{sc:4.6} and Fig.~\ref{fg:slope} we measure the slopes of the \PLs relations of each host as well as the mean slope above and below $P\!=\!10$~d, finding no clear evidence of a break. It is important to note that we have not included Cepheids with $P\!<\!5$~d (either in the anchors or SN hosts) in our analyses.  Such Cepheids could provide additional support for a break; however, Cepheids with overtone pulsations become much more common below this period, obey a different \PLs relation, and may be confused with fundamental Cepheids.  

However, in R16 the baseline was to allow for a break at $P=10$~d, in accord with prior claims of a break in the optical in the LMC \citep[][although no evidence for one was found by R16]{sandage04}. To allow for additional comparison to R16, in Fit~24 we allow for a break at $P\!=\!10$~d with independent slopes above and below this pivot.  The two slopes have a slope difference of $0.10\pm 0.05$ as given in Table~\ref{tb:h0var} which lowers H$_0$ by 0.45~\kms, with the difference largely driven by the ground-based samples from the LMC and SMC, whose Cepheids at lower periods have a shallower slope by $0.12 \pm 0.08$ and $0.34 \pm 0.20$ (respectively).   To further explore the evidence for a break, we expand the sample of Cepheids in M31 by including those from \citet[][sample III]{Kodric:2018} in Fit~28.  Because in M31 the opposite occurs (shorter-period Cepheids have a steeper slope by $0.20 \pm 0.12$), the inclusion of both samples reduces a difference in slope to $0.07\pm 0.04$ and H$_0$ lower than the baseline by only 0.3~\kms.  A free-form analysis of a slope change (allowing each host to have two individual slopes) reduces the sample evidence of a break to $< 1\sigma$, with only three hosts (the LMC, SMC, and M31) providing any significant weight at $P<10$~d (MW Cepheids with {\it Gaia} parallaxes provide no indication of a change in slope). The lack of significant evidence of a break (or change in slope) is why the baseline analysis uses the single-slope parameterization.  Another option here is to exclude Cepheids with $P<10$~d which we do in Fit 25, resulting in an 0.1~unit increase in H$_0$.

\subsection{M31 Variants\label{sc:6.5}}

Fits 27 and 28 exchange the \citet{Li:2021} sample of M31 Cepheids for the PHAT sample from \citet{Kodric:2018} with a ten-fold increase in the number of variables, but with filters that are transformed to the set of three used elsewhere rather than directly observed with them.  The main value of this change is to gain further traction of a possible break at $P\!=\!10$~d as discussed in the prior section.  

\clearpage

\subsection{Variants of Metallicity\label{sc:6.6}}  

In Fit~29 we neglect the metallicity term and find that it has little impact, with H$_0$ rising by 0.3~\kms.  The metallicity term has little impact on H$_0$, as shown in the marginal confidence region of Fig.~\ref{fg:h0zw}.  The reason can be seen in Fig.~\ref{fg:mety}, where the distribution of Cepheid metallicities in SN hosts well matches the range of the three anchors; thus, the metallicity term does not move the SN hosts with respect to the mean of the anchors.  It does, however, provide for consistency of the Cepheid and anchor distances.

Appendix~\ref{sc:appc} gives further details about the Cepheid metallicity scale we adopt, the average of nine recent and well-characterized relations \citep{Teimoorinia:2021}.  In Fit~30, we replace the use of the mean metallicity scale with one of these systems \citep{Pettini:2004}\footnote{The PP04 calibration was suggested by F. Bresolin (priv. comm.) and indicated in \citet{Bresolin:2016} to provide a good match to extragalactic star spectral abundances, making it a good reference system.} based on [O~III] and [N~II] lines, which we show in Appendix~\ref{sc:appc} provides consistent metallicities for MW H~II regions with those at the same radius derived from spectra of MW Cepheids.  

\begin{figure}[b]  
\begin{center}
\includegraphics[width=\textwidth]{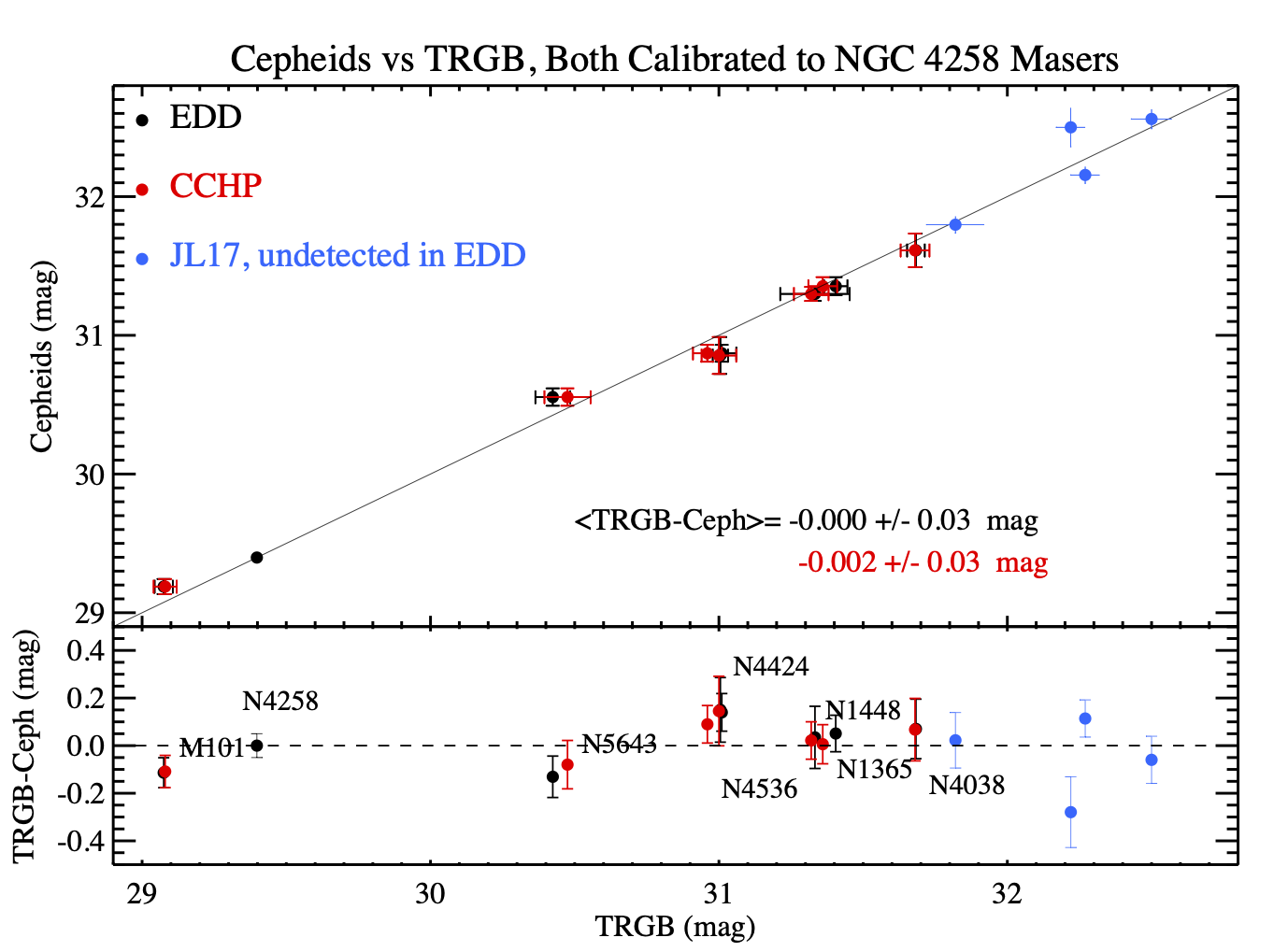}
\end{center}
\caption{\label{fg:trgb} Cepheid distances from this work and TRGB distances from the CCHP (F21) and the EDD \citep{Anand:2021} groups to the same SN hosts calibrated by the same geometric anchor distance to NGC$\,$4258, all measured with {\it HST}.  The seven hosts indicated are all of those measured by these three groups.  In blue we show four more-distant hosts measured by \citet{Jang:2017} for which evidence of a TRGB was not detected by \citet{Anand:2021}.  The Cepheid and TRGB distances are consistent, allowing us to combine them (including for three additional TRGB hosts: NGC$\,$1316, 1404, and 4526) simultaneously in Fits~31 and 32.}
\end{figure}

\clearpage

\subsection{Inclusion of TRGB\label{sc:6.7}}

The TRGB offers an additional, independent rung between anchors and SN hosts \citep[][and references therein]{Freedman:2019,Anand:2021}. The inclusion of TRGB distances can in principle improve the calibration of SN~Ia hosts and add to the number of such objects on the second rung.  However, before including TRGB distances, it is important to determine if they are consistent with those employed here for Cepheids\footnote{It is not possible to make an absolute statement of whether TRGB and Cepheid distances are consistent because this will depend on how each is calibrated and measured.  Here we focus on a specific implementation of each.}.  We use the TRGB samples measured by the CCHP \citep{Freedman:2021} and EDD \citep{Anand:2021} groups.

To determine the consistency of the two distance indicators, we calibrate both using the same geometric anchor, NGC$\,$4258, whose Cepheids and TRGB were both observed with {\it HST}, making this a purely differential comparison to a set of the same SN hosts.  In Fig.~\ref{fg:trgb} we compare the distances from each method for all seven SN~Ia hosts with Cepheid and TRGB measurements which are available from {\it both} CCHP and EDD.  The mean difference between Cepheids and TRGB is $0.000\pm0.030$ and $0.002\pm0.030$~mag for EDD and CCHP, respectively (we have not included the four more-distant SN~Ia hosts measured by \citet{Jang:2017} and included in F19 because \citealt{Anand:2021} could not identify any reliable TRGB for these, but we plot them in Fig.~\ref{fg:trgb}), and conclude these are consistent (see~\S\ref{sc:7.2} for further discussion).  For consistency with the preceding Cepheid-only analyses, we do not include SNe~Ia with TRGB measures which do not pass the SN quality cuts employed above ($\abs{c} < 0.15$ and $\abs{x1} < 2$), which affects four objects (SNe~1989B and 1998bu with $c=0.3$ corresponding to $A_V \approx 1$~mag, and SN~1981D with $c=0.2$, also with high reddening; SN~2007on with $x1 = -2.2$). Ten TRGB distances for 11 SNe~Ia are thus added: the seven hosts with Cepheid distances (and eight SNe) in Fig.~\ref{fg:trgb} (M101, NGC$\,$1365, 1448, 4038, 4424, 4536, 5643) and three additional hosts without Cepheids with four additional SN calibrators (SNe~1980N and 2006dd in NGC$\,$1316, SN~2011iv in NGC$\,$1404, and SN~1994D in NGC$\,$4526).  

As we did for the Cepheids, we use TRGB distances where the anchor and SN hosts made use of the {\it same telescope and instrument} to negate telescope zeropoints.  In this case we first use Table~2 from \citet{Anand:2021}, which includes the calibration based on new observations of NGC$\,$4258 from our program (GO 16198) that also employ the same filters ({\it F606W} \& {\it F814W}), ACS electronics and similar level of CTE as for the SN hosts. We do not include TRGB zeropoints in the Magellanic Clouds or the MW because these have not been measured directly on the {\it HST} system (owing to the impracticable area).
For the joint analysis we include an additional parameter for the TRGB luminosity, $M_I$, which is optimized in the fit.  We note that the TRGB constraint is included as available for a SN host {\it simultaneously} with the Cepheid constraint through the addition of the relation for the $i$-th SN host,
\bq m_{I,\textrm{TRGB},i}=\mu_{0,i}+M_{I,\textrm{TRGB}} , \eq
and the calibrating relation
\bq m_{I,\textsc{n4258}}-\mu_{0,\textsc{n4258}}=\Delta \mu_{\textsc{n4258}}+M_{I,\textrm{TRGB}} , \eq
which adds a single free parameter, $M_{I,\textrm{TRGB}}$.

Including TRGB (EDD), Fit~31 lowers H$_0$ by 0.3~units to \hwtrgb\ and reduces the overall error by 5\%, yielding a value of $M_I=-4.003\pm0.025$~mag similar to that found by \citet{Anand:2021}.
The TRGB SN host and Pantheon SN data yield 71.5~\kmss before the sample is tripled by combining with the SN-Cepheid hosts with the increase in H$_0$ of $\sim1$~\kms, consistent with the shot noise of each subsample as discussed in detail in \S\ref{sc:7.2}.
We note that because the EDD TRGB parameterization includes a color dependence, the value of $M_{I,\rm{TRGB}}$ quoted here corresponds to their fiducial, blue TRGB with $V\!-\!I=1.23$~mag. 
For Fit~32 we replace the EDD TRGB measurements with the CCHP set as given by F19 and F21 for the same SN hosts and by \citet{Jang:2021} for NGC$\,$4258 TRGB\footnote{The \citet{Jang:2021} measurement in NGC$\,$4258 used observations with a different filter, {\it F555W}, and different electronics than the {\it F606W} and the refurbished electronics used for all SN hosts and by \citet{Anand:2021}. More significantly, the state of CTE degradation on ACS was also markedly different between these observations of NGC$\,$4258.   In 2003--2005, a typical TRGB star in the halo would have suffered CTE losses of $\sim 0.04$~mag.  At the time of the SN host observations in 2015--2019, such stars would have lost $\sim 0.14$~mag, more similar to the observations analyzed by 2020 \cite{Anand:2021}. While pixel-based CTE rectification in the STScI pipeline attempts to account for such losses, the differential loss over the 15~yr would be $\sim 0.1$~mag and the uncertainty in the correction would be a sizeable fraction of that, likely a few hundreths of a magnitude. For this reason it is preferred to use data more closely spaced in time.} and without a color dependence to match the CCHP implementation.  This reduces H$_0$ by 0.45~units from the EDD-based result (0.75~units below the baseline).  The difference between the EDD and CCHP-based result is a direct consequence of the 0.04~mag brighter measurement of the tip in NGC$\,$4258 by \citet{Jang:2021} compared to EDD, the significance of which is $1.9\sigma$ as further discussed in \S\ref{sc:7.2}.  The \citet{Jang:2021} measurement in NGC$\,$4258 in concert with the other constraints  yields $M_{I,\rm{TRGB}}=-4.025 \pm 0.023$~mag for the above parameter. (We note that a direct comparison of the $M_{I,\rm{TRGB}}$ parameter to that observed for NGC$\,$4258 requires adding to this $\Delta \mu_{\rm N4258}$ in Equation~(2), which for the baseline is $-0.013$~mag.) The mean result from the two TRGB implementations is \hwtrgbmean, which we adopt as a reference value for TRGB inclusion.

\subsection{Hubble-Flow SN Sample Variants\label{sc:6.8}}

Here we explore fits utilizing different selection criteria for inclusion in the Hubble-flow sample. The goal of the selection of SNe and their hosts for the Hubble-flow sample is to match as well as possible the same criteria used to collect the calibrator sample to guard against the presence and imbalance between samples of additional host or SN characteristics currently known or {\it unknown} that may correlate with Hubble residuals.  The enhanced size of the calibrator sample, now with 42 SNe~Ia, reduces the likelihood of a chance imbalance of such properties.

The host selection for the baseline Hubble-flow samples requires visual identification from the best available optical imaging to be a spiral type in the full range of Sa--Sd, making them likely hosts of massive star formation in the last 0.1~Gyr --- the primary criterion for our targeting them for Cepheid observations.  This excludes highly-inclined hosts ($>75^\circ$) as these are complex targets and were not considered for finding Cepheids nearby.  The host requirements are in addition to the SN~Ia quality cuts which are relatively tight in order to match the calibrators ($\abs{c} \leq 0.15$, $\abs{x1} < 2$), with the first observation earlier than 5~d after maximum light, distance measurement error of $\leq 0.2$~mag, with outliers $>3.5\sigma$ from the Hubble flow removed.  The redshift range, as in R16, is $0.0233 < z < 0.15$.  We refer to this sample as ``LZSPI'' (low-$z$, spiral).   In \S\ref{sc:7.2} we will further analyze the host properties of these samples to confirm their balance in mass, SFR, and specific star-formation rate.  In Appendix~\ref{sc:appa} we compare the colors and light-curve shapes of the SN samples, confirming the similarity of the calibrator and selected Hubble-flow sample.

In Fit~33 we expand the Hubble-flow sample by removing any limitation on the host and including all SNe that pass the Pantheon+ quality cuts (which allow $\abs{c} \leq 0.3$ and $\abs{x1} < 3$), nearly doubling the sample to 482 SNe.  It is noteworthy that there is no change in H$_0$, suggesting that after Pantheon+ standardization, SNe in different host types provide consistent distances.  In Fit~34, we expand the sample to high redshift ($z<0.8$) and more than 1300 SNe, which raises H$_0$ by 0.5~\kms.  The use of such high-redshift SNe requires knowledge of $q_0$ as discussed in \S\ref{sc:5.2} and \S\ref{sc:6.12}. Fits~35 and 36 use these two expanded samples but exclude $z<0.06$ to circumvent concerns about the presence of a hypothesized local void structure.  These raise H$_0$ by $\sim 0.3$~\kms.  

\subsection{Calibrator SN Sample Variants\label{sc:6.9}}

Here we explore variations of the calibrator sample. A complete, volume-limited sample is desirable, as it has the simplest and most unbiased selection function. Such a sample is limited by the volume $z \leq 0.011$ of suitable calibrators between 1980 and 2021 and can be selected by excluding two SNe~Ia from the baseline sample at higher redshifts (SNe~1999dq and 2007A), with results given in Fit~38.  Fit~39 contains the same with the added TRGB distance measures.  Fit~40 limits the hosts to only high-mass galaxies, $\log\,(M/M_\odot) > 10$.  
  
In \S\ref{sc:7} we discuss tests of the Cepheid background.  Here we undertake another test of the background with Fits~41 and 42, which divide the calibrator sample into halves with lower-than and higher-than median background.  The difference in H$_0$ is 0.0~\kms, which is much less than the independent shot noise of each half (1.3\% or 0.9~units for each), and thus there is no indication of a difference in H$_0$ for high and low backgrounds.  Fits~43 and 44 respectively use only the same 19 SNe from R16 or only the 23 SNe~Ia added here, yielding a difference of 0.4~units. Fit~45 includes only the nearer half, defined as having $m_B < 13$~mag (the median of the sample, corresponding to $D < 28$~Mpc), as nearer hosts offer greater spatial resolution, lowering H$_0$ by 0.1~units.
   
\subsection{Excluded SN Survey Variants\label{sc:6.10}}
  
The SN standardized magnitudes have been drawn from the Pantheon+ sample, which is based on more than a dozen past SN surveys.  The Pantheon+ sample recalibrates each survey photometrically to a common reference using the standard-star measurements in the fields of each SN to negate the impact of survey calibration errors.  In addition, most SNe observed in the Cepheid host sample are matched by SNe in the Hubble-flow sample {\it observed by the same SN survey} with no one survey having a dominant share.  As demonstrated by \citet{Brownsberger:2021}, ``gray'' photometric survey errors strongly and beneficially cancel by populating both samples with SNe from the same surveys. \citet{Brownsberger:2021} find that survey miscalibration and incomplete cancellation would affect H$_0$ measured from the survey mix used here at $\sigma=0.15$~\kms\ even for extremely large survey zeropoint errors of $\sim 0.1$~mag, and more likely 0.06~units for realistic survey calibration errors of 0.025~mag.
  
To further explore the sensitivity of H$_0$ to survey errors, Fits~46--54 present the results excluding each of the major surveys contributing to the sample.  While none of these exclusions change the baseline H$_0$ by more than $\sim 0.3$~units, we note that excluding the CSP sample raises H$_0$ by 0.3~units, with most of this (0.2) seen in the change in the Hubble intercept ($a_B$). Using a Hubble-flow sample exclusively from the CSP (while using a mix of surveys for the calibrator sample) yields a lower value of H$_0$ by 0.5~units.  While this is consistent with the sample shot noise, it will reduce H$_0$ for the CCHP results which use only the CSP sample for the Hubble flow.  \citet{Brownsberger:2021} find that this sample asymmetry is expected to produce errors of 0.8~units for realistic calibration errors of 0.025~mag; it is discussed further in \S\ref{sc:7.2}.  

\subsection{SN Fitting Variants\label{sc:6.11}}

Fit~55 changes the way the intrinsic scatter of SN colors is modeled from an empirical approach that includes both a component intrinsic to SNe~Ia and another due to host dust as given by \citet{Brout:2021}, and is further described in the Pantheon+ sample \citep{Brout:2022,Scolnic:2021}. It is a simpler description, where the intrinsic scatter of SNe~Ia is monochromatic (i.e., dispersion only in the luminosity, not the color) from \citet{Guy:2010} and used in the JLA analysis \citep{Betoule:2014} and in the first Pantheon compilation \citep{Scolnic:2018}.  This fit raises H$_0$ by 0.3~units.  The calibrator sample has no preference for either method, yielding similar dispersion between SN and Cepheid distances. 

\subsection{Velocity Variants\label{sc:6.12}}

Here we provide variations related to values of the redshifts as implemented in Equation~(\ref{eq:aB}), which benefit from the combined and improved values from Pantheon+ \citep{Carr:2021}. \citet{Peterson:2021} provide a comprehensive overview and comparative analyses of various predictions of empirical cosmic flows (or peculiar velocities).  They found important improvements to the Hubble-flow residuals by (1) replacing SN host redshifts with their host-galaxy group redshift (when available), and (2) using local density maps to account for motions induced by local gravity. The latter are provided by constrained realizations of the peculiar-velocity field by \citet{Carrick:2015} and  \citet{Lilow:2021} based on 2M++ \citep{Lavaux:2011} and 2MRS \citep{Huchra:2012,Macri:2019}, respectively.  The baseline included both the group-redshift replacement and the 2M++ corrections from \citet{Peterson:2021}.  Fit~56 exchanges the 2M++ values for those from 2MRS, which provide a comparable improvement in residuals and lower H$_0$ by 0.05~units.  A noteworthy change is seen in Fit~57, which forgoes the flow corrections and leaves the redshifts in the CMB frame, reducing H$_0$ by 0.5~units.  However, as shown by \citet{Peterson:2021} for 585 SNe~Ia with $z<0.08$, the tightening of the Hubble diagram (from $\sigma=0.17$~mag to $< 0.15$~mag, or a decrease in $\chi^2$ of 100) gives evidence in favor of these corrections which is too strong to neglect.   Further, the increase in H$_0$ that comes with the decrease in residuals runs counter to the hypothesis that the SN sample lives inside a large-scale void that artificially raises the local value of H$_0$ \citep{Kenworthy:2019}.

While peculiar flows cause a small perturbation in H$_0$ measured from SNe~Ia at $0.0233<z<0.15$, they would produce a greater uncertainty if we forgo the use of SNe and measure H$_0$ directly from only the first two rungs --- that is, from the Cepheid host redshifts (which are not used in the three-rung distance ladder).  The baseline sample host redshifts have a mean of $cz=2000$~km~s$^{-1}$, with many $< 1000$~km~s$^{-1}$. Elsewhere (Kenworthy et al. 2022, in prep.), we present an analysis of H$_0$ from this two-rung ladder which importantly accounts for the spatial covariance of the local peculiar flows, largely limiting the available precision from this route to 3\%--4\% and demonstrating the value of SNe~Ia for the third rung for measuring H$_0$.

Fits~58 and 59 raise $q_0$ from $-0.55$ to $-0.52$ (equivalent to raising $\Omega_M$ in flat $\Lambda$CDM from 0.30 to 0.32) for either the local sample of spiral hosts or the sample with all hosts and for $z<0.8$, with little impact on H$_0$ relative to these samples at $q_0=-0.55$.  In \S\ref{sc:5} we considered a free-form fit for H$(z)$ using $q_0$ as a free parameter simultaneous to the determination of H$_0$.  

\subsection{Optical Wesenheit Variants\label{sc:6.13}}
  
Fits~60--67 use an optical-only Wesenheit magnitude, substituting for Equation~(\ref{eq:wh}), $m^W_I=m_I-R(V\!-\!I)$, and thus discarding the NIR observations.  The \citet{Fitzpatrick:1999} reddening law with $R_V=3.3$ yields $R=1.19$ in the {\it HST} passband system ($m_V$ = {\it F555W}, $m_I$ = {\it F814W}).  The optical Wesenheit has the advantage of lower ``sky'' backgrounds (and their fluctuations) but the disadvantage of higher reddening (and sensitivity to the form of the reddening law).  The baseline fit with the optical Wesenheit yields \hopt, similar to the baseline fit.  However, the optical Wesenheit is somewhat noisier when compared to the SN distances with a relative dispersion of 0.16~mag (vs.~$\sim 0.13$~mag with the NIR data).  We also see larger variations in the anchors and the color variants as seen in Fig.~\ref{fg:vars}, with the scatter among optical-based variants that is three times greater than the NIR-based results and comparable to the statistical uncertainties.  This illustrates the rationale by the SH0ES program for pursuing NIR observations for Cepheids.

Both of these differences are expected consequences of variations in the reddening law in the optical.  For example, for hosts whose Cepheids have a mean $E(V\!-\!I)=0.4$~mag, a difference between a \citet{Fitzpatrick:1999} reddening law with $R_V=2.5$ and $R_V=3.3$ causes a difference in distance of only 0.01~mag for the NIR Wesenheit but 0.09~mag for the optical one, which can explain the aforementioned noise.  In R16 we concluded that future improvements must rely on NIR data until additional studies of variations in reddening laws in the optical were available.  The situation has not improved in that regard. While optical-only Wesenheit data have yielded similar values of H$_0$ \citep{freedman12,Riess:2016}, their larger systematic uncertainties make them unsuitable to pursue the percent-level determination of H$_0$ we approach, and their further analysis is not pursued here.   Nevertheless, none of the optical Wesenheit fits result in a noteworthy change to H$_0$.  

\begin{deluxetable}{llcccccccll}[b]
\tablecaption{H$_0$ Error Budgets (\%), terms approximated from global fit \label{tb:errbudget}}
\tabletypesize{\small}
\tablewidth{0pc}
\tablenum{7}
\tablehead{\colhead{Term} & \colhead{Description} & \multicolumn{3}{c}{{Riess+ (2016)}} & \multicolumn{3}{c}{{Riess+ (2019)}} & \multicolumn{3}{c}{{This work}} \\[-0.2cm]
\colhead{} & \colhead{} & \multicolumn{1}{c}{{\scriptsize LMC}} & \multicolumn{1}{c}{{\scriptsize{MW}}} & \multicolumn{1}{c}{{\scriptsize 4258}} & \multicolumn{1}{c}{{\scriptsize LMC}} & \multicolumn{1}{c}{{\scriptsize{MW}}} & \multicolumn{1}{c}{{\scriptsize 4258}} & \multicolumn{1}{c}{{\scriptsize LMC}} & \multicolumn{1}{c}{{\scriptsize{MW}}} & \multicolumn{1}{c}{{\scriptsize 4258}}}
\startdata
$\sigma_{\mu, \rm anchor}$  &   Anchor distance  & {2.1} & {2.1} & {2.6} & {1.2} & {1.5} & {2.6} & 1.2 & {\ 1.0$^a$} & {\ 1.5$^b$} \\
$\sigma_{{\rm PL, anchor}}$  &  Mean of {\PL} in anchor & {0.1} & \nd &{1.5} & {0.4} & \nd & {1.5} & 0.4 & \nd & {\ 1.0} \\
$R \sigma_{\lambda,1,2}$  & zeropoints, anchor-to-hosts & {1.4} & {1.4} & {0.0} & {0.1} & {0.7} & {0.0} & 0.1 & {\ 0.1$^a$} & {\ 0.0} \\
$\sigma_{Z}$  & Cepheid metallicity, anchor-hosts & {0.8} & {0.2} & {0.2} & {0.9} & {0.2} & {0.2} & 0.5 & {\ 0.15} & {\ 0.15} \\
\tableline
$ $  & subtotal per anchor & {2.6} & {2.5} & {3.0} & {1.5} & {1.7} & {3.0} & 1.4 & {\ 1.0} & {\ 1.8} \\
                     &                             & \multicolumn{3}{p{2.5cm}}{\raisebox{.66\baselineskip}{$\underbrace{\hspace{2.5cm}}$}} & \multicolumn{3}{p{2.5cm}}{\raisebox{.66\baselineskip}{$\underbrace{\hspace{2.5cm}}$}} &
                     \multicolumn{3}{p{2.7cm}}{\raisebox{.66\baselineskip}{$\underbrace{\hspace{2.7cm}}$}} \\
\multicolumn{2}{l}{All Anchor subtotal}  & & {1.6} & & & {1.0} & & & {\ 0.7} \\
\tableline
$\sigma_{{\rm PL}}/\sqrt{n}$  &  Mean of {\PL} in SN~Ia hosts & & {0.4} & & & {0.4} & & & {\ 0.4} \\
$\sigma_{\rm SN}/\sqrt{n}$  &  Mean of SN~Ia calibrators (\# SN) & \multicolumn{3}{l}{\ \ \ \ \,\ \ \ \ \ {1.3 (19)}}&\multicolumn{3}{l}{\ \ \ \ \ \ $\,\,\,$\ \ {1.3 (19)}} &\multicolumn{3}{l}{\ \ \ \ \  \ \ \ $\,\,$ {0.9 (42-46)}}\\
$\sigma_{m-z}$  &  SN~Ia $m$--$z$ relation & & {0.4} & & & {0.4} & & & {\,\,0.4} & \\
$\sigma_{\rm PL}$ & {\PL} slope, $\Delta$log\,$P$, anchor-hosts & & {0.6} & & & {0.3} & & & {\ 0.3} & \\
\tableline
\multicolumn{2}{l}{statistical error, $\sigma_{{\rm H}_0}$}  & & {2.2} & & & {1.8} & & &{\ 1.3} & \\
\tableline
\multicolumn{2}{l}{Analysis systematics$^c$} & & {0.8} & & & {0.6} & & & {\ 0.3} & \\
\tableline
\multicolumn{2}{l}{{\bf Total uncertainty on} $\sigma_{{\rm H}_0}$ [\%]} & & {2.4} & & &{1.9} & & & {\ 1.35} & \\
\tableline
\enddata
\tablecomments{$^a$\citet{Riess:2021}. $^b$\citet{Reid:2019}. $^c$Uncertainties labeled in past analyses as ``systematics'' related to the metallicity scale, Cepheid background/crowding corrections, and SN systematics are formally included here in the covariance matrix in {\bf Fig.~\ref{fg:covarlo}} and thus propagate there as part of the {\it \bf complete uncertainty}. Following past work, we measure the remaining systematic errors as the standard deviation of analysis variants presented in each work as in the dispersion in Fig.~\ref{fg:vars} and as discussed in \S\ref{sc:6.14}.  All terms here are approximations derived from the global fit.}
\end{deluxetable}

\subsection{Summary of Variants and Error Budget\label{sc:6.14}}
  
Our baseline determination of H$_0$ lies 0.2~\kmss (20\% of the uncertainty) below the median of all analysis variants, indicating that it is a good proxy for the set.  In R16 we measured the dispersion of 23 variants and identified that as a systematic error.  In this analysis we have moved previous sources of systematic uncertainty into the covariance matrix to include them formally, and thus most of the variants presented here were intended to gauge sensitivities in the analysis (e.g., excluding a data source) rather than true uncertainties.  Nevertheless, we measure the dispersion of the NIR variants (see Fig.~\ref{fg:vars}) around a $3\sigma$-clipped mean to be 0.3~\kmss and conservatively add this in quadrature as characterizing additional systematic uncertainties to yield a full uncertainty in H$_0$ of 1.04~\kms, or 1.4\%.

\clearpage

\begin{figure}[t]  
\includegraphics[width=\textwidth]{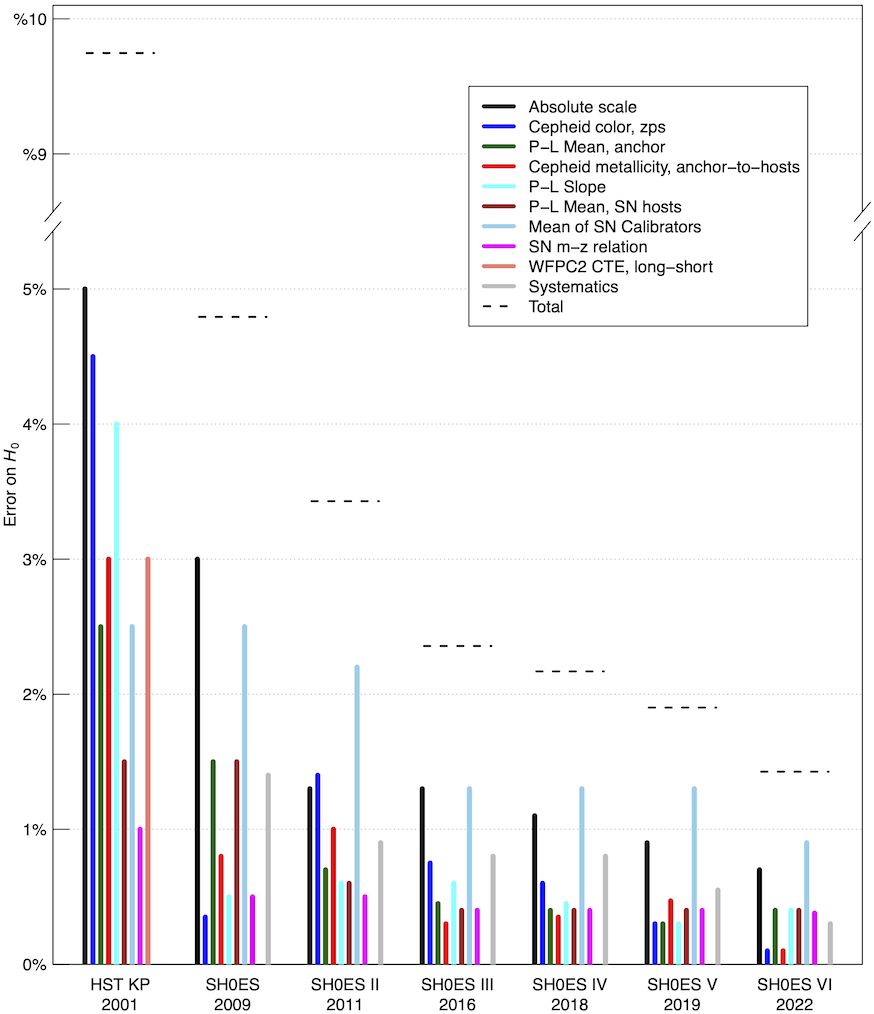}
\caption{\label{fg:errbudget} Comprehensive error budgets for six iterations of the SH0ES measurement of H$_0$ including this one (2022) and the Key Project from 2001 \citep{freedman01}.  The greatest improvement here is realized by the increase in the number of SN calibrators, to decrease all terms to $<1$\%.  The combined error is indicated by horizontal dashed lines.}
\vspace{0.5in}
\end{figure}

\clearpage

None of the variants appear to offer a particularly promising route to solving the Hubble tension, with none shifting the value of H$_0$ much below the full error interval. The lowest value of H$_0$ comes from Fit~65, \hlow, from discarding the NIR Cepheid data and two anchors (leaving only the Milky Way); the highest is from Fit~22, \hhigh, and comes from discarding the use of Cepheid colors to account for extinction.  Both of these fits represent suboptimal accounting of dust.
  
Table~\ref{tb:errbudget} and Fig.~\ref{fg:errbudget} present the full error budget derived (in approximate form) from the global fit and in comparison to prior analyses from the SH0ES team.  The fractional reduction presented here is the largest seen since between \citet{riess11} and \citet{Riess:2016}.

\section{Discussion\label{sc:7}}
     
Here we address a number of considerations using the preceding fits.
    
\subsection{Accuracy of Cepheid Photometry\label{sc:7.1}}

The accuracy of Cepheid photometry is important to the measurement of H$_0$.  The dependence of photometry on {\it calibration} has been negated by the use of the same photometric system throughout the above measurements.  Here we review a number of tests of the relative accuracy for extragalactic Cepheids.

\begin{enumerate}
\item Replication of PSF photometry by others: Measurements of NIR PSF Cepheid photometry using the same raw pixels but different software and methods have been directly compared to those used here by \citet{Javanmardi:2021} in NGC$\,$5584 and by \citet{Yuan:2021_N4051} in NGC$\,$4051, both finding good agreement with reported differences in distance moduli of $0.024\pm0.046$ and $0.00\pm0.04$~mag (respectively)\footnote{Comparing the Cepheids in Table~3 in NGC$\,$5584 in common with \citet{Javanmardi:2021} but revised here since R16, we find they are consistent in the mean with a difference in the Wesenheit of $0.07\pm0.05$~mag; their results are brighter, which if applied to all SN hosts would raise H$_0$ by 2~\kms.}.  While this is short of a full replication of all hosts, it is sufficient to exclude a large methodological error in Cepheid photometry as a primary source of the $\sim 0.2$~mag H$_0$ tension.
    
\item Replication of PSF photometry with apertures: In Appendix~\ref{sc:appb} we use aperture photometry, an independent method which is simple, highly reproducible, and accurate, albeit less precise than the standard approach of using PSFs to model photometry in crowded fields. For this validation we employ aperture photometry for which {\it the background is measured from the {\bf mean} pixel value} (not the mode of the background pixels) in an annulus centered on the Cepheids.  This approach does depend on artificial-star measurements to determine the variable background because the pixels in the annulus include the mean source contribution to the background.  We compare the photometry and find good agreement in their means with a difference (PSF minus aperture) of $0.008\pm0.010$~mag for the Cepheids in SN~Ia hosts and $0.002\pm0.030$~mag for the Cepheids in NGC$\,$4258.
\end{enumerate}

There are a number of strong tests of the accuracy of {\it background} estimates presented here.

\begin{enumerate}
    
\item The Cepheids in NGC$\,$4258 have a similar mean level of crowded backgrounds as in the SN~Ia hosts (see Fig.~\ref{fg:vscrowd}), almost fully negating a systematic underestimate or overestimate of the background on the determination of H$_0$ when NGC$\,$4258 is the sole anchor.  The crowded background is similar because although NGC$\,$4258 is 3--4 times closer than the mean SN host, its Cepheids have been mined from fields which are closer to the dense center by a similar factor.  This results in H$_0$ = \hmaser, similar to the baseline result.
       
\item In Fig.~\ref{fg:crowd} we compare the \PLs relations of Cepheids in the dense, inner (high background) and sparse, outer region (low background) of NGC$\,$4258, finding a negligible difference of 0.01~mag.
  
\item In Appendix~\ref{sc:appb} we regress the background with the distance-ladder fit residuals and find a dependence of $0.010\pm0.014$~mag per magnitude of source background (in the sense of overestimating the background, and consistent with no misestimate).  The background misestimate trends required to explain the tension are strongly excluded as shown in Fig.~\ref{fg:vscrowd}.
  
\item The optical background is nearly an order of magnitude smaller than in the NIR owing to the higher resolution, smaller pixels, and lower flux from red giants.  The baseline results are consistent with those from the optical Wesenheit which do not include the NIR data, H$_0$ = \hsopt.
  
\item There is no significant trend with distance and distance difference between Cepheids and SNe~Ia (zero at $<1.5\sigma$); see Figs.~\ref{fg:ladder} and \ref{fg:comp_ceph}.
    
\item Splitting the calibrator sample by background (Fits~41 and 42) yields no difference in H$_0$.  Splitting in distance (determined by the SN~Ia, Fit~45) yields a difference of 0.1~units.  Both are consistent with no trend based on the shot noise of half the sample (1.0~units).
  
\end{enumerate}

Finally, external to this paper, an additional and unavoidable consequence of the miscalibration of the background, independent of Cepheid mean flux, would be a change in apparent light-curve amplitude.  \citet{Riess:2020} compared the NIR amplitudes of Cepheids in SN hosts and in the MW, found them to be consistent, and provided a quantitative limit of any misestimate of background to be 0.03~mag.  

\subsection{Consistency of TRGB and Cepheid Distance Scales\label{sc:7.2}}
    
In Fig.~\ref{fg:trgb} we presented a comparison of distances measured with Cepheids and TRGB to seven SN~Ia hosts --- the set that allows for a purely differential and direct comparison by employing the same geometric calibration source (NGC$\,$4258), with data in both host and calibrator obtained with the same  telescope ({\it HST}) and setup to negate zeropoint and geometric calibration errors. This comparison further employed the two most widely-used methods for measuring the TRGB, edge-detection (F19) and luminosity-function fit (EDD).  These Cepheids and TRGB measures are consistent with each other, with a mean difference of $-0.002\pm0.03$~mag \citep[CCHP;][]{Freedman:2021} and $0.000\pm0.03$~mag \citep[EDD;][]{Anand:2021}.  With no mean difference between methods as a starting point, we explore the broader question of the sources of difference in the value of H$_0$ from distance ladders measured through the use of either method.

In Table~\ref{tb:trgbdiff} we expand on the results from \citet{Anand:2021}, who compared the value of H$_0$ from EDD, $71.5$~\kms, and CCHP, $69.8$~\kms, derived using the same images to measure the TRGB and the same geometric calibration source, NGC$\,$4258. 
For consistency with the comparison presented by \citet{Anand:2021} of EDD and CCHP which was provided in the units of magnitudes (for differences in $5\,\log\,\Delta$H$_0$) we retain this unit below and in Table~\ref{tb:trgbdiff} in which a difference of 0.03~mag corresponds to a difference of $\sim 1.0$~\kms. 

The largest contribution (0.04~mag of the full 0.05~mag difference) arises from the difference in the calibration of TRGB measured in NGC$\,$4258 applicable (i.e., in relation to) the mean SN host, where CCHP derive $M_{I,\textrm{TRGB}}=-4.05$~mag for a blue TRGB and EDD derive $M_{I,\textrm{TRGB}}=-4.01$~mag, color-corrected to the fiducial, blue TRGB with {\it F606W$-$F814W} = 1.2~mag.  Although additional sources in the LMC and MW have been used to support the TRGB calibration by F21, we use NGC$\,$4258 as the reference here because it is the only source where the TRGB calibration is available directly on the {\it HST} system measured in a manner consistent with the TRGB in SN hosts\footnote{Although we do not recommend mixing photometric systems, to demonstrate the consequence of including the ground-based zeropoints of TRGB as well as the \citet{Jang:2021} measure of the tip of NGC$\,$4258 we can add a strong prior of $M_{I,\textrm{TRGB}}=-4.05\pm0.02$~mag to Fit~31, in which case we find H$_0=72.00 \pm 0.86$~\kms, very close to the mean of the baseline here and the F21 value of H$_0$, simply weighted by the size of each SN sample that passes quality cuts in each set, and a posterior result of $M_{I,\textrm{TRGB}}=-4.040\pm0.015$~mag.}.  The difference in the EDD and CCHP measurement of the tip in NGC$\,$4258 is persistent, having been found for two different fields of NGC$\,$4258 being significant at the $\sim 2\sigma$ level  and is readily evident in Fig.~3 of \citet{Anand:2021}, where the apparent location of the TRGB edge of $m_{\rm F814W}=25.372\pm 0.014$~mag given by \citet{Jang:2021} appears much brighter than the edge highlighted at $25.43\pm0.025$~mag.  A comparison of each group's color-magnitude diagrams (i.e., photometry) might identify the cause of the difference, but these data are only available from \citet{Anand:2021}.   Nevertheless, we offer the option for either TRGB result in Fits~31 and 32.

\begin{figure}[t]  
\begin{center}
\includegraphics[width=0.5\textwidth]{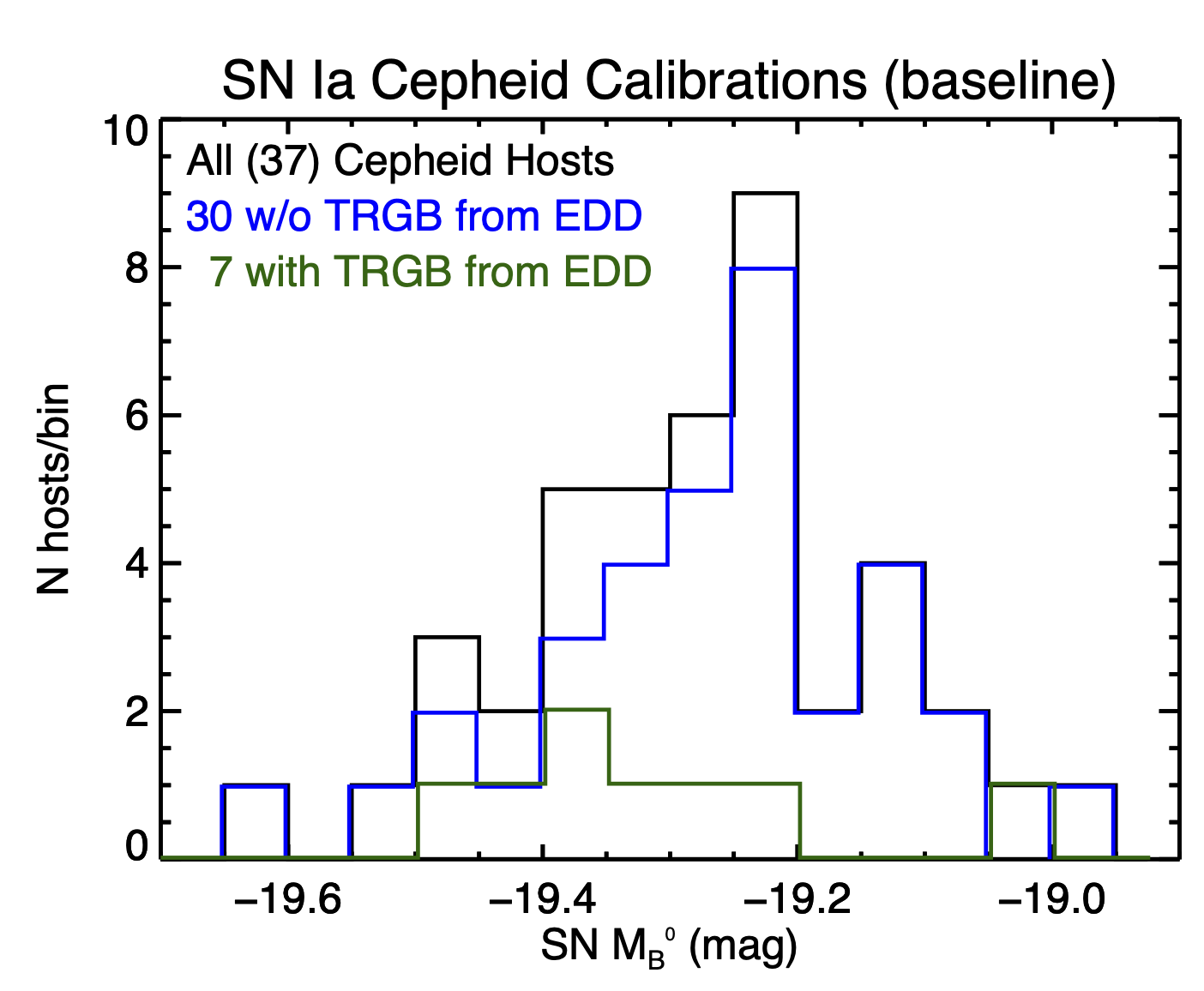}
\end{center}
\caption{\label{fg:compmb} Histogram of the 42 SN~Ia absolute magnitudes of SNe~Ia calibrated from Cepheids in 37 hosts. The 9 SNe~Ia in 7 hosts for which a TRGB distance is available from both EDD and CCHP are shown in green (calibrated here with Cepheids), and the 33 SNe~Ia in 30 hosts without a TRGB measure from both in blue (calibrated here with Cepheids).  The difference in their means of $0.08\pm 0.05$~mag  is consistent with the shot noise of the SN samples as discussed in \S\ref{sc:7.2}, and combined with other differences between the TRGB calibrator set (lines 6--8) produce a net difference in H$_0$ of 1~\kmss ($0.03 \pm 0.05$~mag) as shown in Table~\ref{tb:trgbdiff}, which is less than one $\sigma$ from the shot noise of the two samples.}
\end{figure}

The second half of Table~\ref{tb:trgbdiff} provides the sources of difference between the EDD TRGB analysis (based on 16 SNe~Ia) for which H$_0 = 71.5$~\kmss and our baseline analysis with TRGB (based on 46 SNe~Ia), starting with the same geometric calibrator (NGC$\,$4258).  The primary change is in the makeup of the Cepheid and TRGB study calibrator samples with 34 SNe~Ia added here and 4 subtracted, a net change of 30 objects producing a net increase in H$_0$ of 0.03~mag or 1~\kms.   The first three rows break out changes related to calibrators available from both TRGB analyses which together lower H$_0$ by 0.03~mag.  The first change is due to the addition of SN~2021pit which recently appeared in NGC$\,$1448 and is added here.  SN~2007on has a light-curve shape ($x1=-2.2$) which falls outside the quality range of $|x1| < 2$ imposed in our baseline analysis and is excluded.  There are three SNe~Ia (SNe~1981D, 1989B, and 1998bu) used by CCHP and EDD that are redder ($c>0.15$ and $A_V>0.5$~mag) than our baseline quality range cut and have not been included in this or any past SH0ES analyses.  The next row adds 33 SNe~Ia in 30 Cepheid hosts, tripling the sample  from the preceding line's 13 SNe~Ia in 10 hosts (with EDD and CCHP TRGB measurements and not excluded by SN quality cuts) to 46 SNe~Ia in 40 hosts (for Fit~30, which includes TRGB-only hosts). This step raises H$_0$ by 0.06~mag or 2~\kms.  This difference is fully seen {\it internal} to the set of Cepheid-only SN~Ia calibrations (i.e., independent of absolute anchors or TRGB distances) in the baseline fit as shown in Fig.~\ref{fg:compmb}, which compares the two sets of SNe as measured {\it only by Cepheids}.   The 9 SNe in 7 EDD TRGB hosts are brighter than the 33 SNe~Ia in 30 hosts without EDD and CCHP TRGB by $0.08\pm0.05$~mag as measured by Cepheids, which is consistent at $1.6\sigma$ with the sampling noise of the two sets (0.04~mag and 0.02~mag random error, respectively). As seen in Fits~38--45, this difference does not correlate with potentially relevant changes to the calibrator sample including splitting in distance, background level, or newness of the measurements, so we conclude that this difference of $1.6\sigma$ is a not-unexpected result of the increase in sample statistics.

Combining all of the additions and subtractions to the calibrator sample, the net change between the calibrator samples with both TRGB measures versus the baseline+TRGB here is 0.03~mag and the combined result of all membership changes is less than $1\sigma$ from the shot noise of the two samples.  We note that if we use only the (average) TRGB results of the 13 good-quality SNe, discarding 33 good-quality SNe with only Cepheid measurements, we get $71.0 \pm 2.5$~\kmss as expected (near the mean of CCHP and EDD), but this uses less than a third of the SN sample and has higher shot noise, while still consistent with the baseline.

As the sample size from both methods increases, we would expect them to regress to an increasingly similar mean. Including the two other anchors (MW and LMC) raises H$_0$ by 0.02~mag (and results in $M_{I,\textrm{TRGB}}=-4.00\pm0.024$~mag), though each of the three anchors is consistent with the others at the $0.5\sigma$ level, as shown in Fig.~\ref{fg:anchors}.

Finally, there are two developments based on recent analyses of SNe in the Hubble flow which yield an increase in the value of H$_0$ relative to the CCHP TRGB analysis in F19 and F21, independent of the calibrator sample. The CCHP measurement relies on a {\it calibrator} SN sample from several SN surveys, but selects a {\it Hubble-flow} SN sample from only the CSP SN survey. Regardless of whether the CSP survey is better calibrated than others in an absolute sense, \citet{Brownsberger:2021} show that the use of similar surveys for both SN samples reduces errors in H$_0$ arising from survey miscalibration to 0.1~\kms\ owing to error cancellation, but to only 0.8~\kmss for the CCHP reliance on one survey compared to many.  \citet{Scolnic:2021} find that the CSP sample measures fainter compared to the mean of all other surveys by 0.025~mag, which matches what we find when we remove CSP as shown by Fit~47 in Table~\ref{tb:h0var}, where H$_0$ then increases by 0.4~\kms.  With only 5 of 19 calibrators in F19 observed with the CSP survey, the incomplete cancellation of this survey difference would be 0.6~\kms. In addition, the CCHP analysis of SNe~Ia does not account for cosmic flows expected from local density maps, the best of which (2M++ or 2MRS) as shown by \citet{Peterson:2021} reduce the Hubble diagram $\chi^2$ for $\sim 500$ SNe~Ia by $> 50$~units, decrease the scatter, and raise H$_0$ by $\sim 0.5$~\kms.  These two terms at the bottom of Table~\ref{tb:trgbdiff} should be included in a direct comparison between the results here and those from CCHP.  The combination of both balancing SN survey errors in both samples and accounting for peculiar velocities would raise the CCHP value by 1.1~units and is hard to ignore.   As expected, if we use only CSP for the Hubble flow and also do not account for cosmic flows, the baseline H$_0$ goes down by 1.5~\kms.  

\begin{deluxetable}{lccc}[t]
\tablenum{8}
\tabletypesize{\small}
\tablecaption{Sources of Differences in H$_0$ Between EDD, CCHP, and SH0ES (in mag.)\label{tb:trgbdiff}}
\tablehead{\colhead{Term} & \multicolumn{1}{c}{\hspace{0.5in}$\Delta$F19\ \ \ \ \ \ \ \ \ \ } & \multicolumn{1}{c}{\hspace{0.5in}$\Delta$F21\ \ \ \ \ \ \ \ \ \ } & \multicolumn{1}{c}{\hspace{0.5in}$\Delta$EDD\ \ \ \ \ \ \ \ \ \ }}
\startdata
\multicolumn{4}{c}{EDD vs.~CCHP TRGB} \\
\hline
1. Zeropoint (NGC$\,$4258) & 0.06 & 0.06 & --\\
2. No TRGB detected by EDD (four hosts)\hspace{2in} & 0.00 & 0.00 & --\\
3. NGC$\,$1404 & 0.01 & 0.01 & --\\
4. NGC$\,$5643* & 0.01 & -- & --\\
5. $\Delta m_{\rm TRGB}$* & $-0.03$ & $-0.02$ & -- \\
\hline
\textbf{TRGB subtotal} & \textbf{0.05} & \textbf{0.05} & \textbf{--}\\
\multicolumn{4}{c}{SH0ES Cepheids vs.~EDD TRGB} \\
\hline
6. + SN~2021pit & $-0.02$ & $-0.02$ & $-0.02$ \\
7. $-$ SN~2007on & $-0.01$ & $-0.01$ & $-0.01$ \\
8. $-$ 3 red SNe~Ia & 0.00 & 0.00 & 0.00 \\ 
9. +33 SNe~Ia in 30 hosts & 0.06 & 0.06 & 0.06 \\
\hline
\textbf{Calibrator set change subtotal\ \ \ \ \ \ \ \ \ \ \ \ \ \ } & \textbf{0.03} & \textbf{0.03} & \textbf{0.03}\\
\hline
10. LMC+MW anchors & 0.02 & 0.02 & 0.02 \\
11. Flows & 0--0.02 &  0--0.02  & 0--0.02 \\
12. Hubble-flow surveys & 0--0.02 &  0--0.02  & 0--0.02 \\
\hline
\textbf{Total} & \textbf{$\sim 0.10$} & \textbf{$\sim 0.10$} & \textbf{$\sim 0.05$}\\
\enddata
\tablecomments{``*'' weighted by SN, not by host. $\Delta$F19, $\Delta$F21 = differences between \citet{Freedman:2019} or \citet{Freedman:2021} and EDD, respectively. $\Delta$EDD = differences between EDD and SH0ES. Descriptions of individual entries: (1) measured zeropoint calibration of the TRGB in NGC$\,$4258; (2) \citet{Anand:2021} did not detect the TRGB in four SN host galaxies; (3) inclusion of NGC$\,$1404 (not directly measured by F19; EDD include SN~2007on which F19 included and F21 excluded); (4) inclusion of NGC$\,$5643 (not available in F19); (5) mean difference in measured values for the TRGB of the remaining hosts; (6) SN~2021pit not available in EDD; (7) exclude SN~2007on in NGC$\,$1404 owing to $x1<-2$; (8) exclude 3 SNe with $c>0.15$; (9) addition of 33 SNe~Ia; (10) addition of LMC and MW anchors; (11) corrections for cosmic flows in Pantheon+ raises H$_0$ by 0.02~mag relative to CSP SNe \citep[see][]{Peterson:2021}; (12) EDD used both Pantheon or CSP, the latter lowers H$_0$ by 0.02~mag relative to Pantheon+ average of surveys. The uncertainties in all terms above are 0.01~mag.}
\label{tab:magDifferences}
\end{deluxetable}
      
To summarize, we find that TRGB and Cepheids give consistent distances when using the same anchors and consistent procedures, so that comparisons are meaningful.  There are two main sources of difference in H$_0$ between the CCHP implementation in F21 and the baseline here. 
(1) As given by \citet{Anand:2021}, a (net) difference of 0.04~mag between the EDD and CCHP implementation of TRGB in F21 arises from the difference in the apparent location of the tip in NGC$\,$4258 and is the source of the difference between H$_0 = 69.8$ and 71.5~\kmss (i.e., TRGB-only results) in these two studies. (2) The other, corresponding to a net increase of 1~\kmss (0.03~mag) between the EDD TRGB and the baseline, is due to differences in the SN~Ia calibrator samples including a reduction of 1~\kmss owing to our exclusion of 4 SNe~Ia which fail quality cuts and an increase of 2~\kmss owing to the fainter mean seen for 30 SNe in 33 Cepheid-only hosts compared to 9 SNe in 7 TRGB+Cepheid hosts, a difference seen internal to Cepheid measurements and consistent with a statistical fluctuation due to the combined changes at the $1\sigma$ level (see Fig.~\ref{fg:compmb}). We expect that additional TRGB data will result in these distributions agreeing, as we see no reason for a difference besides shot noise. Relative to F21, the sample difference change is 1.3~\kmss (0.04~mag) with other differences from the baseline owing to the absence of an accounting for peculiar flows as described by \citet{Peterson:2021} and not including SNe from multiple surveys in the Hubble flow to cancel zeropoint differences among the calibrator surveys as described by \citet{Brownsberger:2021}. Combined, these raise H$_0$ by $\sim 1.1$~\kms, and we see no reason not to include these in an SN~Ia-derived measurement of H$_0$ given their strong empirical support.  The combined Fit~30 yields \hwtrgb\ for EDD and \hwtrgbcchp\ for CCHP, and we cite the mean of the two (\hwtrgbmean) as representative of the combination of Cepheids and TRGB.

\begin{deluxetable}{lrrrr} 
\tablenum{9}
\tablewidth{0pc}
\tabletypesize{\small}
\tablecaption{Host Properties of SN~Ia Samples\label{tb:hostprop}}
\tablehead{\colhead{Sample} & \colhead{$N$} & \colhead{Log\,($M/{\rm M}_\odot$}) & \colhead{Log\,SFR} & \colhead{Log\,sSFR}}
\startdata
Baseline Calibrator & 37 & 10.3 SD=0.6 & 0.3 SD=0.5 & -9.8 SD=0.7 \\
LZSPI HF & 276 & 9.8 SD=0.8 & 0.3 SD=0.7 & $-9.3$ SD=0.4 \\
LZBRD HF & 482 & 10.1 SD=0.8 & 0.1 SD=0.9 & $-9.6$ SD=0.7 \\
HZBRD HF & 1354 & N/A & N/A & N/A \\
LZSPI HF $\log\,(M/{\rm M}_\odot) > 10$ & 132  & 10.4 SD=0.5 & 0.4 SD=0.7 & $-9.7$ SD=0.6 \\
\hline
\enddata
\end{deluxetable}

\subsection{Consistency of SNe~Ia on Second and Third Rungs\label{sc:7.3}}
     
In Appendix~\ref{sc:appa} we show that the calibrator SNe~Ia are spectroscopically all prototypical and photometrically have a distribution of light-curve shapes and colors that are well-matched to the selected Hubble-flow sample.  The use of tighter-than-typical quality cuts ($\abs{c}<0.15$, $\abs{x1}<2$) ensures that the sample comparison is insensitive to the standardization method, as it makes little difference in the sample means.
     
In R11 the Hubble-flow sample consisted of SNe~Ia without limitations placed on the properties of their hosts (with about two thirds coming from spirals). \citet{Rigault:2015} suggested that the calibrator sample, all with spiral hosts and thus greater mean SFR, could introduce a bias in H$_0$ if Hubble-flow residuals (even after accounting for an empirical host-mass dependence as done by R11 and R16) presented a residual correlation with host SFR (either the global rate of the host or local to the SN).  R16 addressed this sample difference by including as an analysis variant a Hubble-flow sample composed of only spiral or globally star-forming hosts, a precaution (regardless of whether such a correlation exists) conservatively adopted here as the {\it baseline}.  Because the size of the Hubble-flow sample is so much larger than the calibrator sample, it is sensible to cull the former to match the selection of the latter to control even yet-undiscovered systematics with little cost to the precision of H$_0$.  It is also important to recognize that the size of a correlation of Hubble residuals with a host property {\it depends on the method of SN standardization and the SN sample host selection}, and that it is quite possible for a specific combination of method and sample to show a significant correlation that does not exist for different methods and samples.  

\citet{Jones:2018} measured the correlation of the local and global SFR and specific SFR (sSFR) with the Hubble residuals from Pantheon SNe \citep{Scolnic:2018} used by R16 and found little or no significant correlation with implied corrections (if significant) at the 0.3~\kmss level. Here we use two SN standardization methodologies from Pantheon+, one of which has no correlation with host mass \citep{Brout:2021} in the baseline, and Fit~55 which has a 0.045~mag step at $\log (M/{\rm M}_\odot) = 10$, increasing H$_0$ by 0.3~\kms.  

In Table~\ref{tb:hostprop} we compare the global properties of galaxies in the calibrator and Hubble-flow samples including mean mass, SFR, and SSFR. The baseline Hubble-flow and calibrator samples have the same mean SFR (0.3), consistent with their matched selection (the mean SFR of the early-type hosts in Pantheon+ is $-0.6$).  The mean mass is also similar, with the calibrators higher by 0.5~dex --- a difference smaller than the dispersion of either sample.  In Appendix~\ref{sc:appa} we compare the distributions of host masses for various samples.  To produce a late-type Hubble-flow sample with mass exceeding the calibrator sample, Fit~37 limits the LZSPI to $\log (\rm M/{\rm M}_\odot) > 10$ and lowers H$_0$ by 0.3~\kms.  We conclude that the calibrator and baseline Hubble-flow samples are well matched in mass and SFR.

We would not expect any other host properties, especially any local to the SN which was not a selection criterion, to significantly differ between samples.  For example, \citet{Anderson:2015} measured the relative strength of H$\alpha$ at the sites of 98 SNe~Ia in exclusively late-type, star-forming hosts, including by chance 20 of the 38 selected for Cepheid measurements, and the fraction with detected local H$\alpha$ is similar for calibrators and Hubble-flow hosts (30\% vs.~45\%). Indeed, it would be very hard to understand how a difference in local SN host properties could occur between the two sets of hosts with matched selection. While additional host or SN properties beyond the ones we have used to measure H$_0$ may be used to improve SN distance estimates now or in the future, the matching of SN and host samples employed here and the large sample size for each would mitigate any significant effect on the determination of H$_0$.  

\subsection{State of the Hubble Tension\label{sc:7.4}}

Our baseline determination of H$_0$ is \hsbase\ (with systematics), which exceeds the {\it Planck}$+\Lambda$CDM result by $5\sigma$. It may be of interest and it is straightforward to calculate a {\it combined} value of H$_0$ which is free of measurement interdependencies and has lower uncertainty using additional redshift-magnitude relation data.  We start with the combined Cepheid and TRGB result, \hswtrgbmean\ (or Fits~31 and 32), and to this we add two recent measures which provide enough information that allow us to have a consistent calibration and redshift frame while avoiding double use of data.  \citet{Pesce:2020} measured 6 masers in the Hubble flow; excluding the nearest from that set, NGC$\,$4258 (because its maser distance is used here), and quoting their result in the 2M++ frame (same as our baseline) gives H$_0 = 72.1 \pm 2.7$~\kms.  \citet{Blakeslee:2021} measured the IR surface brightness fluctuation distances with {\it HST} in 63 galaxies calibrated by the TRGB in the 2M++ frame.  Accounting for the small difference in TRGB zeropoint ($M_I=-4.014 \pm 0.025$~mag for the mean of CCHP and EDD) and used there ($M_I=-4.03$~mag) yields H$_0= 74.0 \pm 3.0$~\kms.  The combination of these independent and independently consistent measures gives H$_0= 72.61 \pm 0.89$~\kmss (or H$_0= 72.42 \pm 0.89$~\kmss with the CCHP TRGB in Fit~32 and H$_0= 72.80 \pm 0.89$~\kmss from the EDD TRGB with Fit~31), a local determination with 1.2\% precision which is also $5.2\sigma$ greater than the Planck+$\Lambda$CDM result.  Other combinations may be determined but require care to avoid measurement inconsistencies in calibration, redshift frame, or double use of any data.  
     
There has been a wide variety of ideas proposed to resolve the Hubble tension, including (but not limited to) an episode of scalar-field dark energy before recombination, the presence of additional species of neutrinos (perhaps with interactions), decaying dark matter, the presence of primordial magnetic fields, a changing electron mass, decaying or interacting dark matter, a breakdown of the Friedmann-Lema\^{i}tre-Robertson-Walker metric or general relativity, and so on; we direct the reader to recent reviews, such as \citet{Divalentino:2021} and \citet{Olympics:2021}.  These proposals range from moderately successful to unsuccessful with no clear resolution.  Some of the more successful ideas mitigate the tension through a similar mechanism, such as increasing H$(z)$ in the early Universe so that recombination occurs earlier, thereby shrinking the sound horizon which is the fundamental scale of the CMB (and also of baryon acoustic oscillations).  In many scenarios this will produce additional features in the CMB which are either incompatible with the data or make them appear more plausible \citep{Hill:2021,Poulin:2021}. The presence of unaccounted systematics in early- or late-Universe measurements have also been suggested, but in \S6 we comprehensively reviewed those pertaining to the route presented here with none showing indications of validity.

Both the late- and early-Universe data present formidable obstacles to hypotheses involving ``new physics'' or new systematics owing to the rigor and redundancy of the measurements; any proposals require {\it specificity} to see which may be viable. However, opportunities for progress on this problem exist on many fronts. We anticipate gains from improved characterization of H$(z)$, the use of new facilities to refine the local measurements (e.g., LIGO and {\it JWST}) and the early-Universe measurements (e.g., CMB Stage 4 and the Simons Observatory), neutrino experiments, as well as from new theoretical insights.

\ \par

\section{Conclusions\label{sc:8}}    
 
\begin{enumerate}

\item Our baseline determination of H$_0$ is \hsbase\ (with systematics) from a Cepheid-only calibration of 42 SNe~Ia with good SN data quality, or \hswtrgbmean\ combining Cepheid and TRGB for a total of 46 SN~Ia calibrators with good SN data quality.
   
\item The measurement exceeds the {\it Planck}$+\Lambda$CDM result by $5\sigma$ (one in 3.5 million), making it implausible to reconcile the two by chance.
   
\item An exhaustive study of variations in the analysis and systematic uncertainties including 67 variants of analyses reveals no indication of significant inconsistencies within the measurement or promising sources of unrecognized error.  The dispersion of the 59 NIR variants is 0.3 \kms and is conservatively adopted as an additional systematic uncertainty.  The mean of the variants is 73.25 \kms which is higher than the baseline by 0.2 \kms.

\item We find the dispersion between 42 SN~Ia and Cepheid relative distance measures is $\sigma=0.130$~mag, similar (albeit lower) than the $\sigma=0.135$~mag dispersion of SNe~Ia in the Hubble-flow sample and yielding no evidence of excess noise in Cepheid distance measurements.  
   
\item We find that Cepheid and TRGB distance measures are consistent when starting and ending from the same hosts (i.e., between rung one and two). We highlight a net difference of 1.3~\kmss (or $0.04 \pm 0.02$~mag) between measurements by two groups of the location of the tip in NGC$\,$4258 and resulting calibration of the TRGB (we use the mean of both), and a net 1~\kmss (0.03~mag) higher value of H$_0$ from the change (tripling) of the SN calibrator sample, which is consistent with the $\pm 0.05$~mag shot noise of the subsamples.
   
\item We find that each of the three independent geometric anchors is consistent with the distance predicted by its Cepheids and the other two anchors.  Improvements in the calibration of the Cepheid metallicity dependence have tightened this conclusion.
   
\item The SNe between the second and third rung of the ladder are hosted by galaxies of the same late type (i.e., spiral) with the same or similar mean SFR and mass.  Their color and shape distributions are also highly consistent.  The calibrator set contains a complete sample of all suitable SNe~Ia (i.e., with good data quality) at $z<0.011$ in the last four decades.  We see no indication of differences between the mean properties of the samples' hosts, nor a reason with matched selection that such would exist and impact H$_0$.  
   
\item Extragalactic Cepheids appear to have a uniform relation between period and luminosity consistent with a single slope, and fine structure in their light curves that resembles those in the MW at the same period (i.e., the Hertzsprung Progression).
   
\item The constraint provided by the distance ladder presented here is well approximated by the derived value of H$_0$, except in the case of models which introduce rapid, unexpected, late-time changes in H$(z)$ (relative to either $\Lambda$CDM or low-order fits to H$(z)$), or perhaps some forms of new physics. For such models, we advise replacing H$_0$ with the absolute SN~Ia host distances derived from the first two rungs, the set of SN~Ia standardized magnitudes in these hosts, SN~Ia magnitudes in the Hubble flow, and their covariance, and provide an example which yields a joint constraint of H$_0$ = \hhiz\ and $q_0$ = \qhiz.
   
\item The source of this long-standing, significant discrepancy between the local and cosmological routes to determining the Hubble constant remains unknown.
     
\end{enumerate}

\ \par

\section{Acknowledgments}

We thank Graeme Addison, Fabio Bresolin, George Efstathiou, and Doron Kushnir for helpful conversations related to this work. We are grateful to Peter Challis for sharing an unpublished spectrum of SN~2008fv. The ESA {\it Hubble} office provided most of the color composites shown in Fig.~\ref{fg:hstcol}.
An anonymous referee provided a thoughtful report that led to improvements in this paper.

This research was supported by NASA/{\it HST} grants GO-12879, GO-12880, GO-13334, GO-13335, GO-13344, GO-15145, GO-15146, and GO-15640 from the Space Telescope Science Institute (STScI), which is operated by the Association of Universities for Research in Astronomy, Inc., under NASA contract NAS5-26555. Some of the data presented in this paper were obtained from the Mikulski Archive for Space Telescopes (MAST) at the STScI. The specific observations analyzed can be accessed via this\dataset[DOI]{https://urldefense.com/v3/__http://dx.doi.org/10.17909/fkay-8z97__;!!CrWY41Z8OgsX0i-WU-0LuAcUu2o!lfR95MkkMfsN1T9bn700vLionER7hWKCXsb_XXF3VcPLs8-y_jSkSO7Sy1hTJA$ }.

L.M.M. acknowledges additional support from the Mitchell Institute for Fundamental Physics \& Astronomy at Texas A\&M University. D.O.J. acknowledges support from NASA Hubble Fellowship grant HF2-51462.001 awarded by the Space Telescope Science Institute. A.V.F.'s group at UC Berkeley is also grateful for financial assistance from NSF grant AST-1211916, the TABASGO Foundation, the Christopher R. Redlich Fund, the Miller Institute for Basic Research in Science (in which A.V.F. is a Miller Senior Fellow), and numerous individual donors.

Some of the data presented herein were obtained at the W. M. Keck Observatory, which is operated as a scientific partnership among the California Institute of Technology, the University of California, and NASA; the observatory was made possible by the generous financial support of the W. M. Keck Foundation. We thank Brad Tucker for designing the multislit masks used for the Keck LRIS spectroscopy of H~II regions in the host galaxies.

\clearpage

\begin{center}
APPENDICES    
\end{center}

\begin{appendices}
\section{Properties of Calibrator SNe~Ia\label{sc:appa}}

The members of the SN~Ia calibrator sample were first identified when selecting hosts for observing Cepheids with {\it HST}. We have attempted to follow the criteria for selection given by \cite{riess05} to provide reliable SN~Ia distances: objects observed before maximum light, through low interstellar extinction ($A_V\,<\,0.5$~mag), with modern (i.e., non-photographic) photometry, and with typical light-curve shapes.  Here we analyze the spectra of the calibrators to confirm their initial selection.
      
\subsection{Spectral Properties Using \texttt{DeepSIP}}

We consider two samples of calibrators. The first contains 40 SNe~Ia and is a complete, volume-limited ($z<0.011$) sample of all suitable SNe~Ia seen over the years 1980--2021 in spiral hosts (a requirement for finding Cepheids, with results for H$_0$ provided for these alone in Fit~38). A second sample contains just two SNe~Ia (SN 1999dq and SN~2007A) from a program targeting hosts that are located at greater redshifts and are more luminous, in an effort to reach the Hubble flow directly with Cepheids. By the following quantitative spectroscopic analysis we find that all 42 SNe~Ia are in the normal range.

We obtain spectra directly from the Open Supernova Catalog\footnote{\url{sne.space}} \citep[OSC;][]{openSNe}, which aggregates data from numerous sources \citep[including notable low-redshift SN~Ia spectroscopy releases; e.g.,][]{bsnipI,Blondin2012,Folatelli2013,S20}. From the OSC-retrieved spectra, we select --- per SN~Ia --- the nearest-to-maximum-light spectrum having (i) full coverage\footnote{As in \citet{S20}, we define ``full coverage'' as having a wavelength minimum below 5750~\AA\ and a maximum above 6600~\AA.} of the characteristic Si~II $\lambda 6355$ absorption feature, and (ii) an SNR of at least 10 per pixel\footnote{Ties when two spectra have exactly the same phase are broken by taking the one with broader wavelength coverage.}. If these criteria cannot be met by the available spectra, we reduce the SNR threshold to 5~pixel$^{-1}$; if still no spectra satisfy the criteria, we remove the Si~II $\lambda 6355$ coverage requirement. If no spectra are available even after this relaxation, our automated algorithm flags the SN in question for manual intervention. Following this approach, 37 spectra are obtained with our full criteria satisfied, one is obtained with our most relaxed criteria, and two fail.

We resolve these two failures as follows.
\begin{enumerate}
\item SN~2008fv was spectroscopically classified as a normal SN~Ia in CBET 1522 \citep{Challis:2008}. Though never published, we have obtained this spectrum from the author.
\item SN~2021hpr has three spectra available on the Transient Name Server\footnote{\url{https://urldefense.com/v3/__http://www.wis-tns.org__;!!CrWY41Z8OgsX0i-WU-0LuAcUu2o!lfR95MkkMfsN1T9bn700vLionER7hWKCXsb_XXF3VcPLs8-y_jSkSO707qWM2A$ }}. We take the spectrum that best satisfies our full criteria as stated above.
\end{enumerate}

We also override successful OSC acquisitions in several cases where more suitable spectra (i.e., those that better match our criteria) are available elsewhere. The spectra in our final set have a median phase of $-0.1$~d, with the earliest at $-11.7$~d and the latest at 14.5~d. A full accounting of relevant metadata is available upon reasonable request.

\begin{figure}[t]
\figurenum{A1}
\begin{center}
\includegraphics[width=0.95\textwidth]{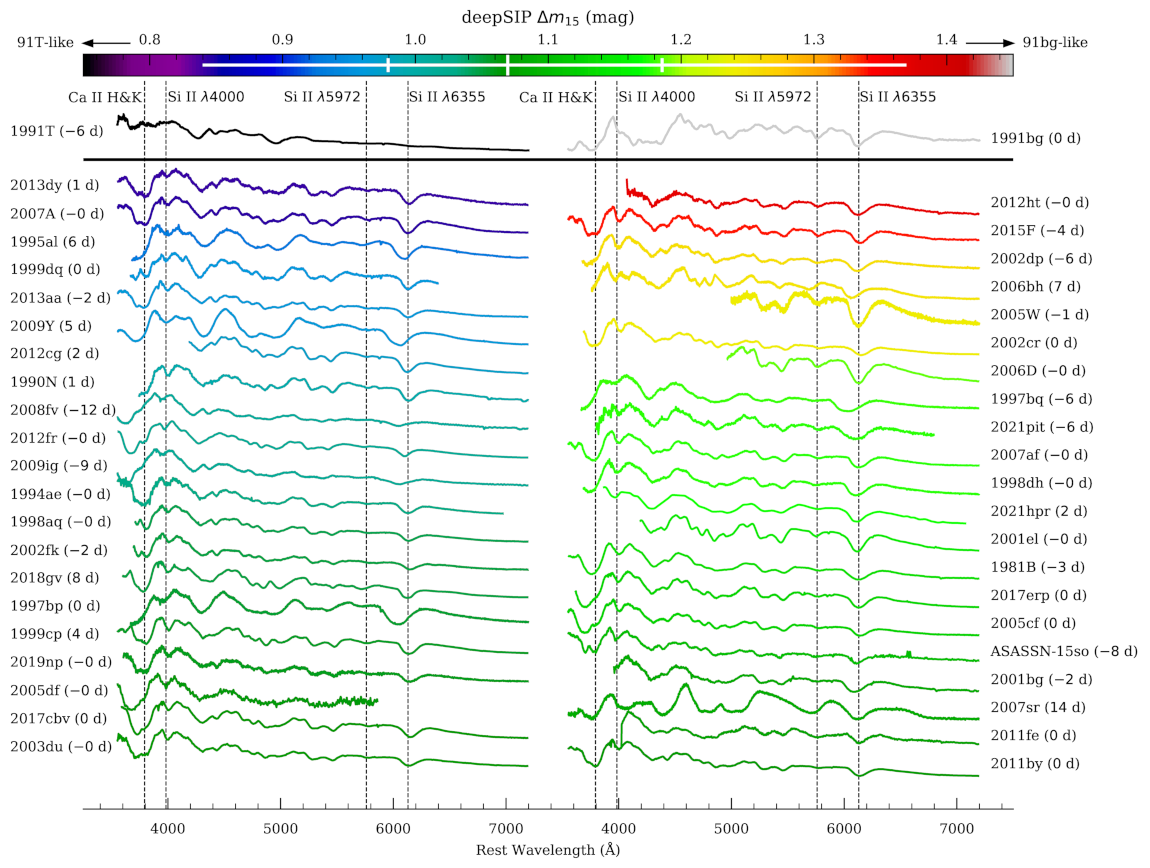}
\end{center}
\caption{\label{fg:spectra} Spectra (scaled $f_\lambda$) of all 42 SNe~Ia in the calibrator sample, color coded by their deepSIP score calibrated to the photometric parameter, $\Delta m_{15}$.  As discussed in the text, we find all calibrator SNe to be in the normal range. Peculiar objects SN~1991T and SN~1991bg bracket the extremes and are shown for comparison.  }
\end{figure}

A key concern in studies that utilize SN~Ia distances (such as this one) is ensuring that the objects used are indeed ``normal'' SNe~Ia in the sense that they can be standardized using width-luminosity relations \citep{Phillips:1993}. Here we take the conservative approach of excluding subluminous SN~1991bg-like \citep{91bg-Filippenko,91bg-Leibundgut} and overluminous SN~1991T-like objects \citep{91T-Filippenko,91T-Phillips}, thereby defining ``normal'' SNe~Ia as those that fall within the central, most well-studied and precise region of the Phillips relation. Moreover, we confirm the normalcy of our selected objects spectroscopically by employing the \texttt{deepSIP} package \citep{deepSIP}, which provides highly effective, trained convolution neural networks that, amongst other things, can (i) classify if a spectrum belongs to an SN~Ia with a rest-frame phase between $-10$~d and 18~d and light-curve shape \citep[parameterized by \texttt{SNooPy}'s $\Delta m_{15}$ parameter; see][for more details]{SNooPy} between 0.85 and 1.55~mag, a conservatively narrow window corresponding to ``normal'' objects, and (ii) predict (with uncertainties) quantitative $\Delta m_{15}$ values. Because we know the rest-frame phase of each spectrum in our sample from the light-curve-derived times of maximum brightness, distinctions made by the aforementioned classifier provide a direct probe if the SNe~Ia in our sample are spectroscopically normal.  The calibrator SNe~Ia have DeepSIP $\Delta m_{15}$ values between 0.84 and 1.37~mag.

We find that only two spectra do not satisfy this test, one of which (SN~2008fv) is expected owing to it being the sole case in our sample where the spectrum is at a phase earlier than $-10$~d. The other case (SN~1997bq) is most certainly spectroscopically normal \citep[see][]{Blondin2012}. The fact that this single false negative is obtained is not unexpected because in developing \texttt{deepSIP}, \citet{deepSIP} tuned the decision threshold of the model with an eye to their subsequent scientific use case in which false positives (i.e., classifying an SN~Ia as normal when it actually is not) represent a far worse error than false negatives (i.e., failing to classify an SN~Ia as normal when it actually is). We visualize the entire spectral sample along with representative near-maximum-light spectra of SN~1991bg and SN~1991T for reference, color coded by \texttt{deepSIP}-predicted $\Delta m_{15}$ value, in Fig.~\ref{fg:spectra}.

\subsection{Photometric Properties}
     
\begin{figure}[h]   
\figurenum{A2}
\begin{center}
\includegraphics[height=0.3\textheight]{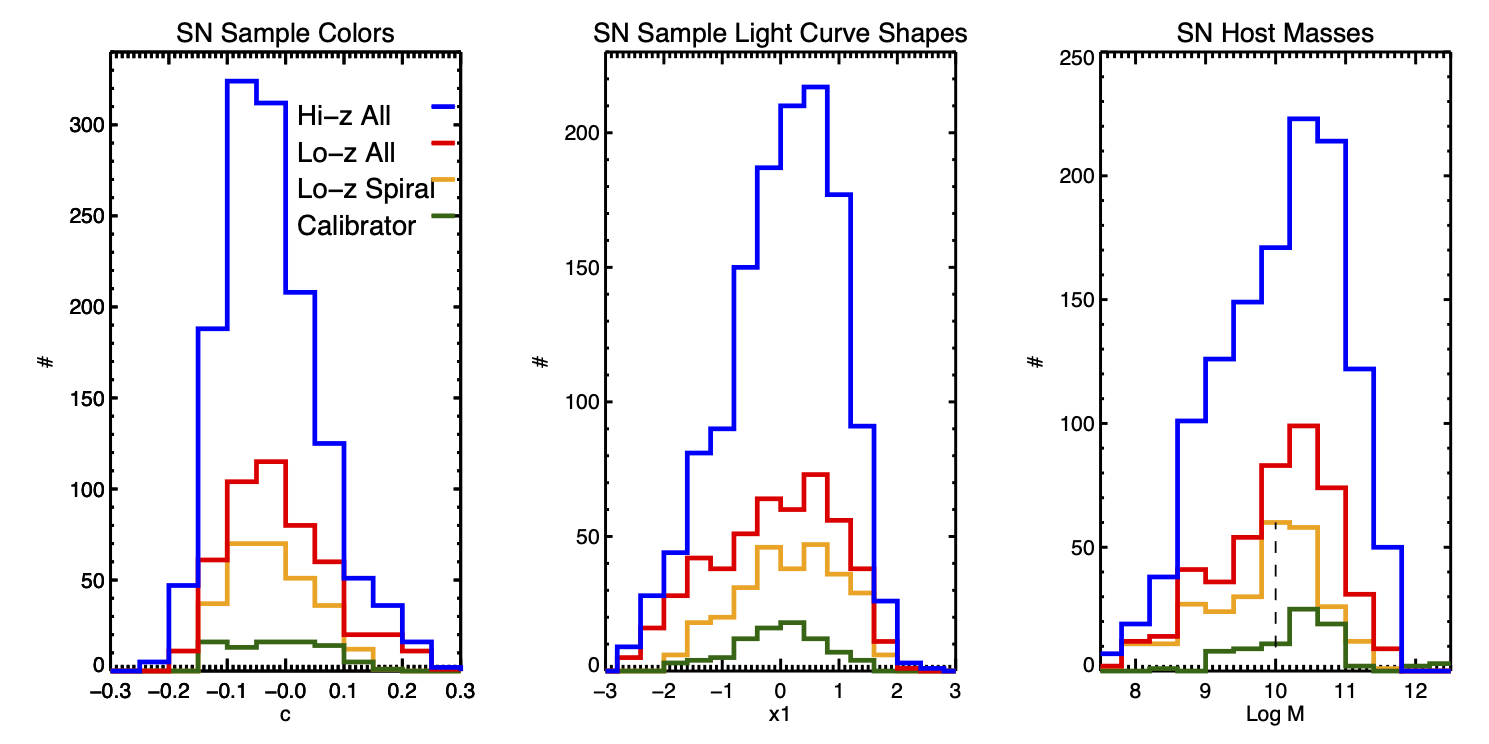}
\end{center}
\caption{\label{fg:snsamp} SN~Ia and host properties for the calibrator sample (green) and three Hubble-flow samples.  The color (left panel) and light-curve shape (middle) distributions are in good agreement as discussed in the text, particularly for the baseline sample (Lo-$z$ Spiral, orange) whose quality cuts limits are $\abs{c} < 0.15$ and $\abs{x1} < 2$.  The right panel shows the host mass including a truncation of the Lo-$z$ sample (dashed line) with values for this and other host properties in Table~\ref{tb:hostprop}.}
\end{figure}

In Fig.~\ref{fg:snsamp} we show the distributions of the SALT~II color ($c$) and shape ($x1$) parameters, as well as the host masses, for the calibrator sample and for three Hubble-flow samples: (1) the baseline $0.0233 < z < 0.15$ spiral sample and the tighter quality cuts $\abs{c}<0.15$, $\abs{x1}<2$; (2) the same redshift range for all host types and the Pantheon+ standard quality cuts $\abs{c}<0.3$, $\abs{x1}<3$; and (3) a sample of all types to $z<0.8$ and standard quality cuts.  It can be seen that the samples are well matched in the mean, with the baseline sample better matched in breadth to the calibrator sample owing to the tighter quality cuts. As a result of this investigation, we conclude that each SN in the calibrator sample is unambiguously normal and thus can be reliably standardized to well match the baseline Hubble-flow sample.  

\section{Independent Tests of Cepheid Photometry\label{sc:appb}}
\subsection{Aperture Photometry}

The photometry of Cepheids presented here is derived from a set of standard procedures referred to in the astronomical literature as ``scene modeling'' or ``crowded-field photometry.''  These methods use an empirical description of the PSF to model unresolved sources (i.e., stellar profiles) by comparing image pixels to a model constructed from the superposition of PSFs, each with its own $X$ and $Y$ coordinate and amplitude as well as the constant level of the background.  The initial set of sources and their positions may be derived from the image or a catalog.  

This approach offers greater precision than fixed-aperture photometry (i.e., summing flux in an aperture), as it can separate the blended flux of distinct sources and additionally improves the SNR by the optimal weighting of source pixels.  Bias resulting from the inability to resolve nearly-coincident background sources from the Cepheid can be determined {\it statistically} by adding artificial stars of known flux to the scene.  This bias is really just the mean level of the fluctuating background. Uncertainties are measured from the distributions of recovered artificial stars.   Frequently used software packages that enable this approach include DAOphot, DoPHOT, DolPhot, HSTphot, and Romaphot.  In general, the use of these different packages has been shown to yield similar results subject to the settings for which these packages are employed.
   
However, it is valuable to have a robust cross-check of {\it the accuracy} of photometry measured with these techniques using a simpler approach that is easily replicated by others without reference to any specific piece of software.  Such a method is aperture photometry which, though less precise, provides a strong test of the accuracy the Cepheid photometry reported above.
   
Since aperture photometry cannot readily separate superimposed sources in dense regions, a few considerations described here are necessary to produce accurate aperture photometry. (1) Small apertures are required, here set to 1.4 drizzled pixels (0\farcs11) in radius, along with an aperture correction derived from our PSF model to estimate the flux outside the aperture. (2) We limit the comparison to 1/3 of the Cepheid sample with the lowest surface brightness, a sample large enough to measure H$_0$ but which still has less background and thus greater precision. (3) We use a {\it mean} of the pixels in an annulus around each Cepheid to determine the sky, rather than the mode or median which are more commonly used in sparse regions.  This last step is {\it very} important for deriving accurate aperture photometry in the presence of a fluctuating background.  In dense regions, the mean of the sky is an unbiased estimator of the level beneath a source, as it includes both the spatially constant background as well as the mean level of superimposed sources. In contrast, the mode or median of the crowded sky or any statistic calculated after fitting and subtracting visible stars necessarily underestimates the sky under the target source where we cannot resolve or subtract blended sources. We assume the position of the Cepheid was previously identified from sparse, high-contrast optical images, as is the case for the SH0ES program.
   
\begin{figure}[t]   
\figurenum{B1}
\begin{center}
\includegraphics[width=0.5\textwidth]{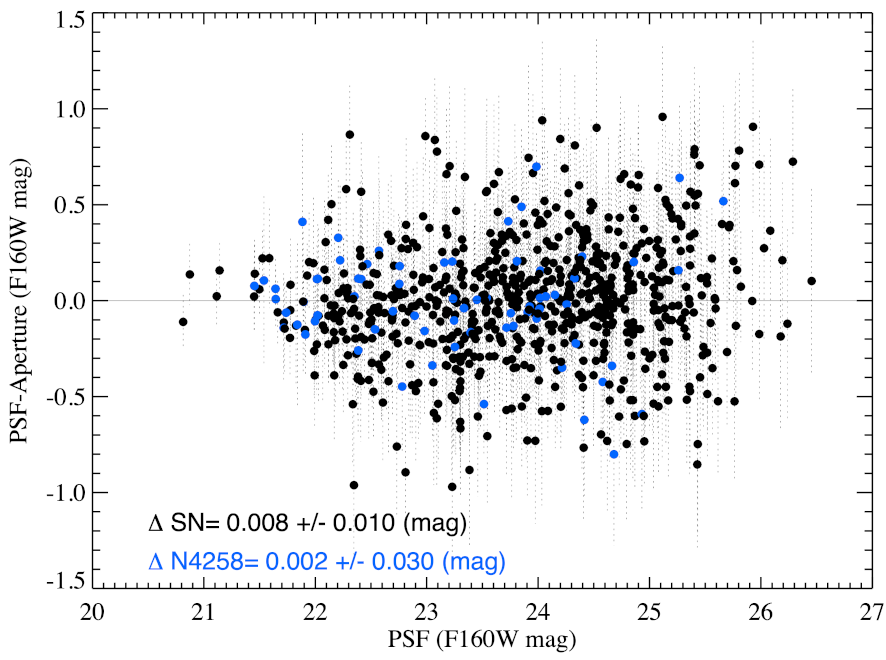}
\end{center}
\caption{\label{fg:apvpsf} Comparison of PSF photometry (with backgrounds from artificial stars) for Cepheids in SN hosts (black) and NGC$\,$4258 (blue) used here and basic small-aperture photometry with backgrounds from the mean pixel in a local annulus as discussed in the text.}
\end{figure}

We measured aperture photometry of the Cepheids using the same images (before removing any background level) used for the PSF-based photometry and the same, fixed Cepheid positions.  The sky value was set to the simple mean of the sky pixels in an annulus between radii of 15 and 25 pixels from each Cepheid\footnote{If anomalously bright Cepheids are removed as outliers from the final sample it is necessary to apply the same threshold to the sky pixels before calculating their mean to avoid a measurement bias.  In practice, it is common to exclude Cepheids brighter than the \PLs by $\leq 3\sigma$ or $>1$~mag to provide a more robust \PLs relation. Thus, we remove sky pixels from the determination of the mean which would cause an outlier of this size, i.e., those $\geq 2.5 \times$ the level of the central Cepheid pixel.  For the images of Cepheids studied here this fraction is typically a few percent of the brightest Cepheids and sky pixels. To avoid a similar bias in PSF photometry, it is necessary to apply this same threshold when measuring the simple mean of artificial stars to determine the mean background correction.}. Fig.~\ref{fg:apvpsf} displays a comparison of the aperture and PSF photometry for the Cepheids as a function of PSF magnitude, showing agreement within the sample mean errors. 

We can make a few additional observations about the comparison that clarifies the relationship between the statistics used to measure the sky in aperture photometry and the crowded background corrections used to account for bias in PSF photometry.  Since PSF photometry sets the sky as a ``floor'' level added to modeled sources, the PSF sky level will be similar to the mode (i.e., the most common value or peak of the distribution of sky annulus pixels) of the sky pixels used in (uncrowded) aperture photometry, the uniform level of pixels without apparent sources.  Therefore, {\it the difference in aperture photometry calculated from a sky level using the mode or mean of the sky pixels will be similar to the crowded background correction in PSF photometry.}  In sparse fields, the mode and mean are equivalent and the crowded background bias is negligible.  

The asymmetric distribution of pixel levels in the sky annulus also explains the useful feature that in log-normal or magnitudes the Cepheid uncertainties which are dominated by the asymmetric distribution of sky pixels are relatively Gaussian to a few standard deviations, as shown in \S\ref{sc:3.3}.  

\subsection{Impacts of Improvements in Photometry}

As discussed in \S3.4, we itemize half a dozen improvements in Cepheid photometry from the last time these Cepheids were measured between $\sim 6$ and $\sim 16$~yr ago.  Matching Cepheids by position, the error-weighted mean of the matched Cepheids are fainter in SN hosts by 0.06~mag and in NGC$\,$4258 by 0.04~mag (a net difference of 0.02~mag when NGC$\,$4258 calibrates the Cepheids and H$_0$). For the LMC, Cepheids became fainter by 0.03~mag between the ground sample used by R16 and their replacement by direct {\it HST} observations in R19.  The mean change in {\it F555W--F814W} was below 0.01~mag for Cepheids in SN hosts and in the outer field of NGC$\,$4258 but 0.06~mag in the inner field relative to the previous calibration undertaken by \citet{macri06} (and provided in H16) which lacked accounting for the crowded background level, a larger effect in {\it F814W} than in {\it F555W}, and pixel-based CTE rectification.  We cannot report the change in photometry for MW Cepheids since we use different MW parallax samples here than by R16.  

For context, R16 propagated a zeropoint uncertainty between ground and {\it HST} system photometry of $\sigma=0.03$~mag, which is the same size as the changed observed for the LMC sample upon direct observation with {\it HST}.  The net change between the Cepheids in the SN hosts and NGC$\,$4258 or the LMC, the quantity which determines H$_0$, is 0.02, comparable to the systematic uncertainty in R16 of 0.026~mag for the NGC$\,$4258 anchor and smaller than the overall uncertainty in H$_0$ of 0.052~mag.  A more noteworthy change appears between the MW-only anchor results where the replacement with the {\it Gaia} EDR3 results plus {\it HST} photometry with the sample from \citet{benedict07} and their ground-based photometry reduced H$_0$ by 0.086~mag, with some of this change related to the aforementioned update of the Cepheid photometry in SN hosts, though the quadrature sum error for only the parallax samples is $\sim 0.06$~mag, making this change not surprising.  For additional context for the LMC, we note that the DEB distance also decreased by $\sim 0.015$~mag and the metallicity term by 0.033~mag (same sense) since R16.  These factors explain the net increase in H$_0$ for the LMC.   

\begin{figure}[t]  
\figurenum{B2}
\begin{center}
\includegraphics[width=0.6\textwidth]{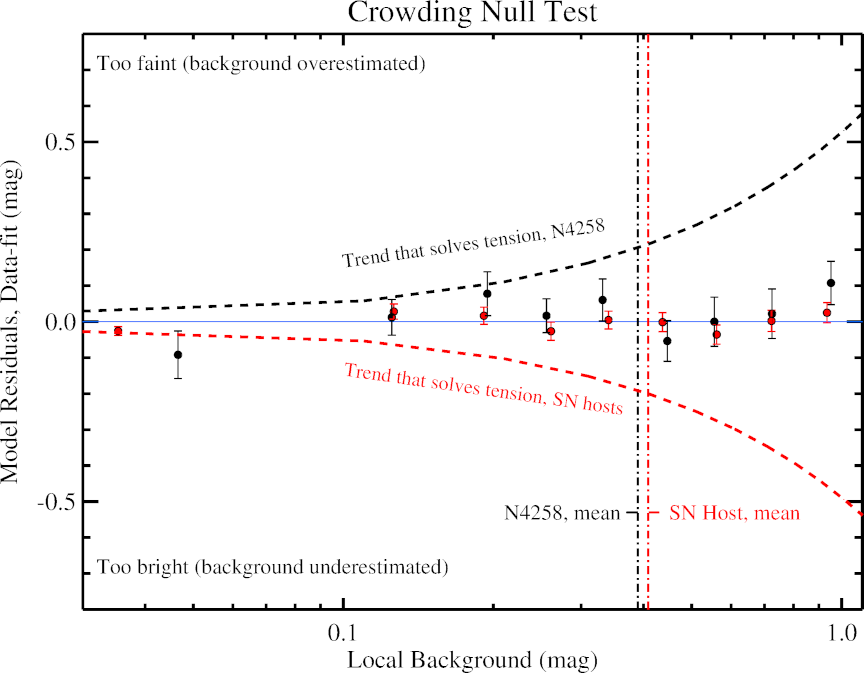}
\end{center}
\caption{\label{fg:vscrowd} Mean differences between Cepheid photometry and distance-ladder fit vs.~the local background measured by artificial stars (in units of the Cepheid magnitudes), binned in equal numbers of Cepheids. Red is for Cepheids in SN~Ia hosts and green for Cepheids in the anchor NGC$\,$4258 with vertical lines indicating the mean background of each set.  An underestimate of Cepheid backgrounds in SN~Ia hosts at a level of 0.2~mag for the mean Cepheid background {\it or} an overestimate of Cepheid backgrounds in NGC$\,$4258 by 0.2~mag for the mean Cepheid (indicated), or half that level for both {\it simultaneously}, would be required to solve the tension, but these possibilities are inconsistent with the data, which are consistent with no overestimates or underestimates for either set.  Furthermore, it is hard to imagine how an overestimate or underestimate would not similarly affect (and therefore cancel) Cepheids in both NGC$\,$4258 and SN~Ia hosts.}
\end{figure}

\subsection{Background Dependence of Fit Residuals}

An accurate assessment of the background flux is critical to the use of standard candles.  The Cepheid backgrounds are determined {\it statistically} by adding and measuring artificial stars in random positions local to the Cepheid scenes as described above. Here we provide an additional null test of the background estimates.

Since the individual backgrounds are determined locally as part of measuring the photometry, they are not part of the model.  A null test of the level of background is to analyze the correlation of background with the model residual of each Cepheid. If backgrounds are significantly, systematically overestimated or underestimated, we would expect a correlation in this space as the residual would be a function of the (misestimated) background.  We show the results of this test in Fig.~\ref{fg:vscrowd} for the baseline fit, binning the residuals in small ranges of background values.  These bins show no significant trend with background.  A linear fit gives a relation with slope $0.015\pm 0.014$~mag per mag of background in the sense of overestimating the background by $0.006\pm0.005$~mag at the mean background level in the SN hosts and in NGC$\,$4258 (as displayed in that figure). The figure also shows the expected trend if the background were underestimated enough to produce a 0.2~mag change in H$_0$ for the mean host (i.e., to solve the tension), which is very far from the data (and would also apply to the Cepheids in NGC$\,$4258 and thus not address the discrepancy for that anchor in Fit~10).

For those who would like to check the accuracy of independently-measured extragalactic Cepheid PSF photometry, we can provide (upon reasonable request) images of hosts with artificial Cepheids of known brightness and position added to the frames. Reproducing the known photometry would be an important test of any photometry algorithm that can be completed blindly and in advance of an independent determination of H$_0$.
      
\section{Cepheid Metallicity\label{sc:appc}}

Cepheid abundances beyond the Magellanic Clouds are generally derived from radial metallicity gradients measured from the ratios of strong emission lines in H~II regions, such as the $R_{23}$ strong-line diagnostic based on oxygen and hydrogen \citep[][hereafter Z94]{zaritsky94}.  The conversion of $R_{23}$ to 12 + log\,[O/H] in the Z94 formula was based on photoionization models which themselves were calibrated at a time when the standard solar metallicity was believed to be 12 + log\,[O/H] $=8.93$ \cite{Anders:1989}. Since \citet{Asplund:2005,asplund09}, the solar abundance was revised downward by 0.24~dex to $8.69\pm0.05$. Over the same period, empirical transformations between strong-line measures like $R_{23}$ were also revised downward by $\sim 0.2$--0.3~dex, making them more consistent with stellar measures referenced to the Sun \citep{Pettini:2004,Kewley:2008,Tremonti:2004,Dopita:2016,Curti:2017}. \citet{Bresolin:2016} have shown good agreement between extragalactic oxygen abundance measurements from young stars and from H~II regions using these more modern systems \citep[particularly for][hereafter, PP04 O3N2; Bresolin, priv. comm.]{Pettini:2004}. Collisionally-excited lines generally underestimate stellar abundances by 0.2~dex for reasons discussed by \citet{Carigi:2019}. Using 30,000--50,000 galaxy spectra from SDSS DR7, \citet{Teimoorinia:2021} derived third-order polynomial conversions between the Z94 scale and 9 different strong-line metallicity calibrations developed since 2004.  

For our measurement of H$_0$, we employ two strong-line abundance systems.  The first is the simple mean of all 9 recent (since 2004) diagnostics of \citet{Teimoorinia:2021}. This revises the Z94 scale down by a mean of 0.28~dex.  The change is not quite constant, as the mean produces flatter galaxy gradients than Z94, revising the Z94 values down an additional 0.05~dex at $>9.0$ and less at $<8.8$ by 0.04~dex. 

The second is the PP04 O3N2 diagnostic which uses the dereddened abundances of [O~III] $\lambda5007$ relative to H$\beta$ $\lambda4861$ and  [N~II] $\lambda6583$ relative to H$\alpha$ $\lambda6563$ to measure the log of the ratio of [O~III] to [N~II], O3N2, with the calibration 12 + log\,[O/H] $=8.73-0.32\times$O3N2. The difference between the Z94 diagnostic and the PP04 O3N2 diagnostic is very nearly constant, reducing the oxygen abundance scale by 0.23~dex (0.235~dex at Z94 $>9.0$ and 0.222 at $<8.8$).  The PP04 conversions of  \citet{Teimoorinia:2021}  are valid for oxygen abundance on the Z94 scale $>8.35$, encompassing all Cepheid hosts, with a typical stated precision of 0.03~dex.  This change not surprisingly mirrors the drop in solar abundance of 0.24~dex from the revisions of \citet{Asplund:2005,asplund09} and the related recalibration of strong-line diagnostics. For this reason it has made little difference for Cepheid measurements whether to use the prior Z94 calibration referenced to the \citet{Anders:1989} solar abundance or the revised (mean of) the recent systems referenced to the \citet{Asplund:2005,asplund09} solar abundance.   

\begin{figure}[t]   
\figurenum{C1}
\begin{center}
\includegraphics[width=0.45\textwidth]{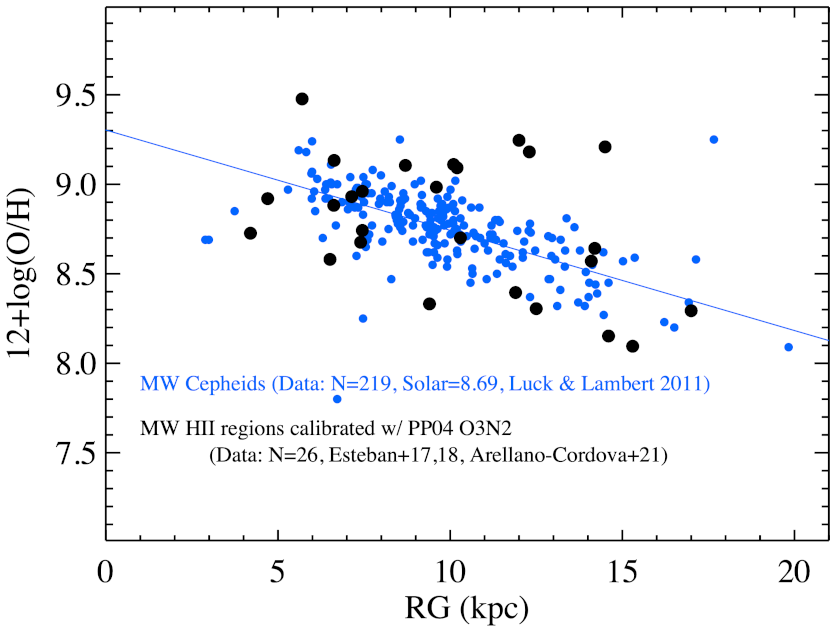}
\end{center}
\caption{\label{fg:mwabund} Comparison of oxygen abundance gradients in the MW inferred by spectra of Cepheids (blue) and from the PP04 calibration of H~II regions (black) with data sources indicated in the plot.  The mean levels are found to be consistent to $\sigma < 0.1$~dex and we propagate systematic differences in scales in the covariance matrix (as discussed in the text). }
\end{figure}

When calibrating Cepheids in SN~Ia hosts to those in NGC$\,$4258, both sets of metallicities are derived using the same strong-line abundance scale so that differences in abundance calibration do not affect the difference in abundance.  However, LMC and MW Cepheid abundances are generally measured through direct spectroscopy of the stars, so when these calibrators are employed it is important to test the consistency of direct and indirect methods for measuring Cepheid abundances. This is most readily accomplished in the MW where both measures are available.  \citet{Luck:2011} provide oxygen abundances for 219 MW Cepheids on the \citet{Asplund:2005} solar calibration of 8.69 from which they measured a gradient of 12 + log\,[O/H] = $9.303 (\pm 0.028)-0.056(\pm 0.003)$~dex~kpc$^{-1}$ as shown in Fig.~\ref{fg:mwabund}.

To compare to these, we derived the strong-line abundances from the three largest samples of MW H~II regions from \citet{Esteban:2017}, \citet{Esteban:2018}, and \citet{Arellano:2021}.  We used the tabulated values of [O~III] $\lambda5007$, H$\beta$, H$\alpha$, and [N~II] $\lambda6583$ in those references to calculate the ratio O3N2 and from that the PP04 calibration to determine 12 + log\,[O/H] as shown in Fig.~\ref{fg:mwabund}. The abundance gradient from the Cepheids as seen in that figure is a good match to the H~II regions.  A formal fit gives the PP04-stellar scale $=0.035\pm 0.06$~dex.  From the \citet{Teimoorinia:2021} transformations we find the mean difference between the PP04 and the 9 recent systems, PP04$-$ave$=0.047$~dex, demonstrating that the average of the 9 systems is right on the stellar scale to a precision of $\sim 0.06$~dex.  In \S\ref{sc:3.5}, Equation (\ref{eq:covmet}), we employ a metallicity covariance in the fit covariance matrix measured as the difference between the mean of the metallicity diagnostics and the PP04 O3N2 diagnostic which propagates both a 0.05~dex scale uncertainty, and also the differences which depend on the range of metallicity and rise to 0.1~dex at 12 + log\,[O/H] $=8.8$ (i.e., [O/H] $=+0.11$) and up to 0.15~dex at 9.0.  We also present the determination of H$_0$ on both reference systems as well for the exclusion of any dependence of Cepheid luminosity on metallicity to bracket the full range. 

\section{Use and Misuse of Wesenheit Magnitudes\label{sc:appd}}

The Wesenheit magnitude used to measure distances in Equation (\ref{eq:wh}) is not a conventional magnitude (i.e., not a ratio of flux to a standard star), but rather a ``dereddened magnitude'' \citep[see Appendix C of][for a review]{madore91}, the ratio two standardized candles with the same fiducial luminosity would be expected to have in the absence of reddening.  The Wesenheit formalism takes advantage of the candles having the same mean luminosity and color (before being reddened by dust) to preserve their difference in distance (in magnitudes) without requiring knowledge of either the (absolute) intrinsic luminosity or color. 

This method, and potential pitfalls with alternative uses of these magnitudes, can be better understood by decomposing the Cepheid color into its constituent parts.  We can define the apparent Cepheid color,
\bq V\!-\!I=(V\!-\!I)_0 + \Delta(V\!-\!I) + E({V\!-\!I}) , \eq 
where $(V\!-\!I)_0 $ is the color (at a given period) in the middle of the Cepheid instability strip,  $\Delta(V\!-\!I)$ is the displacement in color from the midline of the instability strip (with an accompanying displacement in mean luminosity owing to Stefan's law projected into the observed bandpasses, redder is fainter) and is quantified empirically in the Cepheid P--L--C relation.  For a statistical sample of Cepheids beyond period completeness (which provides Cepheids uniformly distributed over the instability strip, mixed by the comparatively small evolution time to cross the strip compared to the larger differences in formation times), the mean $\langle \Delta (V\!-\!I)\rangle =0$~mag.  Hence, the relative difference in distance (in magnitudes) between samples of Cepheids is statistically preserved by subtracting from the apparent magnitudes of each any fixed number, $R$, which multiplies the (same) intrinsic color of the midlines of their instability strips (i.e., $R\,(V\!-\!I)_0$).  If the value of $R$ is chosen to be the reddening ratio derived from an accurate reddening law of dust, then the subtracted term will also remove the absorption by dust (i.e., $R\, E({V\!-\!I})$) from each Cepheid for dissimilar values of $E({V\!-\!I})$.  This is the function of a Wesenheit magnitude.  A ``happy coincidence'' is that Stefan's law is quite parallel to the reddening law (redder is fainter or dustier), so this operation will also reduce the small dispersion due to the width of the instability strip \citep{madore91}.  As given in \S\ref{sc:6.3}, we derived the intrinsic PLC relation from the LMC Cepheids as above to be $\Delta H=0.635 (\pm 0.021)\ \Delta(V\!-\!I)$, the same sense and similar in scale to reddening by dust for which $R \approx 0.4$ .

It is not valid to use a different reddening ratio (law) in different hosts by simply varying the value of $R$ for each host using the same Wesenheit system.  Using a different value of $R$ for different Cepheids (or for Cepheids in different hosts) in Equation (\ref{eq:wh}) without first removing the intrinsic color from the apparent color produces a change to the measured differences in distance of Cepheid samples of size $R_1(V\!-\!I)_0-R_2(V\!-\!I)_0$, where $R_1 \neq R_2$ which is {\it unrelated} to dust but is rather a large, ad-hoc change in the standard candle luminosity itself as a fraction of its color.  Because the typical $(V\!-\!I)_0$ color is a factor of a few times larger than the typical $E({V\!-\!I})$, this change in Cepheid luminosity will dominate the dust correction and produce a bias that is a rather arbitrary fraction of a Cepheid's color.  This is a shortcoming in the analysis presented by \citet{Mortsell:2021}.

It is also important to recognize that Cepheids are well-understood yellow supergiant stars that will only pulsate over a very narrow range in surface temperature of $\log T = 3.7$ ($T=5000$~K), which sets their intrinsic color
 of $(V\!-\!I)_0 \approx 0.8$~mag (at $P=10$~d) at the instability strip of the temperature-magnitude diagram of stars.  This is the surface temperature where its internal structure supports pulsations by the kappa mechanism (change in internal opacity in a deep partial ionization zone from ionization and recombination of He+), so that the value of $(V\!-\!I)_0$ is well constrained by the physics which causes a Cepheid to be formed.  

Thus, varying $R$ in the Wesenheit definition does not appear to follow any specific hypothesis. To consider the hypothesis of different reddening laws in different hosts, it is necessary to first subtract the intrinsic color as described in \S\ref{sc:6.3} and by \citet{Follin:2017}.  

\subsection{Errors in Color and Fitting}

Because the value of $R$ is tightly constrained by the sample of MW Cepheids to $0.363\pm 0.038$ (and better together with the Cepheids in other nearby hosts such as M31 and M101), the error in color ($V\!-\!I$) can be propagated to an error in $H$ through $H \propto R(V\!-\!I)$ neglecting the small uncertainty in $R$ and added in quadrature in the covariance matrix of errors.  However, this is no longer true if $R$ is poorly constrained by the data and the errors in color are large, as would be the case if we tried to determine $R$ {\it independently for each individual SN host} after subtracting the intrinsic color.  Because the measurement errors in color are significant, to determine individual host values of $R$ without bias it is necessary to consider errors in both axes ($H$-band mag and $V\!-\!I$ color) to optimize $R$ as discussed for a similar problem by \citet{Tremaine:2002}, or follow the constrained approach of \citet{Follin:2017}.

\begin{figure}[b]   
\figurenum{D1}
\begin{center}
\includegraphics[width=\textwidth]{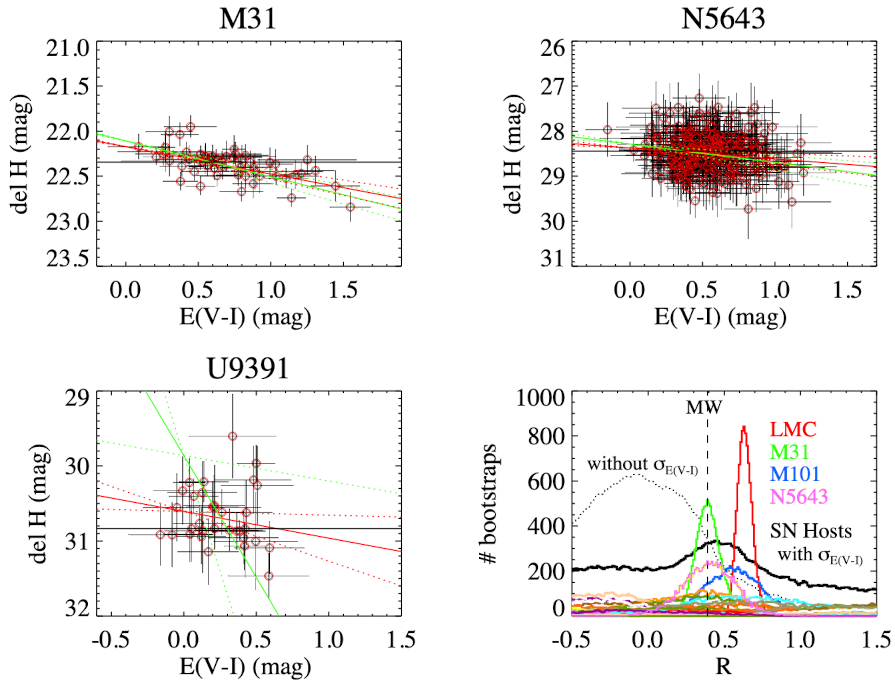}
\end{center}
\caption{\label{fg:twodcolor} Measuring the reddening ratio $R$ in each host. For each host we subtract the intrinsic period--color relation as given in the text to provide $E(V\!-\!I)$ and correct the $H$-band magnitudes for the \PLs relation (period and metallicity).  The ideal case, M31, has a large span of extinction (because it is highly inclined) and small measurement errors (because it is nearby) to produce a good constraint on the slope, $R$, whether we consider errors in only the dependent variable (red line and dotted for uncertainty) or both axes (green line).  $R=0$ is shown as a black line.  NGC$\,$5643 is one of the two nearest SN~Ia hosts and has the most Cepheids; it yields some constraint on $R$ along with M101. In contrast, the rest of the SN hosts have data similar to UGC$\,$9391, which produces an uninformative constraint on $R$ and shows a large difference in its value and uncertainty by neglecting uncertainties in both axes.  The lower right shows the bootstrap resampling for all hosts, with a different color for each, showing that a meaningful constraint is only seen for the few nearest cases with the others relatively flat.  The measurable cases are consistent with the MW $R$ shown as a black dashed line, as are the sum of all SN hosts.  Neglecting errors in $E(V\!-\!I)$ underestimates $R$ and its uncertainty, as discussed in the text.}
\end{figure}

Unfortunately, beyond the nearest few galaxies, the data are quite inadequate for constraining $R$ owing to the combination of a small range of color and comparatively large color uncertainties.  Fig.~\ref{fg:twodcolor} shows the available Cepheid data for a nearby host with considerable measurable extinction (M31), one of the two nearest SN~Ia hosts (NGC$\,$5643) for which the data are just sufficient, and for a typical SN~Ia host (UGC$\,$9391) where the data are insufficient.  Accounting only for uncertainty in the dependent variable produces a biased result and underestimates the uncertainty when the data quality is low, as is readily seen in Monte Carlo simulations compared to the input value of $R$.

In these cases the breadth of colors is largely attributed to the measurement errors rather than simply extinction. Accounting for uncertainty in both axes, we find that the value of $R$ is quite unconstrained beyond the nearest few hosts.  A bootstrap resampling of the data shows that the constraints on $R$ become uninformative beyond the nearest two SN~Ia hosts with a mean uncertainty for the rest of $\sim 1.5$.  The lower-right panel of Fig.~\ref{fg:twodcolor} shows the likelihoods for different host $R$ values.  We find that the nearest few large spirals have $R$ consistent with the MW, but there is no meaningful constraint or well-defined peak for the rest.  We can combine all such constraints as shown in the figure with a result that is unsurprisingly not well-defined.   \citet{Mortsell:2021} and \citet{Peri:2021} did not include the color errors, which our analysis shows leads to an underestimate of the values of $R$ and their uncertainties for nearly all SN hosts.  From a Bayesian perspective, the addition of a free parameter, $R$, for each host ($\sim 40$ free parameters) is not justified by the improvement in the fit over the global $R$, as also suggested in the analysis from \citet{Mortsell:2021} and \citet{Follin:2017}.  An informed approach to model individual host $R$ values without so many poorly-constrained free parameters can be found in \S\ref{sc:6.3}, Fit~22, and \citet{Hahn:2021}.

The EDA framework also provides estimates of the mean extinction seen midway through the SN hosts (i.e., the average location of a Cepheid) in our NIR bandpass from the same host properties used to determine $R$ above but now including the host inclination.  These values range from $A_H= 0.07$~mag (NGC$\,$4424 which at type Sa is the earliest spiral in our sample) to 0.30~mag with a mean of 0.18~mag and a dispersion of 0.05~mag.  This mean is similar to the empirical result of the product of the SN host mean $E(V\!-\!I)$ and $R$ which yields $\langle A_H\rangle=0.14$~mag, indicating that the level of extinction inferred from the Cepheid reddening is the expected amount\footnote{A more detailed comparison would require modeling of the Cepheid selection in the presence of variable depth of dust, and is beyond the scope here.}.

\end{appendices}

\clearpage 

\bibliographystyle{apj}
\bibliography{shoes6}
\end{document}